\documentclass[12pt]{iopart}
\usepackage{setspace} 
\usepackage{graphicx,psfrag}
\usepackage{amssymb}
\usepackage{iopams}
\usepackage{soul}
\usepackage{subfigure}
\usepackage{esint}
\usepackage{cite}

\usepackage{wasysym}
\usepackage{textcomp}
\usepackage[dvipsnames]{xcolor}
\begin{document}

\title{Active colloids in the context of chemical kinetics}

\author{G. Oshanin$^1$, M. N. Popescu$^{2,3}$ \& S. Dietrich$^{2,3}$}
\address{$^1$ Laboratoire de Physique Th\'{e}orique de la Mati\`{e}re Condens\'{e}e, UPMC,
CNRS UMR 7600, Sorbonne Universit\'{e}s, 4 Place Jussieu, 75252 Paris Cedex 05, France}
\address{$^2$ Max-Planck-Institut f\"ur Intelligente Systeme, 
Heisenbergstr. 3, D-70569, Stuttgart, Germany}
\address{$^3$ IV. Institut f\"ur Theoretische 
Physik, Universit\"at Stuttgart, Pfaffenwaldring 57, D-70569 
Stuttgart, Germany}
\eads{\mailto{oshanin@lptmc.jussieu.fr}, \mailto{popescu@is.mpg.de}, \mailto{dietrich@is.mpg.de}}

\begin{abstract}
We study  a mesoscopic model of a chemically active colloidal particle which 
on certain parts of its surface promotes chemical reactions in the surrounding solution. 
For reasons of simplicity and conceptual clarity, 
we focus on the case in which 
only electrically neutral species are present in the solution and 
on  chemical 
reactions which are described by first order kinetics.
Within a self-consistent 
approach we explicitly determine the steady 
state product and reactant number density fields around the colloid as functionals 
of the interaction potentials of the various molecular species in solution with 
the colloid. By using Teubner's reciprocal theorem, this allows us to compute and 
to interpret -- in a transparent way in terms of the classical Smoluchowski theory of 
chemical kinetics -- the external force needed to keep such a catalytically active 
colloid at rest (\textit{stall} force) or, equivalently, the corresponding velocity 
of the colloid \textit{if} it is free to move. We use the particular case of 
triangular-well interaction potentials as a benchmark example for applying 
the general theoretical framework developed here. For this latter case, we derive 
explicit expressions for the dependences of the quantities of interest on the 
diffusion coefficients of the chemical species, the reaction rate constant, 
the coverage by catalyst, the size of the colloid, as well as on the parameters of the interaction
potentials. These expressions provide a detailed picture of the phenomenology associated with 
catalytically-active colloids and self-diffusiophoresis.
\end{abstract}


%\keywords{abc}


%\noindent\MNP{Note: new references, \cite{Howse2014}, \cite{Uspal2016} and \cite{Graaf}, were added and the rest re-indexed accordingly.}

%\submitto{\jpa}
%\tableofcontents

\maketitle

%\tableofcontents

\section{\label{intro}Introduction}

The reduction in length scales brought about by lab-on-a-chip applications 
has raised a number of challenging issues. One of them is how to enable small 
objects to perform autonomous, directional motion (so-called self-propulsion) 
in a liquid environment
\cite{Ismagilov2002,Paxton2004,ozin2005,Paxton2006,Lauga2009,solovev2009,
mirkovic2010,Ebbens2010}. For particles of micrometer size or smaller viscous and surface
forces dominate the effects of inertia. Therefore, in order to achieve motion for 
them, new strategies have to be developed, which differ from those applicable for 
macroscopic objects \cite{Paxton2006,Lauga2009,Ebbens2010,Golestanian2005}.

\subsection{Active particles}

Currently two main routes are followed towards this goal. The first one is based on 
mimicking the ingenious mechanical locomotion strategies of natural micro-organisms 
such as \textit{E. Coli} or \textit{Spiroplasma} \cite{Lauga2009}. These objects move 
by undergoing deformations of their bodies in a cyclic, non-reciprocal manner and by 
exploiting the anisotropy of the hydrodynamic drag. A thorough, extensive discussion 
of the developments in this area can be found in the recent reviews by Lauga and 
Powers \cite{Lauga2009}, Elgeti, Winkler, and Gompper \cite{Gompper2015}, Z\"ottl and Stark \cite{zoettl2016}.

A promising alternative is that of employing catalytically activated chemical reactions 
to extract ``chemical'' free energy from the surrounding liquid environment and to 
transform it into mechanical energy (i.e., motion of microparticles), resembling the way 
in which biological molecular motors function. One class of mechanisms 
to achieve this builds on ideas borrowed from classic phoresis, i.e., colloid motion in 
externally maintained gradients of certain thermodynamic fields \cite{Anderson1989}. 
The basic idea to achieve \textit{self}-phoresis is to use particles which are partially 
covered by a catalyst in a well controlled way, designed such that symmetry breaking 
occurs and the particle is endowed with a ``polar'' axis. The catalyst activates a chemical 
reaction in the surrounding 
solution \cite{Ismagilov2002,Ebbens2010,Golestanian2005,Catchmark2005}, generating 
gradients of reaction products across the surface of the colloidal particle. Owing to the
solvent mediated, effective interactions between the reaction product molecules and the 
surface of the particle, within a thin interfacial layer a gradient in the osmotic 
pressure along the surface of the particle emerges. This leads to hydrodynamic flow of the 
solution around the particle and sets it into motion\footnote[1]{We note that 
there are also other mechanisms to convert chemical free energy into mechanical motion, 
such as electrochemical pumping employed by bi-metallic rods used in early 
experiments dealing with active colloids \cite{Paxton2004,Paxton2006,Paxton2006a}, 
thermally induced demixing of a binary liquid mixture near its critical temperature 
\cite{Bechinger2012}, or the ''micro-jet-engines'' which rely on a bubble-pumping mechanism 
\cite{solovev2009,Sanchez2011,Sanchez2011jacs,Sanchezjacs2011_b,Solovev2010,
Wang2012,Wang2011}. At low Reynolds numbers, the latter has similarities with 
the
mechanisms discussed theoretically in Ref. \cite{Lauga2010}. Yet another example
is that of propulsion through pressure waves generated via catalyst activated
chemical reactions in the surrounding fluid \cite{Felderhof2010}.}. Furthermore, in addition 
to the mechanical swimmers and chemically powered active colloids discussed here,
self-thermophoresis \cite{Sano2010} and photophoresis in combination with radiation-pressure
\cite{Cichos2013} have recently been proposed as alternative self-propulsion mechanisms. 
(In the scenario above the requirement, that only a portion of the surface is catalytically active, 
is necessary for particles having axial- and fore-aft symmetry. For a shape missing the 
latter, self-phoresis may occur even if the whole surface of 
the particle is covered by catalyst. For example, for a uniform reaction rate (i.e., a constant 
flux of product molecules) across the surface of the particle, geometrical shape asymmetry 
is sufficient to induce an inhomogeneous number density of reaction products along the 
surface and therefore propulsion \cite{Brady2014,Michelin2015}.) 

Several proof-of-concept proposals of such chemically self-propelled objects, employing 
either self-phoresis
\cite{Catchmark2005,Kline2005,Fournier-Bidoz2005,Paxton2006a,Hong2007,Howse2007,
Erbe2008,Laocharoensuk2008,Baraban2012,Sen2011} or bubble-pumping
\cite{solovev2009,Sanchez2011,Sanchez2011jacs,Solovev2010,Wang2012,Wang2011}
mechanisms, have already been successfully tested experimentally. Because of the
intrinsic non-equilibrium character and due to the subtle combination of physics and 
chemistry behind this type of self-induced motility, many of the experimental
\cite{Paxton2006,Ebbens2010,Paxton2006a,Howse2007,solovev2009,Sanchez2011,Wang2012,
Wang2011,Ajdari2006} and theoretical  
\cite{Golestanian2005,Golestanian2007,Julicher2009,Ruckner2007,Tao2008,Popescu2010,
Popescu2011,Ajdari2006} studies have so far been focusing on systems which can be approximated as 
being \textit{unbounded}.\footnote[2]{See, however, the recent Refs.
\cite{Popescu2009,Crowdy2013,Kaiser2012,Kaiser2013,Lefauve2014,Menzel2013,Berke2008,
Pototsky2012,Pototsky2013,tenHagen2009,Lee2013,Takagi2014,Marsden2014,Volpe2011,
Elgeti2013,Hsu2010,Volpe2013,vanTeeffelen2009,Brotto2013,Keh2013,Spagnolie2012,
Ishimoto2013,Uspal2015a,Uspal2015b,Wurger2014,Lauga2011,Stone2014,Alvaro2016,Howse2015,
Simmchen2016,MPD16,koplik2016,Uspal2016} which address the issue of confinement 
effects on self-propulsion.}
Recent experimental work has addressed new issues, such as improving the
directionality of the motion against Brownian diffusion either by using external 
magnetic fields and paramagnetic core particles
\cite{Kline2005,Erbe2008,Solovev2010,Sanchez2011,Wang2011}, or by exploiting the 
emergence of surface-bounded steady states of motion as they can occur in the vicinity of hard 
walls \cite{Howse2015,Simmchen2016}, or by using a
collection of chemically active particles as an artificial, 
non-biological system in order to study chemotaxis \cite{Hong2007}.

One interesting theoretical approach consists of describing an active particle 
in terms of an effective Langevin equation 
\cite{Lowen2011jpcm,Lowen2011pre,Stark2011pre,Stark2011prl,Stark2012pre,Stark2012prl}. 
This has been, however, critically examined experimentally by Bocquet et al who have
pointed out that in these out-of-equilibrium systems the notion of an ``effective
temperature'' and the mapping onto systems described by equilibrium statistical mechanics
are problematic \cite{Boquet2010,Boquet2012}. The intriguing question concerning the
extent to which similarities in collective behavior between active colloids and 
interacting Brownian particles can be captured by mappings with effective thermodynamic
parameters has been recently scrutinized in Ref. \cite{Cates2013}. This latter study
established criteria under which certain models of ``run-and-tumble'' particles are
equivalent with the models of active particles developed in Refs.
\cite{Lowen2011jpcm,Lowen2011pre}. However, it is important to note that 
in general, such systems cannot be described in terms of an equation 
of state \cite{Tailleur2014_1,Tailleur2014_2}.

The hydrodynamic interactions between active swimmers have been shown to give 
rise to a rich behavior, such as synchronization of the swimmers 
\cite{Yeomans2007,Yeomans2011}. Accordingly, studies of simple models for the collective
dynamics of active particles are enjoying considerable attention 
\cite{Lowen2012prlA,Lowen2012prlB,Lowen2012PNAS,Boquet2012,Stark2012pre,
Stark2011prl}.

Returning to the framework of ``single-colloid'' systems and the emergence of
self-phoresis, we note that a microscopic modeling of the processes in the interfacial
region, explicitly taking into account the molecular structure of that region
around a self-propeller, has been performed in Refs. \cite{Ruckner2007,Tao2008}, and more
recently in Ref. \cite{maldarelli}, by using molecular dynamics simulations. This approach
allows one to keep track of the molecular details for all species at the expense, however, 
that only relatively small systems and short time scales can be accessed. A different 
strategy, pursued in Refs. \cite{Julicher2009,Golestanian2005,Golestanian2007,Popescu2010,Popescu2009,Graaf}, 
is based on a continuum description of these active particles, transferred from the 
classical theory of diffusiophoresis \cite{Anderson1989}. Although this approach involves 
numerous assumptions, which are discussed in detail in 
Refs. \cite{Julicher2009,Popescu2010,Popescu2009,Seifert2012,Graaf}, a qualitative agreement
between the theoretical predictions and available experimental results has been reported
\cite{Howse2007,Erbe2008}. On the other hand, a strong sensitivity of the self-propulsion velocity 
even on tiny
traces of salt in the solution \cite{Howse2014} seems to indicate a more complex picture of the motion mechanism.
Furthermore, for a ``dimer'' model of an active particle 
(an active spherical colloid and an inert one of a different radius, connected via an 
infinitely thin rigid rod) quantitative agreement has been reported between this continuum 
description and the results of particle based numerical simulations \cite{Kapral2015}.
Recently, a theoretical formulation of self-diffusiophoresis within the framework of 
classical non-equilibrium thermodynamics has been proposed \cite{Julicher2009,Seifert2012}. 
The influence of the details of the chemical reaction and of the transport of the reactants 
and products on the emerging self-phoretic motion has also been studied recently 
\cite{Ebbens2012_soft,Golestanian2012,Seifert2012,Kapral2011_JCP,Kapral2011_AngChem}. 

In the context of chemically active particles, the question of the detailed structure 
of the non-equilibrium steady-state density distributions around active particles, which is
central to the present study, has been investigated in Refs. \cite{Ruckner2007,Tao2008}, as
well as recently in Refs. \cite{Seifert2012,Seifert2012b,Kapral2011_JCP,Koplik2013}. 
Focusing on spherical colloids with catalytically active spherical caps, in Refs.
\cite{Seifert2012,Seifert2012b,Koplik2013} the number density fields have been calculated
for various types of interactions (hard core, exponentially decaying, power law resembling
van der Waals interactions). These calculations have been carried out as perturbation series 
in the Peclet number, which for these active colloids is typically small (see, c.f., Sec.
\ref{model}), and in the range of the interactions between the reactive species and the
colloid at which these are significant with respect to the thermal energy (e.g., more 
than a percent of the latter). These ranges are assumed to be short compared to the size 
of the colloid. (If this assumption does not hold, carrying out numerical simulations 
seems to be the only option available \cite{Seifert2012,Koplik2013}.)  In principle, 
this systematic expansion allows one to estimate the accuracy of the approximation caused 
by truncating the expansion at a certain order, but it has the drawback of a dramatic 
increase in the difficulty of obtaining closed form expressions beyond the first or 
second term in the expansion. For the dimer-sphere model with one catalytic sphere rigidly 
connected to an inert one, in Refs. \cite{Ruckner2007,Tao2008,Kapral2011_JCP} a Boltzmann-like
\textit{ansatz} was employed for the non-equilibrium distribution of reactants and
products. The prefactor of these distribution was chosen to be the corresponding solution
of the steady-state diffusion equation in the absence of interactions (other than the
impenetrability of the \textit{active} sphere) and with the surface of the catalytic
sphere acting as a sink for the reactants and a source for the products. Under the
assumption that the influence of the inert sphere on the diffusion is negligible, the
reactant and product distribution profiles have been calculated. Obviously, in this case
the accuracy of these approximations cannot be assessed intrinsically but only \textit{a
posteriori} via comparison with results from experiments or numerical simulations. For
example, in Ref. \cite{Kapral2011_JCP} it has been shown that such \textit{approximate
analytic} results are in good agreement with data obtained from direct MD simulations. We
note here, however, that as discussed in Ref. \cite{Popescu2011} it is important to
account for the distortion of the density distributions due to the impenetrability of
the inert sphere -- in particular if the active sphere is only partially covered by
catalyst and the two spheres do not touch -- in order to fully understand how the
inert sphere facilitates or hinders the emerging motion of the dimer.

\subsection{Outline of the paper}

Here we study in detail a mesoscopic model of an active colloid with a catalytic 
patch on its surface (see Fig. \ref{fig1}), so that the reactants, diffusing in a
(\textit{chemically passive}) solvent, react at this part of the surface by 
converting themselves into (diffusive) reaction products. The initial distribution 
of the reactants 
is taken to be homogeneous, but in the course of time and near the colloid a depletion
zone for the reactants can emerge. We first focus on a simple reaction process in which 
a reactant $A$ converts itself into a single product molecule $B$, but further on we
consider also the more general dissociation reaction $A \to B + C$, in which two product
molecules $B$ and $C$ emerge upon contact of an $A$ molecule with the catalytic patch. We
note that this latter process directly connects with, e.g., the dissociation of hydrogen
peroxide into water and oxygen which has often been used in experimental realizations
of chemically automotive particles (see, e.g., Refs.
\cite{Paxton2006,Howse2007,Baraban2012}.) 

We focus on understanding the steady-state structure of the number density fields 
of the product(s) and reactants near the active colloid by taking into account 
that reactants and products may have different diffusion coefficients and different 
interaction potentials with the colloid. Distinct from the related previous
approaches in Refs. 
\cite{Ruckner2007,Tao2008,Seifert2012,Seifert2012b,Kapral2011_JCP,Koplik2013}
discussed above, in order to calculate these non-equilibrium steady-state distributions 
we adopt an idea proposed previously and studied in the context of calculating chemical 
reaction rate constants \cite{shoup2}. By replacing therein the position dependent sink 
(for the reactant) and source (for products) boundary conditions on the surface of the 
active particle with certain effective ones, a \textit{completely analytical} calculation 
of these steady-state density profiles can be carried out for quite general types of 
interactions between the diffusing species and the colloidal particle. These results allow 
us to compute that \textit{part} of the force experienced by the colloidal particle which 
is solely due to the self-generated non-equilibrium spatial distributions of the reactive 
species. (We note that, on the other hand, the same interactions generate a force of the 
colloid acting on the reactive species which via their interaction with the solvent, 
encoded \textit{inter alia} in the viscosity of the solution, transmit body forces acting 
on the solvent, inducing hydrodynamic flow and thus in turn a hydrodynamic force on the 
colloid (see, c.f., Sec. \ref{force}). At steady-state, and in the absence of external 
forces or torques acting on the colloid or on the fluid, this second \textit{part} of the 
force experienced by the colloidal particle exactly cancels the first one, and therefore 
the net force on the colloid vanishes.) This part of the force can be used to calculate 
either the hydrodynamic flow of the solution (i.e., the pumping strength) induced by the 
active colloid (if the colloid is spatially fixed, e.g., by optical tweezers) or the 
velocity of the colloid, if the colloid is free to move. Our analysis is reliable as long 
as in the steady state the diffusion of the reactants and product molecules -- here and 
in the following referred to also as ``solutes'' -- and the hydrodynamics in the system 
are such that the Peclet number (i.e., the ratio of the displacement of the solutes due 
to convection to that due to diffusion) and the Reynolds number (i.e., the ratio of the 
inertial to the viscous contribution to the hydrodynamic flow of the solution) are small 
(see Sec. \ref{model}). This implies that we disregard any change of the density 
distributions due to the flow of the solution. In this case, the diffusive and the 
convective (i.e., hydrodynamic) transport are effectively decoupled\footnote[9]{An 
analysis of the motion of a self-propelled colloid accounting for a non-zero Peclet number 
can be found in Refs. \cite{Lauga2013,Michelin2014}.}. The dependence of these quantities 
(i.e., the velocity of the colloid, if it is free to move, or the hydrodynamical pumping 
of the solution generated by an immobilized colloid) on the size of the colloid emerges 
naturally from the interplay between the reaction and diffusion constants, generalizing 
the results reported in Refs. \cite{Golestanian2012,Manjare2014}. Therefore these results 
for the non-equilibrium spatial distributions of the reactive species -- besides being of 
interest in their own right -- can be immediately adopted in order to calculate the 
steady-state velocity of the self-propelled particle.

As a by-product of our analysis, in the case of triangular-well interaction potentials 
between the reactants and the colloidal particle we obtain several new results for 
effective reaction constants which, within the framework of the classical Smoluchowski 
theory of chemical kinetics \cite{smoluchowski}, describe the reaction rates occurring 
at particles with inhomogeneous catalytic activity at their surface.

In the context of active particles, the detailed calculations and analyses noted above are motivated, \textit{inter alia} by the following points:
\begin{itemize}
 \item Most of the experimental studies of active particles have employed 
 microscopy in order to trace the particle motion and to extract the corresponding 
 velocity. However, carrying out such studies for a three-dimensional, bulk motion of 
 active particles and determining their velocity $V_{free}$ is very challenging. 
 In Sec. \ref{connect_move} we show that 
 measurements of the stall force, e.g., by employing an optical trapping of the particle, 
 provide the same information as the knowledge of the velocity $V_{free}$ in the bulk.
 In Sec. \ref{force} the results of such measurements are related to detailed 
 properties of the system, such as the coverage by catalyst, the interaction potentials between the particle and the molecular species, the rate of reaction, and the diffusion constants.
 \item In a stall configuration (``pumping'', i.e., although the particle is motionless, 
 the fluid is in motion because of the reactions at the surface of the particle and of 
 the interactions between molecular species and the particle), the stall force is 
 balanced by a contribution solely due to the interactions between molecular species 
 and the particle and a hydrodynamic contribution due to the flow. The latter can be inferred 
 eventually by mapping the flow around the particle using particle image velocimetry. 
 If it is possible to independently map the distribution of chemical species around 
 the immobilized colloid (see, e.g., Ref \cite{Espandiu2013}), the expressions for 
 $F_{chem}$ in Secs. \ref{connect_move} and \ref{disc} in connection with specific 
 models for the interaction potentials allow one to estimate the parameters 
 characterizing the potentials.
 \item If the interaction potentials between the various molecular species and the 
 particle are known, measurements of the stall force for particles of different radii 
 $R$ provide the means, via the predicted dependence of the stall force on the radius 
 (Sec. \ref{force}), for a critical test of the model assumptions for the mechanism 
 of reaction and motion. Furthermore, by repeating such experiment at various 
 temperatures of 
 the system, and thus eventually exploring the crossover regime between the 
 kinetically-controlled and the diffusion-controlled ones, the confidence limits of the 
 model can be tightened.
\end{itemize}

The outline of the paper is  as follows. In Sec. \ref{model} we define our model for a
translationally and rotationally immobile, chemically active colloid  immersed in an 
unbounded system with randomly dispersed diffusive reactants. Section \ref{conc_distr} is 
devoted to calculate the stationary density profiles of mutually non-interacting $A$ 
(reactant) and $B$ (product) molecules around the immobile colloid with a reactive patch 
covering partly its surface. 
  There we also describe the self-consistent approximation used to determine
 the coefficients in the expansions of the density profiles in terms of eigenfunctions. 
 Extending our approach towards the general case of a dissociation reaction (such as, e.g., $A \to B + C$) is straightforward (see \ref{dissoc_reaction}).
 In Sec. \ref{force} we first 
calculate $F_{chem}$ -  that part of the force exerted on the immobile colloid which is due to its 
molecular interactions with the solvent and the solute molecules (i.e., species $A$, $B$, etc.), 
%and $C$, \GO{if present}), 
as well as that due to the hydrodynamic flow of the solution induced by the 
spatially non-uniform density profiles of the reactant and the products. Furthermore, 
by employing the reciprocal theorem in the presence of distributed body forces on the fluid 
\cite{KiKa91,Teubner1982}, we 
% Moreover, we show 
% how by using the reciprocal theorem due to Teubner \cite{Teubner1982} one can 
obtain an explicit expression for the self-propulsion velocity ${\bf V}$ of the colloid if it is free to move. Moreover we also derive an explicit expression for the stall force necessary to immobilize the active colloid which otherwise would move freely. 
%Here, we also present 
In Sec. \ref{disc} these results are 
%employed to explicit results for the force exerted on an immobile colloid and 
%for the self-propulsion velocity of a freely moving colloid 
%in the presence 
discussed further for the particular case of triangular-well interaction potentials. 
There, we present explicit expressions for the chemical force, the stall force, 
and the self-propulsion velocity as functions of the parameters of the 
triangular-well interaction potentials, the diffusion coefficients, the radius 
of the colloid, the size of the catalytic patch, and the chemical rate constants. The 
dependences on these parameters are discussed in detail and 
we highlight 
several 
counter-intuitive effects, such as that 
(i) 
for certain parameters $F_{chem}$ and ${\bf V}$ may have different signs; (ii) 
${\bf V}$ is a non-monotonic function of the value of the interaction potential 
at the surface of the colloid, and (iii) the maximal self-propulsion velocity, 
as function of the size of the catalytic patch, is in general attained at 
coverages by catalyst being smaller than one half. We summarize the results in 
Sec. \ref{concl} along with our conclusions. Certain important derivations are 
collected in the Appendices A - D. In Appendix E we provide a summary of our 
notations which is meant to ease following the text.

\section{\label{model}Model}

\begin{figure}[!t]
\begin{center}
\includegraphics[width = .4 \textwidth]{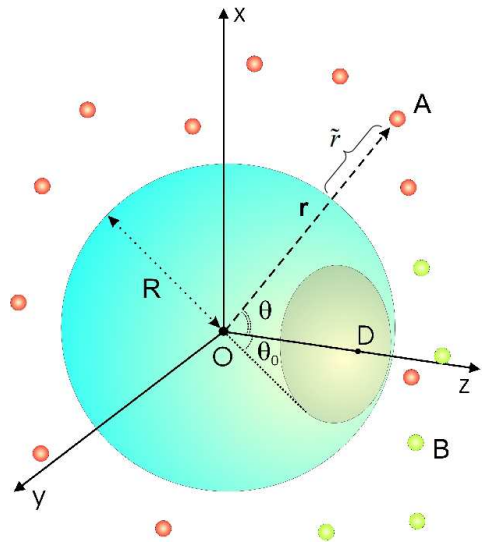}
\caption{An immobile spherical colloid of radius $R$ is centered and fixed at the 
origin $O$. A spherical cap with opening angle $\theta_0$ and centered around the 
point $D$ acts as a catalytic ``patch'' on the surface of the colloid. The $z$-axis 
runs through $D$. $A$ and $B$ are the reactant and product molecules, respectively. 
In view of their low number densities, they are considered to be pointlike. The 
position vector $\mathbf{r}$ with $|\mathbf{r}| = r$ of a molecule $A$ forms an angle 
$\theta$ with the symmetry axis. The molecule $A$ has a distance $\tilde r$ from the 
surface of the colloid so that $\tilde r = r-R$. The solvent molecules are not shown; 
they form a homogeneous background acting as a heat bath. The solvent is passive with 
respect to the chemical reaction.
\label{fig1}
}
\end{center}
\end{figure}

We consider a macroscopically large reaction bath of volume ${\cal V}$ in which 
a spherical colloidal particle of radius $R$ is immersed. We choose the origin of 
the coordinate system to be at the center of the colloid (see Fig. \ref{fig1}). We 
consider two situations: either an immobile 
colloid (e.g., held by an optical tweezer or a very thin tethered spring), or an 
unconstrained colloid which at steady state moves with velocity $\mathbf{V}$ through the
solution. (In the absence of thermal fluctuations, due to the axial symmetry of the system
the axis of the colloid does not rotate.) In the latter case, the coordinate system is
attached to (and thus co-moving with) the colloid.
%%%%%%%%%%%%%%%%%%%%%%%%%%%%%%%%%%%%%
%%%%%%%%%%%%%%%%%%%%%%%%%%%%%%%%%%%%%
The reaction bath consists of solvent molecules $S$, reactant molecules $A$, 
diffusing with diffusion coefficient $D_A$, and (after some time) product molecules $B$, 
diffusing with diffusion coefficient $D_B$. The reactant molecules $A$ (molecular mass
$m_A$), the product molecules $B$ (molecular mass $m_B$), and the solvent molecules 
(molecular mass $m_S$) are much smaller than the colloid, so that the solution can
be approximated as a continuum, apart from density oscillations in the molecular vicinity 
of the colloid surface. The number densities (number of molecules per volume) of $A$ 
and $B$ molecules are taken to be very small compared with the number density of the 
solvent; therefore we assume that there are no $AA$, $BB$, and $AB$ interactions. The
interactions between the $A$ ($B$) molecules and the solvent are encoded, together
with those among the solvent molecules, \textit{inter alia} in the viscosity $\mu$ of the
solution treated as a continuous medium; they enter as well, e.g., via the
Stokes-Einstein relation, into the corresponding diffusion coefficient $D_A$ ($D_B$). The
solution is assumed to behave as an incompressible, Newtonian fluid.

A spherical cap of the surface of the colloid particle, lying within the angular region 
$0 \leq \theta \leq \theta_0$ centered around D (see Fig. \ref{fig1}), has the following
``catalytic'' properties: As soon as an $A$ molecule approaches the surface of the colloid 
at any point belonging to the catalytic patch (CP), within a minimal distance $a$ (where 
$a \ll R$ is of the order of the size of the $A$ molecule), it undergoes with probability 
$p$ an irreversible reaction (conversion) into a product molecule $B$:
\begin{eqnarray}
\label{reaction}
A + {\rm CP} \stackrel{p}{\longrightarrow} B + {\rm CP}\,.
\end{eqnarray}
The probability $p$ characterizes the elementary reaction act: $p = 0$ means that no 
reaction occurs, while for $p \to 1$ one has a ``perfect'' reaction, i.e., $A$ 
turns into $B$ upon any encounter with the surface. We assume that the catalyst is neither 
consumed by the reaction in Eq. (\ref{reaction}), nor temporarily ``passivated'' by an
intermediate product, and thus at any time it can participate actively in an unlimited 
number of chemical transformations. The solvent, acting as a heat bath, serves as a
source or a sink of heat if the catalytic reaction is endothermic or exothermic, 
respectively.  
For reasons of conceptual clarity, we consider only the 
case that all species in the solution ($S$, $A$, $B$, etc.) are electrically neutral. The extension to the case of ionic species, which requires to consider charge conservation and its coupling to mass transport, eventually can be 
treated similarly to the cases discussed in, e.g., 
Refs. \cite{Seifert2012,Seifert2012b,Graaf,Howse2014}.

In what follows we shall also consider a more general reaction scheme, corresponding to  a 
catalytically induced dissociation reaction of the form
\begin{equation}
\label{react_dissoc}
 A + {\rm CP} \to B + C + {\rm CP} \,
\end{equation}
in which the reactant $A$, upon contact with the catalytic patch, breaks up into a 
pair of products $B$ and $C$ (with the latter two, in the general case, being different 
from the solvent molecules) created at a certain distance apart of each other. 
The diffusion coefficient of the $C$ molecules will be denoted as 
$D_C$. Such a reaction is inspired by Pt-catalyzed dissociation of hydrogen 
peroxide into water and oxygen molecules, which has been used in many 
experimental realizations of self-propelling particles \cite{Ebbens2012_soft}. 
In the case of peroxide decomposition, if the solvent $S$ is water the 
product molecules $C$ are also water molecules so that the reaction produces 
excess solvent. The excess amount of solvent, which emerges in the course of this
reaction is small because it is controlled by the concentration of $A$ molecules 
(hydrogen peroxide) which is small.
In this particular case the reaction scheme 
in Eq. (\ref{react_dissoc}) reduces to the simpler one in Eq. (\ref{reaction}).

The initial state of the system, i.e., before the immersion of the chemically active 
colloid (or before switching on its catalytic activity), is that of a well stirred 
reaction bath: the $A$ molecules are uniformly distributed within the reaction bath with 
mean number density $n_0^{(A)} = n_0$. (But in the course of time, upon turning on
the reaction, a non-uniform concentration profile can emerge.) Initially the $B$ %\GO{(and $C$)}
molecules are absent in the system. The bath is in contact with a reservoir of $A$
molecules which maintains this density $n_0^{(A)} = n_0$ at distances far away from
the colloid at all times. 
We furthermore assume that the reaction bath is also in contact with a perfect sink of 
$B$ 
%\GO{(and $C$)} 
molecules so that $n_0^{(B)} = 0$
% \GO{(and $n_0^{(C)} = 0$)} 
 far away from the colloid at all times; together 
with the reservoir for $A$ molecules, this ensures that a steady state can be maintained 
in the bath if the reaction is active. As mentioned before, the mean number densities 
of the $A$ and $B$ 
%(and $C$) 
molecules are assumed to be, at all times, much smaller than the mean 
number density $n_0^{(S)}$ of the solvent molecules; therefore the latter is 
assumed to 
be at all times approximately equal to its initial value $n_0^{(S)}$ before the reaction
is turned on. The colloid is impermeable to any of the molecules ($A$, $B$, 
%\GO{$C$, if present,} 
or $S$) in 
the mixture.

The solvent molecules interact among themselves and with the $A$ and $B$ 
%\GO{ (and $C$, if present)}  
molecules.
As noted above, these interactions are, \textit{inter alia}, accounted for by the
viscosity $\mu$ of the mixture and by the diffusion constants $D_A$ and $D_B$ 
%(and $D_C$)
. They also 
interact with the colloid via a radially symmetric interaction potential 
$\tilde\Phi_S(\tilde r)$ (per molecule of solvent), in addition to the hard core repulsion 
accounting for the impermeability of the colloid. Due to the spherical symmetry, for a 
molecule located at $\mathbf{r}$ this distance from the colloid is given by $\tilde{r} = 
r-R$ (see Fig. \ref{fig1}) and thus for convenience we introduce the radially symmetric 
potential $\Phi_S(r): = {\tilde \Phi}_S(r-R)$. We account for the hard-core repulsion via 
$\Phi_S(r \leq R) = \infty$. (Note that in the latter condition the molecules are implicitly 
treated as pointlike particles because of their size being much smaller than $R$.) As 
already noted in the Introduction, we assume that the solvent is passive with respect to 
the reaction and that it acts as a macroscopic heat bath, which keeps the temperature 
constant even in the presence of the reactions.

Similarly, the $A$ and $B$ 
%(and $C$) 
molecules interact with the colloid via an interaction
potential $\tilde \Phi_{A,B}(\tilde{r})$ (per molecule), respectively, in addition to the
hard core repulsion which makes the colloid impermeable for the molecules in the mixture.
We introduce the radially symmetric potentials 
$\Phi_{A,B}(r): = \tilde \Phi_{A,B}(r-R)$
and account for the impenetrability of the colloid via 
$\Phi_{A,B}(r \leq R) = \infty$.
Thus we disregard the possibility that the catalytic patch introduces an angular
dependence on $\theta$ of any of the interaction potentials 
$\Phi_{A,B,S}(r)$ between
the molecules and the colloid, as well as the influence of the eventual size 
disparity
between the three types of molecules on the location of the ''hard wall`` surface of the
colloid (see also the succinct discussion below). For reasons of having simple 
notations later on (see, c.f., Sec. \ref{conc_distr}), we also introduce here the 
interaction potentials of the $A$ and $B$ 
%\GO{and $C$} 
molecules with the colloid relative to the 
corresponding potential for the solvent molecules weighted by the ratio of the 
mass of the 
$A$ ($B$) molecules to the mass of the solvent molecules (see \ref{derivation}):
\begin{equation}
\label{def_rel_pot}
W_{A,B} := \Phi_{A,B}(r) - \frac{m_{A,B}}{m_S}\,\Phi_{S}(r)\,,~~r > R\,.
\end{equation} 
For the case of the reaction in Eq. (\ref{reaction}), mass conservation obviously requires 
$m_A = m_B$.
%\GO{for the reaction in Eq.(\ref{reaction}) or $m_A = m_B+m_C$, in case of the %reaction scheme in Eq.(\ref{react_dissoc}))}. 
However, in order to keep the formulation consistent with the more general 
reaction processes, involving more than one species of product molecules, which are 
considered later on, we maintain the labels $A$ and $B$ 
%\GO{(and $C$)} 
even for this simple case.

Before proceeding with the analysis of the dynamics of the system, it is important 
to clarify the choice of the position of the hard wall. In general this is a complicated issue for  
interfaces involving more than two molecular species.  
Additionally, the divergences of the interaction 
potentials (and of their derivatives) 
caused by molecular hard cores
in the limit
$\tilde r \to 0$ significantly complicate the 
mathematical arguments given below (see, c.f., Sec. \ref{conc_distr}). 
This leads us to use the following approach.
Within a point-particle description of all molecular 
species, the radius $R$ of the colloid, and thus the position of the hard wall, 
is fixed by the most outward layer of molecules composing the colloid. We assume 
that 
each of the molecular pair  interactions between a molecule from the colloid and a 
molecule of species $j$ ($j = A,B,S$) is described well by a Lennard-Jones 6-12 potential. This involves the 
parameters $\sigma_j$ (corresponding to the  molecular diameter) and $\epsilon_j$ (corresponding to the characteristic energy, in units 
of the thermal energy $k_B T$,  where $k_B$ denotes the Boltzmann constant and $T$ the 
absolute temperature), and leads to  the ''exact`` interaction potentials $W(r)$ 
which
 (all of 
them) diverge as $r \to R$. 
Subsequently,
the radius $R$, and therefore the position of the hard 
wall, is redefined as $R \to R + \epsilon$, with $0 < \epsilon \ll R$ of the order of 
a molecular size. These truncated potentials are bounded at $r = R$ (with $R$ interpreted 
as the shifted radius), at the expense that $W_j(R)$ now turn into
phenomenological, 
effective parameters, and that the results depend on the choice of $\epsilon$. The 
arbitrariness can be reduced by further modifying the shape of the potentials $W_j(r)$, 
e.g., by adjusting the point at which their tails are truncated  while requiring 
that the excess adsorption of each component, which is an experimentally measurable quantity, is 
the same for the shifted and modified potentials and for the exact ones. On one 
hand this does not eliminate the dependence on the choice of $\epsilon$, on the other hand 
it provides a prescription based on which, in principle, this dependence can be 
explored systematically. While this approach cannot be justified rigorously, we note 
that such a procedure has been employed recently with satisfying results for describing 
equilibrium properties, such as phase coexistence and interfacial properties, by using 
triangular-well potentials, ''mimicking`` (in the above sense) Lennard-Jones interactions 
\cite{fowler1965,Barcenas2015,Sengupta2016}.

Focusing on the reaction scheme given in Eq. (\ref{reaction}), the goal is to determine the spatial distributions of the $A$ and $B$ molecules 
once a steady state has been attained and to study the resulting chemical force $\mathbf{F}_{chem} = 
\mathbf{F}_{chem}(\theta_0, D_A, D_B, R, p; [W_A,W_B],\mu)$ exerted on 
the immobile
colloid by the inhomogeneously distributed $A$ and $B$ molecules, the velocity 
$\mathbf{V}$ (as a function of the same parameters) if the colloid is free to move, 
and 
the stall force $\mathbf{F}_{ext}$ necessary to immobilize a freely moving colloid.  
%, as well as to determine
% the hydrodynamic flow of the solution induced by these distributions.
%A detailed derivation of these properties will be presented in the subsequent 
%sections. 
(It is straightforward to consider the more general dissociation reaction in Eq.
(\ref{react_dissoc}); therefore we shall  merely present the 
corresponding results without providing their derivation.)

\section{\label{conc_distr}Steady state distribution of reactant and product 
molecules}

Within the framework of the classical theory of linear non-equilibrium 
thermodynamics \cite{Mazur_book,Landau_book,Seifert2012,Koplik2013}, and under the 
assumptions of small number densities, negligible cross-diffusion, and small 
P{\'e}clet numbers of the $A$ and $B$ molecules, the dynamics of the number density 
distributions are governed by the diffusion equations (see \ref{derivation})
\begin{equation}
\label{final_diffusion}
 \frac{\partial n_j}{\partial t} =   
D_j \nabla \cdot \left[e^{-\beta W_j(r)} \nabla\left(e^{\beta W_j(r)} n_j \right)\right]
\,,~j = A,\,B\,.
\end{equation}
At steady state, the time derivatives in Eq. (\ref{final_diffusion}) are zero, 
and the resulting partial differential equations are to be solved subject to appropriate 
boundary conditions.

The coordinate system (see Fig. \ref{fig1}) is chosen such that the $z$-axis 
passes through the center of the patch (point $D$). The system exhibits azimuthal 
symmetry 
around the $z$-axis and therefore the densities $n_{A,B}(\mathbf{r},t)$ of the $A$ and
$B$ molecules, respectively, are independent of the corresponding azimuth angle $\phi$
around the $z$-axis.

\subsection{\label{az} Steady state distribution of the reactant}

In terms of spherical polar coordinates and by dropping the dependence on the azimuthal 
angle $\phi$, one finds that at steady state the number density $n_A$ of the reactant 
$A$ molecules fulfills (see Eq. (\ref{final_diffusion}))
\begin{eqnarray}
\label{gen}
0 &=&
\frac{D_A}{r^2} \frac{\partial}{\partial r}
\left(r^2 \frac{\partial n_A}{\partial r}\right) + \frac{\beta D_A}{r^2}
\frac{\partial}{\partial r} \left(r^2 n_A \frac{d W_A}{d r}\right) + \nonumber\\
&+& \frac{D_A}{r^2 \sin\theta} \frac{\partial }{\partial \theta} \left((\sin \theta)
\frac{\partial n_A}{\partial \theta}\right)\,,
\end{eqnarray}
where $\theta$ is the polar angle measured with respect to the $z$-axis (see 
Fig. \ref{fig1}). For Eq. (\ref{gen}) to be well defined for all $r > R$ we 
restrict our analysis, as discussed in Sec. \ref{model}, to interaction potentials $W_A(r)$ which are finite for all 
$r > R$, have a finite first derivative for all $r > R$, and vanish sufficiently 
fast as $r \to \infty$ (see also below and, c.f., Sec. \ref{force}). 

Equation (\ref{gen}) is to be solved  subject to the boundary condition
\begin{equation}
\label{infinity}
\left. n_A \right|_{r \to \infty} = n_0\,,
\end{equation}
which stipulates that the solute number density $n_A$ approaches the reservoir number 
density $n_0$ for $r \to \infty$, and subject to an effectively reflective boundary 
condition valid at the non-catalytic part of the surface of the colloidal particle 
($r = R,\, \theta_0 < \theta \leq \pi$):
\begin{equation}
\label{00}
j_A := - D_A \left. \left(n_A' + \beta n_A W_A' \right)\right|_{r = R} = 0\,, \,\,\, \theta_0 < \theta
\leq \pi \,,
\end{equation}
where $j_A := \mathbf{e}_r \cdot \mathbf{j}_A$ denotes the radial 
component  of the  current of $A$ molecules at the point $\theta$ on the surface 
$r = R$ of the particle, $W_A' = d W_A/dr$, and $n_A' = \partial n_A(r,
\theta,t)/\partial r$ (see Eq. (\ref{final_diffusion})), with the convention 
that all the evaluations at $r = R$ are interpreted as evaluating at 
$r = R + \epsilon \to R$.

Next we turn to the boundary condition at the catalytic part of the colloid surface which 
accounts for the conversion of $A$ molecules into $B$ molecules. Under general conditions 
such a conversion can be taken into account by requiring that at any point within the area 
of the catalytic patch ($r = R,\,0 \leq \theta \leq \theta_0$) the  current of $A$ molecules 
normal to the surface equals the rate of their annihilation due to the chemical 
reaction, i.e., by imposing the so-called radiation \cite{col} (or imperfectly absorbing) 
boundary condition
\begin{equation}
\label{7}
j_A := - D_A \left.
\left(n_A' + \beta n_A W_A'\right)\right|_{r = R} = -
\left. \kappa \, n_A\right|_{r = R}\,, \,\,\, 0 \leq \theta \leq \theta_0 \,,
\end{equation}
where $\kappa > 0$, such that the right hand side has the proper meaning of an ''annihilation`` term. (Eq. (\ref{7}) can be alternatively interpreted as a 
current ''into the particle``  on the left-hand side (lhs), thus reducing the number density of $A$ 
species in solution, which is given by $\kappa n_A$ on the right-hand side (rhs).) Note that in Eq.(\ref{7}) $j_A$ varies as function of $\theta$.
Before proceeding it is useful to comment on the proportionality factor $\kappa$ (with the 
units of length/time). This factor can be expressed \cite{berd} as the ratio of the 
elementary reaction act constant $K$ (volume times number of acts per unit of time within 
this volume) and the surface area of the catalytic patch:
\begin{equation}
\label{constantK}
\kappa = \frac{K}{4 \pi R^2 f_g} \,,
\end{equation}  
where $f_g=f_g(\theta_0) = \sin^{2}(\theta_0/2)$ is the so-called \textit{g}eometric 
steric factor characterizing the fraction of the surface of the particle covered by the 
catalytic patch. In turn, the elementary reaction act constant is given by 
$K \equiv  W_0 V_a$ \cite{berd}, where $W_0$ is the rate describing the number of reaction 
acts per unit of time within the volume $V_a$ of the reaction zone. In the present system the 
reaction occurs within a segment of a spherical shell given by $R \leq r \leq R + a$ and 
$0 \leq \theta \leq \theta_0$ with $a$ the minimal distance (Fig. \ref{fig1}). This volume 
is $V_a = 4 \pi R^2 a f_g$, $a \ll R$. Thus the proportionality factor is given by
\begin{equation}
\label{constantk}
\kappa = W_0 a
\end{equation}
and is independent of $\theta_0$. Via $W_0$ this factor depends, however, on the 
probability $p$ of the $A \longrightarrow B$ conversion because $W_0 \propto 
p/[(1 - p) \tau]$, where $\tau$ is a typical time spent by an $A$ molecule within the
reaction zone \cite{redner_book}. (A full equation for $W_0$ is not available.) This 
implies that for $p = 0$ (i.e., no reaction) one has $\kappa = 0$ and the rhs of 
Eq. (\ref{7}) vanishes, i.e., in this limit the reflecting boundary condition for the 
normal current is recovered. For $p \to 1$ (i.e., the reaction is perfect and $A$ turns 
into $B$ upon any encounter with the surface) one has $\kappa \to \infty$. Since we have 
assumed that the gradient of the interaction potential is bounded for $r \to R$, the 
lhs of Eq. (\ref{7}) is bounded. This can be reconciled with $\kappa \to \infty$ on 
the rhs only if in this limit $\left. n_A \right|_{r = R} \to ~0$ such that 
$0 < \kappa \left. n_A \right|_{r = R} < \infty$. Thus in this limit one finds the perfect 
sink boundary condition. For $\kappa < \infty$ (i.e., $p < 1$), the boundary condition in 
Eq. (\ref{7}) is therefore equivalent to the physical assumption that not all encounters 
of the $A$ molecules with the catalytic part of the surface of the colloid lead to a
reaction event.

For systems which are catalytically active as a whole (i.e., $\theta_0 = \pi$) and in
the absence of the interaction potential (apart from the hard-core repulsion) between the 
colloid and the $A$ molecules, such a boundary condition has first been  proposed
by Collins and Kimball \cite{col} (see also Ref.~\cite{weiss}) as a generalization of the
conventional Smoluchowski theory \cite{smoluchowski} which stipulates a perfect sink 
boundary condition. Furthermore, for $\theta_0 = \pi$ an exact calculation of the 
steady-state diffusion problem defined by Eqs. (\ref{final_diffusion}), (\ref{00}), 
and (\ref{7}) is possible for quite general forms of the interaction potential $W_A(r)$ 
as well as for a spatially varying diffusion constant $D_A(r)$ \cite{shoup,shulten77}.

In the absence of interactions (apart from the hard-core repulsion) between the
colloid and the $A$ and $B$ molecules, the solution of the diffusion equation
(Eqs. (\ref{gen}) - (\ref{7})) subject to such \textit{mixed} boundary conditions (i.e.,
different boundary conditions are used on different parts of the boundary of the domain of
the equation) has been studied by several groups in the past. This started with the
seminal work by Solc and Stockmayer \cite{solc1,solc2},  who developed an approximate
method to treat this mathematical problem. (For this case of $W_{A,B} = 0$, also several other approximations have been  proposed (see, e.g., Refs. \cite{solc2,samson,berd}).)
Moreover, an exact solution based on the so-called dual series relations is available
\cite{tray,tray2}. However, this procedure is not only cumbersome, but it also does not
render a transparent, analytical form of the solution and thus masks the physical
content of the results. 

In what follows we resort to an approximate approach as the one proposed in 
Ref. \cite{shoup2}. This choice is motivated by the fact that the approach developed 
in Ref. \cite{shoup2} is rather straightforward and is known to yield -- in 
the absence of interactions -- results which are in a good agreement with 
numerical solutions of the original problem \cite{tray}.  Here, we generalize this 
approach by taking into account the interactions between the molecules and the 
colloid, which are absent in the original work \cite{shoup2}. As a byproduct of our 
analysis, this allows us to obtain several new results concerning the effective 
reaction constants and density profiles for systems with heterogeneous reactivity 
in the presence of interaction potentials.

In accordance with Ref. \cite{shoup2}, the boundary condition in Eq. (\ref{7}) 
is replaced by 
\begin{equation}
\label{1}
- D_A \left.
\left( n_A' + \beta n_A W_A'\right)\right|_{r = R} = - Q, \,\,\, 0 \leq \theta \leq \theta_0 \,,
\end{equation}
where $Q > 0$  is a "trial" constant, independent of $\theta$, 
which is determined self-consistently by
requiring that the \textit{former} condition given by Eq. (\ref{7}) holds \textit{only} on
average over the region $0 \leq \theta \leq \theta_0$ of the catalytic patch, i.e.,
\begin{equation}
\label{rbcc}
\left.
 D_A \int_0^{\theta_0}  d\theta \, \sin \theta
\left( n_A' + \beta n_A W_A'\right)\right|_{r = R}=
\kappa  \int_0^{\theta_0}  d\theta \,
\sin\theta  \, n_A(r = R,\theta)\,.
\end{equation}
Equations (\ref{rbcc}) and (\ref{1}) imply (see also Eq. (\ref{7})) 
\begin{equation}
\label{Q_eq}
 Q (1 - \cos \theta_0) = \kappa  \int_0^{\theta_0}  d\theta \,  
\sin\theta \, n_A(R,\theta)\,.
\end{equation}
Since via Eq.~(\ref{1}) $n_A(r,\theta)$ depends on $Q$, Eq. (\ref{Q_eq}) represents
an implicit equation for $Q(\kappa)$ (see below). 

We seek the solution of Eq. (\ref{gen}), which is the steady-state solution of 
Eq. (\ref{final_diffusion}), via the \textit{ansatz} 
\begin{equation}
\label{sol}
 n_A(r,\theta) = n_0 \,e^{- \beta W_A(r)}
\left[ 1 + \sum_{n = 0}^{\infty} a_n g_n(r)
P_n\left(\cos \theta \right) \right]\,,
\end{equation}
where $a_n$ are dimensionless coefficients to be determined from the boundary conditions 
and $P_n\left(\cos \theta \right)$ is the Legendre polynomial of order $n$.
%, with 
%$P_{-1}(\cos\theta) = 1$, $P_0(\cos\theta) = 1$, $P_1(\cos\theta) = \cos\theta$,  
%$P_2(cos\theta) = (3 \cos^2\theta - 1)/2$, $\dots$, and the orthogonality relation $\int \limits_0^\pi d\theta \sin\theta
%P_n(\cos\theta) P_m(\cos\theta) = \delta_{n,m}/(n+1/2)$. 
We note that the first term on 
the rhs of Eq. (\ref{sol}) is the \textit{equilibrium} Boltzmann 
distribution of $A$ molecules around the colloid (see Eq. (\ref{infinity})), e.g., 
in the absence of the chemical reaction or for an infinitely fast diffusion of  $A$ molecules (i. e., $D_A = \infty$) in which case all coefficients 
$a_n$ vanish (see below). Therefore the second term represents the whole
out-of-equilibrium contribution to the concentration distribution due to the chemical 
reaction and due to a finite value of the diffusion constant $D_A$ of  the $A$ molecules.

Upon inserting Eq. (\ref{sol}) into Eq. (\ref{gen}), one finds that the former is the 
solution of the latter if the functions $g_n(r)$ solve 
\begin{equation}
\label{fp}
 g''_n(r) + \left(\frac{2}{r} - \beta W'_A(r) \right) g'_n(r) -
\frac{n (n + 1)}{r^2} g_n(r) = 0 \,. 
\end{equation}
Moreover, in order that $n_A(r,\theta)$ given by Eq. (\ref{sol}) satisfies the
boundary condition in Eq. (\ref{infinity}), the solutions $g_n(r)$ of Eq. (\ref{fp}) are
subject to the boundary condition that they vanish for $r \to \infty$. For each $n$ the
solutions $g_n(r)$ of Eq. (\ref{fp}) are defined up to multiplicative constants. Without
loss of generality, as explained below, we fix these constants by choosing the
normalization $g_n(r= R) = 1$. This renders the coefficients $a_n$ to be unique.

For $n = 0$, from Eq.~(\ref{fp}) one finds 
\begin{equation}
\label{1000}
 g_0(r) = R_D \, \int^{\infty}_r du \frac{e^{\beta W_A(u)}}{u^2}\,
\end{equation}
with $g_0(r \to \infty) = R_D/r$, where $R_D$ is related to the so-called Debye radius
(see Ref. \cite{debye}):
\begin{equation}
\label{raddeb}
 R_D = \left( \int^{\infty}_R du \frac{e^{\beta W_A(u)}}{u^2}\right)^{-1} \,;
\end{equation}
note that $R_D = R$ for $W_A = 0$. For $n \geq 1$, Eq. (\ref{fp}) cannot be solved
analytically for a generic interaction potential $W(r)$. However, for $r \to \infty$
the interaction potentials of interest for our system decay $\sim 1/r$ or faster so that
$\beta W'_A(r) \ll 1/r$ for $r \gg R$. Accordingly, at large distances $r \gg R$ 
Eq. (\ref{fp}) reduces to the differential equation satisfied by the radial functions of
the Laplace equation. This ensures that for any $n \geq 1$ Eq. (\ref{fp}) admits a
solution $g_n(r)$ with the asymptotic behavior $g_n(r \gg R) \sim r^{-(n+1)}$ and thus
obeying the boundary condition to vanish for $r \to \infty$.

Combining the boundary conditions in Eqs. (\ref{00}) and (\ref{1}) and inserting therein
the expansion given by Eq. (\ref{sol}) leads to
\begin{equation}
\label{b}
 \sum_{n=0}^{\infty}
 a_n g'_n(R) P_n(\cos \theta) = \frac{Q \,e^{\beta W_A(R)} }{ D_A \, n_0} \,
\Theta(\theta_0 - \theta),
\end{equation}
where $\Theta(x)$ is the Heaviside step function ($\Theta(x) = 1$ for $x \geq 0$ and zero
otherwise). Multiplying both sides of Eq. (\ref{b}) by $(\sin \theta) \,P_m(\cos
\theta)$, integrating them over $\theta$ from $0$ to $\pi$, and using  the orthogonality
of the Legendre polynomials one obtains
\begin{eqnarray}
\label{z1}
 a_n &=&
\frac{Q \, e^{\beta W_A(R)}}{2 \, D_A \, n_0}\, 
\frac{\phi_n(\theta_0)}{g'_n(R)}, \, n \geq 0,
\end{eqnarray}
where
\begin{equation}
\label{phi}
 \phi_n(\theta_0) =
P_{n - 1}\left(\cos \theta_0\right) - P_{n + 1}\left(\cos \theta_0 \right)\,, \,
n \geq 0\,,
\end{equation}
with the standard convention $P_{-1}\left(\cos \theta_0\right)= 1$. We note that because 
$a_n \sim 1/g'_n(R)$, one has $a_n g_n \sim g_n/g_n'$. Thus in the final result in
Eq. (\ref{sol}) the normalization  amplitude of $g_n$ drops out and $n_A(r,\theta)$ is 
independent of the choice for this normalization, as it should. 

After substituting $a_n$ from Eq. (\ref{z1}) into Eq. (\ref{sol}), inserting the resulting
expression into Eq. (\ref{Q_eq}), and taking advantage of the relation
\begin{equation}
\label{phi_n_def}
 \phi_n(\theta_0) = (2n+1) \int^{\theta_0}_0 d\theta \, \sin\theta \, P_n(\cos\theta) \,,
\end{equation}
one finds the following explicit expression for $Q$:
%%%%%%%%%%%%%%
\begin{equation}
\label{inverse1}
Q = \frac{4 \kappa n_0  D_A e^{-\beta W_A(R)} \phi_0(\theta_0) }{
\left(4 D_A  \phi_0(\theta_0)  - \kappa \sum_{n=0}^{\infty} 
\frac{\phi_n^2(\theta_0) g_n(R)}{(n +1/2) g'_n(R)} \right)} \,,
\end{equation}
%%%%%%%%%%%%%
where $\phi_0(\theta_0) = 1 - \cos\theta_0 = 2 \sin^2(\theta_0/2)$. In the following, due 
to the normalization choice for the functions $g_n(r)$ the ratio $g_n(R)/g'_n(R)$ inside 
the sum will be replaced by $1/g'_n(R)$. This concludes the calculation of the distribution 
$n_A(r,\theta)$ of reactant molecules (Eqs. (\ref{sol}), (\ref{fp}), and (\ref{z1}) - 
(\ref{inverse1})).  

Upon closing this subsection we emphasize that within the expansion in Eq. (\ref{sol}) the radial and angular forms of the functions are exact.  
Only 
the coefficients $a_n$ are determined approximately, using a generalized self-consistent approach 
\cite{shoup2}. In the following, we shall evaluate  
 the contributions to the force $F_{chem}$ 
exerted on the immobile colloid (and, if it is free to move, to its velocity $V$) stemming from the interactions of the colloid 
 with the $A$ molecules, the spatial distribution of which is given by Eq. (\ref{sol}) with the coefficients $a_n$ defined by Eqs. (\ref{z1}) and (\ref{inverse1}) (and similarly due to the interactions with the $B$ and $C$ molecules, the distributions of which will be determined in Subsec. \ref{products} and \ref{dissoc_reaction}). To this end, one has to integrate the distribution in Eq. (\ref{sol}) multiplied by the gradient of the interaction potential in order to determine ${F}_{chem}$, and multiplied by the gradient of the interaction potential and by the axially symmetric part of the velocity field in order to calculate $V$, which will be carried out in Sec. \ref{force}. We proceed by showing 
that the contributions to ${F}_{chem}$ and $V$
due to the interactions of the colloid with the $A$ molecules are given by  $F_{chem} = (4 \pi n_0/3) a_1 I_A$ and $V = (2 n_0/9 \mu R) a_1 {\cal J}_A$, where $I_A$ and ${\cal J}_A$ are \textit{exact} functional expressions of the interaction potentials
and only the coefficient $a_1$,  which due to the symmetry of the system is the only relevant one,  
is determined approximately.  

The accuracy of the approximation for the coefficient $a_1$ can be estimated from the following simple argument. Suppose that one is able to solve the  mixed boundary problem defined by Eqs. (\ref{gen}), (\ref{infinity}), (\ref{00}) and (\ref{7}) exactly, and hence, to 
obtain the exact expressions for the expansion coefficients $a_n$ in Eq. (\ref{sol}). 
Then, inserting this expansion into Eq. (\ref{7}), multiplying both sides by $\sin(\theta) \cos(\theta)$ and integrating over $\theta$ from $0$ to $\pi$ (recall that $j_A : \equiv 0$ for $\theta_0 < \theta \leq \pi$), one obtains the following relation between the exact coefficient $a_1$ and the exact current $j_A$ (which in general is a function of $\theta$):
\begin{equation}
\label{1963}
a_1 = - \frac{3}{4} \frac{e^{\beta W_A(R)}}{n_0 D_A g'_1(R)} \int^{\theta_0}_0 j_A \sin(2 \theta) d\theta \,.
\end{equation}
On the other hand, the relations in Eqs. (\ref{z1}) and (\ref{phi_n_def}) provide the following approximate expression for $a_1$:
\begin{equation}
a_1 = \frac{3}{4} \frac{e^{\beta W_A(R)}}{n_0 D_A g'_1(R)} \, Q \, \int^{\theta_0}_0 \sin(2 \theta) d\theta \,.
\end{equation}
Therefore the accuracy of estimating the coefficient $a_1$ via using the self-consistent approach \cite{shoup2} turns out to be  the same as the one associated with
approximating 
the integral on the rhs of Eq. (\ref{1963})
as
\begin{equation}
\label{approx}
- \int^{\theta_0}_0 j_A \sin(2 \theta) d\theta  \approx Q \int^{\theta_0}_0\sin(2 \theta) d\theta  \,. 
\end{equation}  
On physical grounds, one may expect that $j_A$  depends very weakly on $\theta$ within the interior part of the catalytic patch away of its periphery.  Hence, for the major part of the integration interval the current $j_A$ will factor out from the integral rendering Eq. (\ref{approx}) to be an equality (compare with Eqs. (\ref{7}) and (\ref{1})).
An appreciable dependence of the current on $\theta$ may appear only in the vicinity of $\theta = 0$ and $\theta = \theta_0$, where $j_A$ drops to zero. 
Therefore one can
 expect that the self-consistent approximation in Ref. \cite{shoup2} will provide an accurate estimate for $a_1$ provided that $\theta_0$ is not too small.  
Remarkably, even in the limit $\theta_0 \to 0$, in which 
one may expect the most significant deviations,
 the self-consistent approximation is surprisingly reliable,
  as evidenced by the numerical analyses in Ref. \cite{tray} and more recently in Ref. \cite{greb}, which studied the so-called narrow escape problem in the presence of long-ranged interactions 
 with the confining boundary using essentially the same approximate approach as the present one. It was shown in Ref.\cite{greb} that the self-consistent approximation captures 
 adequately even the dependences of 
 those terms, which dominate
in the limit $\theta_0 \to 0$, on all  
 pertinent parameters and it only slightly underestimates the numerical factors. 
 
 Finally, in Sec. \ref{disc} we shall  show that for the particular case $\theta_0= \pi/2$ (i.e., for the so-called Janus colloids) and for short-ranged 
 triangular-well interaction potentials, our general results reproduce 
 the known limiting forms of
 the self-propulsion velocity as obtained  in Ref. \cite{Golestanian2012} 
 within an approach based on the concepts of the effective
Derjaguin length and the phoretic slip.

\subsection{Chemical kinetics interpretation: effective reaction constants}

In order to translate our results into the usual nomenclature of chemical kinetics, we
follow Refs. \cite{shoup,shoup2} and introduce an ``effective'' reaction constant 
$K_{eff}$, which accounts for the combined effect of the nonzero reaction probability at
the catalytic patch and the diffusive transport of the reactants to the patch. 
$K_{eff}$ is defined as the number of $A$ molecules, which are distributed according to
$n_A(r,\theta)$ and which would flow, per time, through the surface $r = R$ if the colloid was
permeable, divided by $n_0$. Using Eqs. (\ref{00}), (\ref{7}), and (\ref{rbcc})
this takes the form
\begin{eqnarray}
\label{constant}
 K_{eff} &=& \frac{2 \pi R^2 D_A}{n_0} \int_0^{\pi} d\theta \, \sin \theta \, 
\left.
\left( n_A' + \beta n_A W_A'\right)\right|_{r = R} \nonumber\\
 &=& \frac{2 \pi R^2 \phi_0(\theta_0)}{n_0} \, Q \,.
\end{eqnarray}
We note that the coefficients $a_n$ in Eq. (\ref{z1}) can be expressed in terms of this
effective reaction constant as
\begin{equation}
\label{an}
a_n = \frac{K_{eff}}{K_S} \, \frac{e^{\beta W_A(R)}}{R \, g_n'(R)} \, 
\frac{\phi_n(\theta_0)}{\phi_0(\theta_0)} \,, \,\,\, n \geq 0\,,
\end{equation}
where $K_S = 4 \pi D_A R$ is the Smoluchowski constant \cite{smoluchowski}. $K_{eff}$ 
and $K_S$ have the units of volume per time, as $K$ does.

By replacing $Q$ with the expression in Eq. (\ref{inverse1}), and by noting that 
Eq. (\ref{1000}) implies $R_D = - R^2 g'_0(R) e^{-\beta W_A(R)}$, Eq. (\ref{constant})
can be cast into the physically intuitive form
\begin{equation}
\label{inverse}
 \frac{1}{K_{eff}} = \frac{1}{K^*} + \frac{1}{K_{SD} \, f_{dc}(\theta_0)} \,,
\end{equation}
where \textit{dc} stands for diffusion controlled. The Smoluchowski-Debye constant
$K_{SD}$ \cite{debye} (compare Eq.~(\ref{raddeb}))
\begin{equation}
\label{sd}
 K_{SD} = 4 \pi D_A R_D
\end{equation}
equals the flux (divided by $n_0$) of diffusive molecules, the density of which at $r \to
\infty$ is kept fixed and equal to $n_0$, through the hypothetical surface of an immobile, 
perfectly absorbing sphere interacting with the molecules via a radially symmetric 
potential $W_A(r)$. The quantity
\begin{eqnarray}
\label{K*_def}
K^*  &=&  2 \pi R^2 \phi_0(\theta_0) \kappa e^{- \beta W_A(R)} =  \nonumber\\
&=& 4 \pi R^2 f_g \kappa e^{- \beta W_A(R)} = K \,e^{- \beta W_A(R)} 
\end{eqnarray}
is an \textit{effective} constant for an elementary reaction act factored into a term 
$K$ (see Eq. (\ref{constantK})), which depends only on the reaction kinetics $\kappa$ and the
geometry (via $\theta_0$) of the catalytic patch, and a term which depends only on the 
interaction potential. In the \textit{diffusion-controlled} limit $K^* \to \infty$ 
(in the sense that $K^* \gg K_{SD} f_{dc}(\theta_0)$, but  not necessarily infinitely large, see 
below), the quantity
\begin{equation}
\label{ster}
f_{dc}(\theta_0) = 2 \phi_0^2(\theta_0) /\left(g_0'(R) 
\sum_{n=0}^{\infty} \frac{\phi_n^2(\theta_0)}{(n+1/2) g'_n(R)} \right)
\end{equation}
plays the role of an \textit{effective} steric factor: it shows how the reaction rate in 
Eq. (\ref{sd}) is reduced effectively due to the fact that the catalytic patch occupies
only some part of the colloid surface. As will be seen below, it shows a different
angular behavior as compared with the purely geometric steric factor $f_g =
\sin^2(\theta_0/2)$. We emphasize that the full functional form of $W_A(r)$ enters both
into $K_{SD}$ (see Eqs. (\ref{sd}) and (\ref{raddeb})) and into $f_{dc}$ (via $g_n(r)$;
see Eqs. (\ref{ster}) and (\ref{fp})) whereas $K^*$ depends only on $W_A(r = R)$ 
(Eq. (\ref{K*_def})). (Here we recall  that the argument $r = R$ must be interpreted as 
$r = R + \epsilon \to R$ and that we assume the potential and its derivatives to be bounded 
at $r = R + \epsilon$.)

Equation (\ref{inverse}), which resembles the law of addition of inverse resistances, 
bears out the combined effect of two rate-limiting steps: the random, diffusive search 
of the $A$ molecules for the catalytic patch and the subsequent elementary reaction
act. Such a form allows one to distinguish easily between the so-called 
\textit{diffusion-controlled} limit, in which the time needed by the $A$ molecule to 
diffuse to the catalytic patch is  the rate-limiting step (with the elementary reaction
contribution $1/K^*$ itself being negligible in comparison), and 
\textit{kinetically-controlled} reactions for which the opposite holds, i.e., $K^* \ll K_{SD} f_{dc}(\theta_0)$. In the following, 
we shall denote this limit symbolically as $"D_A \to \infty"$, which does not imply, however, that $D_A$ is infinitely large, but it only means that the latter inequality holds. 

One can readily check that in the absence of interaction potentials, i.e., for 
$W_A(r) \equiv 0$, the result in Eq. (\ref{inverse}) reduces to (see Eq. (\ref{K*_def}))
\begin{equation}
\label{constantsls}
 \frac{1}{K_{eff}^{(0)}} = \frac{1}{K} + \frac{1}{K_S f_{dc}^{(sls)}(\theta_0)} \,,
\end{equation}
with $K_{eff}^{(0)} \equiv K_{eff}[W_A(r) \equiv 0]$. This is the result presented in 
Ref. \cite{shoup2} and earlier, within the framework of a different approximate approach, 
in Ref. \cite{solc2}. In this case the steric factor in Eq. (\ref{ster}) attains the 
form
\begin{equation}
\label{ster1}
 f_{dc}^{(sls)}(\theta_0) = 2 \phi_0^2(\theta_0) /\sum_{n=0}^{\infty} 
\frac{\phi_n^2(\theta_0)}{(n+1/2) (n+1)} \,, \,\,\, W_A(r) \equiv 0\,,
\end{equation}
where the superscript ``sls'' is a  tribute to  Shoup, Lipari,  and Szabo, 
who 
reported  this result in Ref. \cite{shoup2}. The steric factor $f_{dc}^{(sls)}(\theta_0)$ is a 
monotonically increasing function of $\theta_0$, interpolating between 
$f_{dc}^{(sls)}(\theta_0 =0) = 0$ and $f_{dc}^{(sls)}(\theta_0 = \pi) =1$ (see, e.g., 
Fig. 3 in Ref. \cite{tray}). 
These limits can be understood as follows. Since for $n \geq 0$ the Legendre polynomials 
have defined parity, $P_n(x) = (-1)^n P_n(-x)$, and satisfy $P_n(1) = 1$, it follows that 
for $\theta_0 \to \pi$ one has $\phi_n(\pi) = 2 \,\delta_{0,n}$ for $n \geq 0$ 
(Eq. (\ref{phi})). Thus for $\theta_0 \to \pi$ in the sum in the denominator only 
the term $n=0$ survives and therefore $f_{dc}^{(sls)}(\theta_0 = \pi) = 1$. The limit 
$\theta_0 \to 0$ is more involved because $P_n(1) = 1$ and thus $\phi_n(\theta_0 = 0)
= 0$ for all $n \geq 0$. However, noting that to first order in $(1- \cos\theta_0)$
one has $\phi_n(\theta_0 \to 0) = (2n+1) (1-\cos\theta_0)$ (Eq. (\ref{phi_n_def})), the
steric factor $f_{dc}^{(sls)}$ behaves as $f_{dc}^{(sls)}(\theta_0 \to 0) \propto 
1/\sum_{n=0}^{\infty} \frac{(2n+1)^2}{(n+1/2)(n+1)}$ and thus vanishes in the limit 
$\theta_0 \to 0$ due to the divergence of the series in the denominator. The exact 
asymptotic behavior in this latter limit was analyzed thoroughly in Refs. \cite{shoup2} 
and \cite{tray}, and more recently, in Ref. \cite{greb} with the result that 
in the
 limit $\theta_0 \to 0$ the effective steric 
factor behaves as $f_{dc}^{(sls)}(\theta_0 \to 0) = (3 \pi/32) \,\theta_0$. The geometric
steric factor $f_g = \sin^2(\theta_0/2)$ behaves as $f_g(\theta_0 \to 0) = (\theta_0)^2/4$
and hence is much smaller than $f_{dc}^{(sls)}$. This implies that in the limit 
$\theta_0 \to 0$ the first term on the right-hand side of Eq. (\ref{inverse}) becomes the
dominant one ($K^* \sim (\theta_0)^2$ while $K_{SD} \, f_{dc}(\theta_0) \sim \theta_0$
for $\theta_0 \to 0$), so that the reaction becomes kinetically controlled. 
Interestingly, in the case in which precisely one half of  the 
particle is covered by catalyst, i.e., for a so-called Janus particle, the value 
$f_{dc}^{(sls)}(\theta_0=\pi/2) \approx 0.706$ substantially exceeds $1/2$, which 
is the value of the geometric steric factor for the same coverage.

In the case $W_A \neq 0$, without the explicit dependence of $g_n$ on 
$W_A(r)$ being available in closed form for arbitrary $\theta_0$,  one
cannot state much concerning the behavior of the steric factor in Eq. (\ref{ster}), except
that (based on the same argument as above) for arbitrary potentials $W_A(r)$ it equals $1$
for $\theta_0 = \pi$, i.e., $f_{dc}(\theta_0 = \pi) = 1$ if the whole surface of the particle
is catalytic. In the  opposite limit of having no catalytic properties, i.e., $\theta_0 \to 0$, 
some general statements concerning the form of the coefficients 
belonging to the leading terms 
where made recently \cite{greb}.  
%and we could not establish the behavior of $f_{dc}(\theta_0 \to 0)$ for general $W_A(r)$. 
%Here we only note that we expect that $f_{dc}(\theta_0 \to 0)$ also vanishes in this limit. 
%We shall verify this expectation for the specific cases to be considered in the following 
%sections.\MNP{\footnote{\MNP{We remark that if the dependence on $n$ exhibited by the asymptotic behavior of
%$g_n(r)/g'_n(r) \sim r/(n+1)$ at large $r \gg R$ (see the text following Eq.
%(\ref{raddeb})) would hold also for $r \gtrsim R$, then indeed the behavior of 
%$f_{dc}(\theta_0 \to 0)$ would be the same as that of $f_{dc}^{(sls)}(\theta_0 \to 0)$ and 
%therefore $f_{dc}(\theta_0)$ would vanish in the limit $\theta_0 \to 0$, too.}}}

As discussed above, in the case that the catalytic patch covers the entire surface of 
the colloid, i.e., if $\theta_0 = \pi$, one has $f_{dc}(\theta_0 = \pi) = 1$ for arbitrary
$W_A(r)$. In this case one therefore recovers the classic result \cite{shoup} (see
also Refs. \cite{oshan,oshan1})
\begin{equation}
\label{shou}
 \frac{1}{K_{eff}} = \frac{1}{K^*} + \frac{1}{K_{SD}} \,, \,\,\, \theta_0 
= \pi \,, \,\,\, W_A(r) \neq 0 \,.
\end{equation}
If in addition there is no interaction potential $W_A(r)$ one has $K^* = K$ and 
$R_D = R$. In this case Eq.~(\ref{shou}) reduces to the celebrated relation
\begin{equation}
\label{Collins_Kimball}
 \frac{1}{K_{eff}^{(0)}} = \frac{1}{K} + \frac{1}{K_{S}}\,,\theta_0 = \pi\,,~ W_A(r) 
\equiv 0\,,
\end{equation} 
due to Collins and Kimball \cite{col}. The relation in Eq. (\ref{Collins_Kimball}) can be 
derived also from microscopic stochastic dynamics which allows one to identify the
elementary reaction act constant $K$ through the reaction probability $p$ \cite{ben}. 

In the diffusion-controlled limit corresponding to $K^* \to \infty$ (and with $D_A$ fixed, 
such that the molecular diffusion is the rate-limiting step), the effective reaction 
constant $K_{eff}$ is given by (see Eq.~(\ref{inverse}))
\begin{equation}
\label{difcont}
 K_{eff}(K^* \to \infty) = K_{SD} \, f_{dc}(\theta_0) \,.
\end{equation}
Combining Eqs. (\ref{constant}) and (\ref{difcont}), in the diffusion-controlled limit
the factor $Q$ can be written as 
\begin{equation}
\label{Q_diff_contr}
 Q(K^* \to \infty) = 
\frac{n_0 K_{SD}}{2 \pi R^2} \frac{f_{dc}(\theta_0)}{\phi_0(\theta_0)}\,.
\end{equation}
In this case, the coefficients $a_{n}$ in Eq. (\ref{an}), which enter into the series 
representation of the number density profile of the reactant (Eq. (\ref{sol})), can be 
written as
\begin{equation}
\label{an1}
a_n(K^* \to \infty) = \frac{R_D}{R} \, 
\frac{e^{\beta W_A(R)}}{R \, g'_n(R)} \, 
\frac{\phi_n(\theta_0)}{\phi_0(\theta_0)} \, f_{dc}(\theta_0) \,, 
\,\,\, n \geq 0\,.
\end{equation}

Finally, for completeness we note that according to Eq. (\ref{an}) in the 
kinetically-controlled limit, in which $K^*$ is fixed while $D_A \to \infty$, such that 
$K_{eff} \to K^*$, the coefficients $a_n$ reduce to 
\begin{equation}
\label{an2}
 a_n(D_A \to \infty) = \frac{K}{K_S} \, \frac{1}{R \, g_n'(R)} \, 
\frac{\phi_n(\theta_0)}{\phi_0(\theta_0)} \,, \,\,\, n \geq 0\,.
\end{equation}
Since $K_S = 4 \pi D_A R$, in this limit the coefficients $a_n$ decay with $D_A$ as 
$a_n \sim 1/D_A$ so that $n_A$ converges to the equilibrium distribution
$n_A(r) = n_0 e^{-\beta W_A(r)}$ (see Eq. (\ref{sol})) 
if $D_A \to \infty$. In this extreme limit $n_A$ does not depend on the angle 
$\theta$, and thus Eq. (\ref{1}) turns to an exact expression.  If the diffusion constant of the $A$ molecules is finite, 
(recall that the kinetically-controlled limit should be interpreted as 
$K^* \ll K_{SD} f_{dc}(\theta_0)$, and thus the coefficients $a_n \neq 0$ are nonzero, albeit small in 
magnitude), the chemical reaction enforces an angular dependent, out-of--equilibrium 
steady state for the distribution of the reactant molecules.

%\MNP{when $D_A \to \infty$. In this limit $n_A$ does not depend on the angle 
%$\theta$, and thus Eq. (\ref{1}) is exact; nevertheless, if the diffusion constant(s) 
%of the reaction product(s) remains finite, a non-equilibrium steady-state distribution 
%of the reaction products emerges (see below) and the phenomenology discussed in, 
%c.f., Sec. \ref{force} remains relevant, though no longer dependent on $n_A$. 
%Physically, however, the kinetically-controlled limit should be interpreted as 
%$K^* \ll K_{SD}$, and thus the coefficients $|a_n| \neq 0$ (albeit small in 
%magnitude), i.e., the chemical reaction enforces an angular dependent non-equilibrium 
%steady state for the distribution of the reactant molecules.}

\subsection{Steady state distributions of the reaction products}
\label{products}

For $r > R$ the local number density of the reaction products $B$ obeys the differential 
equation
\begin{eqnarray}
\label{gen1}
0 &=& \frac{D_B}{r^2} 
\frac{\partial}{\partial r}
\left(r^2 \frac{\partial n_B}{\partial r}\right) + \frac{\beta D_B}{r^2}
\frac{\partial}{\partial r} \left(r^2 n_B \frac{d W_B}{d r}\right) + \nonumber\\
&+&
\frac{D_B}{r^2 (\sin \theta)} \frac{\partial }{\partial \theta} \left((\sin \theta)
\frac{\partial n_B}{\partial \theta}\right) \,,
\end{eqnarray}
which has the same form as Eq. (\ref{gen}). Equation (\ref{gen1}) is to be solved
subject to a sink boundary condition at macroscopic distances from the colloid:
\begin{equation}
\label{infinity1}
 \left. n_B \right|_{r \to \infty} = 0,
\end{equation}
which has to be complemented by the reaction boundary condition across the catalytic patch 
and the zero current boundary condition across the non-catalytic part of the surface, 
similar to those for the $A$ molecules. Within the mean-field approximation as used above, 
these latter two conditions can be combined into the equation
\begin{equation}
\label{6}
 D_B \left. \left( n_B' + \beta n_B W_B'\right)\right|_{r = R}
= - Q \, \Theta(\theta_0 - \theta)\,,
\end{equation}
where we have used the fact that the creation of a $B$ molecule is tied to the 
annihilation of an $A$ molecule so that here the current is opposite to the one 
on the right hand side of Eq. (\ref{1}). (With $Q > 0$, the signs in 
Eq. (\ref{6}) correspond, as they should, to molecules of species $B$ being ``released'' into the solution.) We seek the stationary solution of 
Eq. (\ref{gen1}) via the ansatz
\begin{equation}
\label{sol1}
 n_B(r,\theta) = n_0 \, e^{- \beta W_B(r)} \,
\sum_{n = 0}^{\infty} b_n \, j_n(r) \, P_n\left(\cos \theta\right) \,,
\end{equation}
where $b_n$ are dimensionless coefficients to be determined from the boundary condition
in Eq. (\ref{6}). This ansatz solves the stationary Eq. (\ref{gen1}) provided the functions
$j_n(r)$ are those solutions of
\begin{equation}
\label{fp1}
 j''_n(r) + \left(\frac{2}{r} - \beta W'_B(r) \right) j'_n(r) -
\frac{n (n + 1)}{r^2} j_n(r) = 0
\end{equation}
which vanish for $r \to \infty$. As for the functions $g_n(r)$ in Eq. (\ref{fp}) we choose
the normalization $j_n(r = R) = 1$. For the interaction potential $W_B(r)$ we require
similar properties as for $W_A$, i.e., being continuous and bounded, with continuous and
bounded first derivative, and vanishing for $r \to \infty$ (see the previous subsection).
The potential $W_B(r)$ does not comprise the hard-core repulsion, which is taken into
account by the boundary condition in Eq. (\ref{6}). Following the same steps as for the
case of the reactants and recalling the relation between $Q$ and $K_{eff}$ in 
Eq. (\ref{constant}), we obtain the coefficients $b_n$ in the series representation 
of $n_B(r)$: 
\begin{eqnarray}
\label{gensolute}
 b_n &=& - \frac{Q \,  e^{\beta W_B(R)}}{2 \, D_B \, n_0} \, 
\frac{\phi_n(\theta_0)}{j'_n(R)} \nonumber\\
 &=& - \frac{K_{eff}}{4 \, \pi \, D_B \, R} \, 
\frac{e^{\beta W_B(R)}}{R \, j'_n(R)} \, \frac{\phi_n(\theta_0)}{\phi_0(\theta_0)}\,,
~~n \geq 0 \,.
\end{eqnarray}
In the diffusion-controlled limit corresponding to $K^* \to \infty$ the latter equation 
reduces to
\begin{equation}
\label{soluteprofile2}
 b_n(K^* \to \infty)  = - \frac{D_A }{D_B} \, \frac{R_D}{R} \, 
\frac{e^{\beta W_B(R)}}{ R \, j'_n(R)} 
\frac{\phi_n(\theta_0)}{\phi_0(\theta_0)} \,  f_{dc}(\theta_0)
\,, \,~ n \geq 0\,.
\end{equation}
We note that, via Q (see Eq. (\ref{inverse1})) in the boundary condition in Eq. (\ref{6}),
$n_B(r,\theta)$ depends on the characteristics of the $A$ particles such as $n_0$, $D_A$, and
$W_A(r)$. On the other hand, $n_A(r,\theta)$ is independent of the characteristics of the $B$
particles; this is due to the absence of interactions between $A$ and $B$ molecules. In
Eq.~(\ref{soluteprofile2}) $R_D$ and $f_{dc}(\theta_0)$ are the expressions given by Eqs.
(\ref{raddeb}) and (\ref{ster}), respectively. Thus they are determined by $W_A(r)$ only
and are independent of $W_B(r)$. 

In the kinetically-controlled limit, corresponding to $D_{A} \to \infty$ with 
$K^*$ fixed such that $K_{eff} \to K^*$, we have
\begin{eqnarray}
\label{gensolute19}
 b_n(D_A \to \infty) =  - \frac{K^*}{4 \, \pi \, D_B \, R} \, 
\frac{e^{\beta W_B(R)}}{R \, j'_n(R)} \, \frac{\phi_n(\theta_0)}{\phi_0(\theta_0)} \,, 
\,\,\,n \geq 0 \,.
\end{eqnarray}
In this limit the coefficients $b_n$ are independent of $D_A$ and nonzero.

Finally, we note that the distribution of the product molecules $C$, in the case of the reaction described in Eq. (\ref{react_dissoc}), can be calculated in a similar way. We relegate these straightforward 
calculations to \ref{dissoc_reaction}.

\section{\label{force} The force exerted on an immobile colloid and the velocity of a force-free colloid}

In this section we consider the two contributions to the force exerted on 
the colloid due to the chemical reaction. They stem (i) from the 
hydrodynamic flow of the mixture, driven by the inhomogeneous distribution 
of the components and their interactions with the colloid, and (ii) from the 
interactions with the reactants, products, and solvent molecules. For brevity, 
we focus on the $A + {\rm CP} \to B + {\rm CP}$ reaction (Eq. (\ref{reaction}));
the generalization to the case of the catalytically induced dissociation
(Eq. (\ref{react_dissoc})) is straightforward and we will merely list the corresponding results.%\newline 

\subsection{Hydrodynamics of the solution}

The conservation of momentum for the mixture requires that the barycentric velocity
$\mathbf{u}$ satisfies the Navier-Stokes equations \cite{happel_brenner_book}. At 
steady state and under the assumption that the Reynolds number $Re = R \,\rho 
\,U_0/\mu$ is very small (which typically is the case for the flows generated by 
catalytically active colloids 
\cite{Golestanian2005,Popescu2009,Seifert2012,Koplik2013}), for an incompressible 
Newtonian fluid of spatially and temporally constant viscosity $\mu$ and mass 
density $\rho$, subject to the force density $\tilde\mathbf{f}(\mathbf{r})$, the Navier-Stokes equations 
are replaced by the Stokes equations \cite{happel_brenner_book}:
\begin{eqnarray}
\label{St_eq} 
 \nabla \cdot \hat\mathbf{\Pi} &=& - \tilde\mathbf{f} \Rightarrow
\mu \nabla^2 \mathbf{u} = \nabla P - \tilde\mathbf{f}\,,\nonumber\\
 \nabla \cdot \mathbf{u} &=& 0\,.
\end{eqnarray}
In these equations ${\tilde\mathbf{f}}$ denotes the \textit{external} force 
density acting 
on the mixture as a body force (such as, e.g., gravity or the forces due to the 
interaction of the molecules in the mixture with the colloid) while 
$\hat\mathbf{\Pi}$ is the Newtonian stress tensor
\begin{equation}
\label{pressure_tensor_Pi} 
 \hat\mathbf{\Pi}_{i,j} = - P \,\delta_{i,j} + 
\mu \left(\frac{\partial u_i}{\partial x_j} + \frac{\partial u_j}{\partial x_i} 
\right)\,.
\end{equation}
The scalar pressure field $P(\mathbf{r})$, which is equal to 1/3 of the trace of the
stress tensor, plays the role of an auxiliary field\footnote[5]{$P(\mathbf{r})$ is obtained as the solution of the Poisson equation 
$\nabla^2 P = \nabla \cdot \tilde\mathbf{f}$, which follows by taking 
the divergence of the 
first equation in Eq. (\ref{St_eq}), subject to the boundary condition that 
$P(r \to \infty) = P_0$, where $P_0$ is the  (spatially constant) bulk value
of the pressure. 
% Since the pressure field $P$ entering Eq. (\ref{St_eq}) is defined up to an 
% additive constant, $P_0$ can be chosen to be zero. 
Since no other boundary 
condition is imposed on $P$, the remaining integration constants are determined, 
after solving for $\mathbf{u}$, by requiring that $\nabla \cdot \mathbf{u} = 0$ is 
satisfied.} 
ensuring that the velocity field $\mathbf{u}$ obeys the 
incompressibility condition $\nabla \cdot \mathbf{u} = 0$.

Under the assumption that the mass density $\rho$ of the solution is 
spatially uniform, gravity plays no role here. Thus for the present system 
the force density $\tilde\mathbf{f}$ acting on the solution is solely due 
to the interactions of the colloid with the molecules in the mixture:
\begin{eqnarray}
\label{force_dens}
 \tilde\mathbf{f} &=& 
-\left(n_A \nabla \Phi_A + n_B \nabla \Phi_B + n_S \nabla \Phi_S \right) 
\nonumber\\
&=& - \frac{\rho}{m_S} \,\nabla \Phi_S -\left( n_A \nabla W_A + n_B \nabla W_B
\right) \nonumber\\
 &=:& - \frac{\rho}{m_S} \,\nabla \Phi_S + \mathbf{f}\,.
\end{eqnarray}
Here we have used the definition (Eq. (\ref{def_rel_pot})) of the interaction 
potentials $W_{A,B}$ and the fact that, by definition, $n_S = (\rho - n_A m_A - n_B 
m_B)/m_S$ (Eq. (\ref{mass_density})). Since $\rho$ and $m_S$ are constants, 
and noting that $\Phi_S$ depends only on $r$, the term $(\rho/m_s) \Phi_S$ can be 
included in the definition of the pressure $P$ (i.e., the isotropic part of the stress 
tensor). Thus the force exerted by the colloid on the small volume element 
$\delta \cal V$ of the solution is given by $\mathbf{f}\,\delta \cal V$, with 
$\mathbf{f} = -\left( n_A \nabla W_A + n_B \nabla W_B \right)$, and $\mathbf{f}$ 
replaces $\tilde\mathbf{f}$ in Eq. (\ref{St_eq}). 

The solution $\mathbf{u}(\mathbf{r})$ of Eq. (\ref{St_eq}) is subject to 
appropriate boundary conditions. At the surface of the colloid we assume the 
usual no-slip boundary condition to hold. Here, it is  convenient to 
consider the general case that the colloid is in motion with velocity 
$\mathbf{V} = V \mathbf{e}_z$; the case of an immobile colloid is obtained 
by setting $\mathbf{V} = 0$. In the laboratory (fixed) system of reference, 
the no-slip boundary condition on the surface of the colloid then takes the 
form 
\begin{equation}
 \label{no_slip_BC}
 \left.\mathbf{u}\right|_{r = R} = \mathbf{V}\,. 
\end{equation}
Far away from the colloid the mixture is taken to be at rest, i.e., 
\begin{equation}
 \label{quiet_BC}
 \left|\mathbf{u}(r \to \infty)\right| = 0\,.
\end{equation} 

From Eqs. (\ref{St_eq}), (\ref{force_dens}), (\ref{no_slip_BC}), and 
(\ref{quiet_BC}), with the densities $n_A$ and $n_B$ computed according to the 
steps described in Sec. \ref{conc_distr}, 
in principle
the hydrodynamic flow 
$\mathbf{u}(\mathbf{r})$ is obtained (in practice  
may be in a very involved way) as a function of the yet unknown constant 
velocity $\mathbf{V}$ of the colloid. The additional equation needed in order 
to complete the calculation follows from the condition that the colloid is in 
steady-state motion, which implies a vanishing net force acting on the colloid. 
Therefore one has
\begin{equation}
\label{force_balance}
 \int_{r = R} dS \, \hat\mathbf{\Pi} \cdot \mathbf{e}_r + \int_{\cal V} \, d^3 \bf{r} \, (-\mathbf{f})
 + \mathbf{F}_{ext} = 0\,,
\end{equation}
where the first term is the hydrodynamic force $\mathbf{F}_{hyd}$ acting on 
the colloid due to the flow of the surrounding solution, and the second term 
accounts for the force $\mathbf{F}_{chem}$ 
on the colloid due to the interaction with 
the molecules in solution. Since the colloid exerts 
a force $\mathbf{f} \, d\cal V$ on the volume element $d\cal V$ of the solution, due to Newton's third law an 
 opposite force of equal magnitude is exerted by $d\cal V$ on the colloid. 
The last term $\mathbf{F}_{ext}$ is the  sum of 
all external forces (e.g., an optical trapping) acting on the colloid. The 
typical set-up is that the external forces are given and the quantity of 
interest is the velocity of the colloid $\mathbf{V}$, which is determined from
Eq. (\ref{force_balance}). We are interested in: \newline
\indent\textbf{(i)} the particular case of force free motion, i.e., finding the 
velocity $\mathbf{V}$ if $\mathbf{F}_{ext} = 0$; \newline
\indent and\newline
\indent\textbf{(ii)} the external force $\mathbf{F}_{ext}$ needed to immobilize 
the colloid, so that $\mathbf{V} = 0$.\newline
These two quantities are calculated in, c.f., Subsec. \ref{connect_move} by using 
the reciprocal theorem \cite{Teubner1982,KiKa91} which allows one to by-pass the 
issue of solving the complex hydrodynamics problem laid out above.

Before proceeding with these analyses, a succinct discussion of each of the terms 
in Eq. (\ref{force_balance}) in terms of their experimental accessibility is in order. 
Obviously, $\mathbf{F}_{ext}$ is
that contribution which
 is  simplest to access as it accounts for 
external, prescribed forces acting on the colloid. The first and second terms 
account for contributions which seem to be difficult, if not impossible, to 
separate from each other, and thus \textit{de facto} cannot be accessed directly. The first term, 
$\mathbf{F}_{hyd}$, requires knowledge of the hydrodynamic flow in order to be able to
compute the stress tensor $\hat\mathbf{\Pi}$. (Here it is important to note 
that for the system under study the pressure part -- including the term 
$\sim \Phi_S$ -- of the stress tensor $\hat\mathbf{\Pi}$ is isotropic and 
therefore does not contribute to $\mathbf{F}_{hyd}$.) 
In principle this can be achieved,
although being technically very challenging, by using methods such as particle image 
velocimetry, as shown in Ref. \cite{Drescher2010} for swimming microorganisms. 
The second term, which we consider to be the ``chemical'' contribution $\mathbf{F}_{chem}$, can be 
calculated if the distributions of the molecular species can be measured (assuming 
the potentials $W_{A,B,\dots}$ to be known). While one can think of methods such as 
fluorescence spectroscopy to determine the spatial distribution of chemical species, if 
they are fluorescent, it is likely that in general it will be very difficult, if 
not impossible, to determine experimentally 
the dependence of these 
distributions on $r$ and $\theta$. 
However, as we show in the following subsection, the problem can be reduced to 
that of knowing only their first moment rather 
than the whole distribution, which may turn out to be a significant step towards 
rendering $\mathbf{F}_{chem}$ experimentally accessible.

\subsection{\label{force_def} Force contribution $\mathbf{F}_{chem}$ due 
to the anisotropic distributions of reactants, products, and solvent}

Once the steady state spatial distributions of reactants ($A$) and products ($B$) are 
known, the force $\mathbf{F}_{chem}$ they exert on the immobile colloid due to the interaction 
potentials $W_A$ and $W_B$ can be computed. Since the potential $\Phi_S$ is 
radially symmetric and the mass density $\rho$ is uniform due to incompressibility, 
for a spherical colloid the first term in Eq. (\ref{force_dens}) does not contribute to 
the force $\mathbf{F}_{chem} = -\int_{\cal V} d^3 \bf{r} \, \mathbf{f}$, where $\cal V$ denotes the
(macroscopic) volume of the reaction bath. Furthermore, because of the axial symmetry of the
system, only the $z$-component of this force is non-zero so that $\mathbf{F}_{chem} = F_{chem} \mathbf{e}_z$.
According to 
Eqs. (\ref{force_dens}), (\ref{sol}), and (\ref{sol1}) this component is given by
\begin{eqnarray}
\label{F_x}
 F_{chem} &=& -\mathbf{e}_{z} \cdot \int_{\cal V} d^3 \mathbf{r} \, \mathbf{f} = 
- \left[ \int_{\cal V} d^3\mathbf{r} \,\left(- n_A \nabla W_A  - n_B \nabla W_B\right)
\right] 
\cdot \mathbf{e}_z
\nonumber\\
 &=& 2 \pi \int^{\infty}_R dr \,r^2 \int^{\pi}_0 d\theta \sin \theta
\left(n_A W_A' + n_B  W'_B \right) \cos \theta \nonumber\\
 &=& \frac{4 \pi n_0}{3} \left(a_1 I_A + b_1 I_B\right)
\end{eqnarray}
where 
\begin{eqnarray}
\label{iab}
 I_A &=& \int^{\infty}_R dr \,r^2 g_1(r) \,W_A'(r) \,e^{- \beta W_A(r)}\,,
\nonumber\\
 I_B &=& \int^{\infty}_R dr \,r^2 j_1(r) \,W_B'(r) \,e^{- \beta W_B(r)}\,, 
\end{eqnarray}
and $a_1$ and $b_1$ are given, for arbitrary $K^*$, by Eq. (\ref{an}) and
Eq. (\ref{gensolute}), respectively. We note that the above expressions for $I_A$ and $I_B$ are exactly valid, while the expression in Eq. (\ref{F_x}) is approximate, because the factors $a_1$ and $b_1$ are determined  by using an approximate, self-consistent approach (see the discussion at the end of Subsec. \ref{az}).

Since $W_A$ and $W_B$ vanish for $r \to \infty$, 
in this limit the terms $\beta W_A'$ and $\beta W_B'$ multiplying $g'_n$ in
Eqs. (\ref{fp}) and (\ref{fp1}) decay more rapidly than the term $2/r$ therein and 
thus represent subdominant contributions. A straightforward perturbation theory
analysis then shows (see also \ref{append2}) that for $r \to \infty$ the leading
asymptotic decay of both $g_1$ and $j_1$ is $\sim 1/r^2$. Consequently, the
integrals in Eqs. (\ref{iab}) are finite. In the limiting case that  on the 
surface of 
the particle no reaction takes place ($K^* \to 0$), Eq. (\ref{inverse}) implies 
that
$K_{eff} = 0$ and therefore $a_1 = b_1 = 0$ [see Eqs. (\ref{an}) and 
(\ref{gensolute})].
Thus, as expected, at the steady state and in the absence of a chemical reaction
one has $F_{chem} = 0$. Obviously this result holds even if the bulk number density of $B$
molecules is non-zero (i.e., maintained at $n_0^{(B)}$ by a reservoir, as for species
$A$). Moreover, for $K > 0$, $W_A(r) = W_B(r)$, and $D_A = D_B$, i.e., if  the reaction
amounts to just a re-labeling of the molecules, one has $a_1 = - b_1$ (see Eqs.
(\ref{an}) and (\ref{gensolute})) while $I_A = I_B$. Therefore, as expected, Eq.
(\ref{F_x}) predicts that also in this case one has $F_{chem} = 0$.

In the case of diffusion-controlled reactions, i.e., for $K^* \to \infty$ (see 
Eqs. (\ref{an1}) and (\ref{soluteprofile2})) Eq. (\ref{F_x}) reduces to the simpler form
\begin{equation}
\label{mm}
 F_{chem} = \frac{4 \pi n_0 R_D}{3 R^2}  \, \Omega^{(AB)} \, \Psi(\theta_0) 
\,,~K^* \to \infty \,,
\end{equation}
where 
\begin{equation}
\label{Omega}
 \Omega^{(AB)} = \frac{e^{\beta W_A(R)}}{g'_1(R)} I_A -
\frac{D_A}{D_B} \, \frac{e^{\beta W_B(R)}}{j'_1(R)} I_B
\end{equation}
depends on both interaction potentials and on the ratio of the diffusion coefficients 
$D_A/D_B$, but it is independent of the patch size (characterized by $\theta_0$).
On the other hand, the function $\Psi(\theta_0)$ absorbs the whole dependence of 
the force on $\theta_0$:
\begin{equation}
\label{Phi}
 \Psi(\theta_0) = \frac{f_{dc}(\theta_0) \, \phi_1(\theta_0)}{\phi_0(\theta_0)}\, \geq 0 \,,
\end{equation}
where $f_{dc}(\theta_0)$ is given by Eq. (\ref{ster}) (and thus it is determined by $W_A$), 
while $\phi_0(\theta_0) = 2 \sin^2(\theta_0/2) \geq 0$ and $\phi_1(\theta_0) = (3/2)
\sin^2(\theta_0) \geq 0$ (Eq.~(\ref{phi})) are functions of $\theta_0$ only.

Several general conclusions can be drawn from Eq.~(\ref{mm}), i.e., in the 
diffusion-controlled limit, upon setting  formally $K^* = \infty$:
\begin{itemize}
\item If the $A$ and the $B$ particles have the same interaction potentials with
the colloid, i.e., $W_A(r) \equiv W_B(r)$, so that $I_A = I_B$ and $j_1(r) = g_1(r)$ (since
$g_1$ and $j_1$ fulfill the same differential equation with the same boundary condition to
vanish for $r \to \infty$), but have different diffusion coefficients, $D_A \neq D_B$, one 
has (Eqs. (\ref{an}), (\ref{gensolute}), and (\ref{F_x})) $F_{chem} \sim (1/D_A)(1 - D_A/D_B)$. 
Therefore its sign depends on the ratio of the diffusion coefficients. This holds beyond the 
diffusion-controlled limit $K^* = \infty$ (Eq. (\ref{F_x})). 

\item For a completely catalytic colloid ($\theta_0 = \pi$) or a chemically inert
one ($\theta_0 = 0$) one has $\Psi(\theta_0 = 0, \pi) = 0$. For $\theta_0 = \pi$, 
$\phi_0(\pi) = 2$ is nonzero, while $\phi_1(\pi) = 0$ and the effective steric factor attains 
the value $f_{dc}(\pi) = 1$ (see the discussion following Eq. (\ref{ster1})). 
For $\theta_0 = 0$ this is the case because the ratio $\phi_1(0)/\phi_0(0) = 3$ is finite, 
while $f_{dc}(0)$ is expected to vanish (see the discussion following Eq. (\ref{ster1})).
Therefore, as expected, in these cases there is no force $F_{chem}$ acting on the particle 
because the system exhibits full radial symmetry. We remark, however, that our model is a 
mean-field one, so that one can not rule out the possibility that there will be a non-zero 
force due to fluctuations in spatial distributions of the reactants and products 
(see, e.g., the situations discussed in Refs. \cite{Lauga2013,Buyl2013}).

\item Since $\Psi(\theta_0)$ is continuous and $\Psi(0) = \Psi(\pi) =  0$, there
exists a catalyst coverage characterized by $0 < {\tilde\theta}_0 < \pi$ for which 
$|F_{chem}|$ is maximal; ${\tilde\theta}_0$ is a functional of $W_A(r)$ and $W_B(r)$.

\item $F_{chem}$ can be positive or negative, or equal to zero, depending on the relations 
between $W_A(r)$ and $W_B(r)$ and those between the diffusion coefficients. But the 
expression in Eq. (\ref{Omega}) is too complex to allow one to obtain analytically exact 
criteria for certain properties of $F_{chem}$.
\end{itemize}

\noindent In the opposite limit, i.e., in the kinetically-controlled regime ($D_A \to
\infty$, $K_{eff} \to K^*$ fixed), one has
(Eqs. (\ref{an2}), (\ref{gensolute19}), (\ref{F_x}), (\ref{Omega}),
(\ref{constantK}), and (\ref{K*_def}))
\begin{eqnarray}
\label{kinetic}
 F_{chem} &=& \frac{n_0 \, K^* \, \cos^2 (\theta_0/2)}{R^2} 
\left(\frac{I_A \, 
e^{\beta W_A(R)}}{D_A \, g_1'(R)} - \frac{I_B \, e^{\beta W_B(R)}}{D_B \, 
j_1'(R)} \right) \nonumber \\
 &=& \pi n_0 \frac{a W_0}{D_A} 
e^{- \beta W_A(R)}\Omega^{(AB)} (\sin \theta_0)^2
\,,
\end{eqnarray}
which shows that the force $F_{chem}$ vanishes in perfectly stirred systems, 
in which both $D_A$ and $D_B$ tend to infinity. For $D_{A,B}$ very large but 
finite, it is obvious that all the above bullet points apply to this case, too. 
Moreover, in this case the strength of the force attains its maximum at 
$\theta_0 = \pi/2$, i.e., for a symmetric Janus colloid. 

Lastly, we note that the result in Eq. (\ref{F_x}) can be straightforwardly generalized 
in order to cover the dissociation reaction described in Eq. (\ref{react_dissoc}). For this case we find that
the force $F_{chem}$ exerted on the 
colloid by the ``frozen'' spatial distributions of the $A$, $B$, $C$, and $S$ 
molecules due to the interaction potentials $\Phi_A$, $\Phi_B$, $\Phi_C$, and 
$\Phi_S$ (compare Eq. (\ref{force_dens})), respectively, is given by 
(compare Eq. (\ref{F_x}))
\begin{eqnarray}
\label{F_xABC}
 F_{chem} &=& - \left[ \int_{\cal V} d^3\mathbf{r} \,\left(- n_A \nabla W_A  -
n_B \nabla W_B - n_C \nabla W_C \right) \right] \cdot \, \mathbf{e}_z
\nonumber\\
 &=& \frac{4 \pi n_0}{3} \left(a_1 I_A + b_1 I_B + \gamma_1 I_C\right)\,,
\end{eqnarray}
where, for arbitrary $K^*$, 
$a_1$, $b_1$, and $\gamma_1$ are given by the corresponding
Eqs.~(\ref{an}), (\ref{gensolute}), and (\ref{cn}), respectively, 
the
expressions $I_A$ and $I_B$ are defined in Eq. (\ref{iab}), and
\begin{eqnarray}
\label{ic}
 I_C = \int^{\infty}_R dr \,r^2 \, h_1(r) \,W_C'(r) \,e^{- \beta W_C(r)}\,.
\end{eqnarray}
(Similarly as the expressions for $I_A$ and $I_B$, the expression $I_C$ for the force integral is exact, too.)

Accordingly, using Eqs. (\ref{an1}), (\ref{soluteprofile2}), and (\ref{soluteprofile4})
one finds for the diffusion-controlled regime (compare Eq. (\ref{mm}))
\begin{equation}
 F_{chem}(K^* \to \infty) = 
\frac{4 \pi n_0 R_D}{3 R^2}  \, \Omega^{(ABC)} \, \Psi(\theta_0) \,.
\end{equation}
$\Omega^{(ABC)}$ depends on the relative interaction potentials $W_A$, $W_B$, and
$W_C$, as well as on the ratios $D_A/D_B$ and $D_A/D_C$ of the diffusion coefficients: 
\begin{equation}
\label{OmegaABC}
\Omega^{(ABC)} = \frac{e^{\beta W_A(R)}}{g'_1(R)} I_A -
\frac{D_A}{D_B} \, \frac{e^{\beta W_B(R)}}{j'_1(R)} I_B - \frac{D_A}{D_C} \, 
\frac{e^{\beta W_C(R)}}{h'_1(R)} I_C\,.
\end{equation}
But $\Omega^{(ABC)}$ is independent of the catalytic patch size characterized by
$\theta_0$. As before, the whole dependence on $\theta_0$ is captured by the factor
$\Psi(\theta_0)$ defined in Eq. (\ref{Phi}).

For the kinetically-controlled regime one has (compare Eq. (\ref{kinetic}))
\begin{equation}
\label{kinetic2}
F_{chem}  = \pi n_0 \frac{a W_0}{D_A} e^{- \beta W_A(R)}\Omega^{(ABC)} (\sin \theta_0)^2 \,.
\end{equation}
The force $F_{chem}$ vanishes in perfectly stirred systems, in which $D_A$, $D_B$, and $D_C$ tend to
infinity.

\subsection{\label{connect_move} The velocity of a force-free catalytically-active colloid
and the force needed 
to immobilize it }

At steady state and upon neglecting the thermal fluctuations giving rise to
rotational diffusion of the symmetry axis in $z$ direction, the results obtained 
so far can be straightforwardly employed to determine the velocity 
$\mathbf{V} = V {\mathbf{e}}_z$ of an active colloid in 
``force free'' (i.e., $\mathbf{F}_{ext} = 0$) motion, or the external force  
$\mathbf{F}_{ext} = F_{ext} \mathbf{e}_z$ needed to immobilize  the active colloid (i.e., 
$\mathbf{V} = 0$).

The calculation is based on employing the generalized reciprocal theorem due to 
Teubner (see 
 \ref{reciproc_theorem}) \cite{Teubner1982,KiKa91,Seifert2012,Seifert2012b}. 
This states that for two incompressible Newtonian flows characterized by
$\{\mu,\mathbf{u},\hat\mathbf{\Pi}\}$ and $\{\mu',\mathbf{u}',
\hat\mathbf{\Pi}'\}$, respectively, and solving the Stokes equations (Eq.
(\ref{St_eq})) for force densities $\mathbf{f}$ and $\mathbf{f}'$, 
respectively, within the domain $\cal{D}$ exterior to a closed surface 
$\partial \cal{D}$, whereby $\hat\mathbf{\Pi}$, $\mathbf{u}$, and $\mathbf{f}$,
as well as the primed ones, vanish sufficiently fast with increasing distance from 
$\partial \cal{D}$, the following relation holds \cite{Teubner1982}:
\begin{equation}
 \label{Teubner}
\mu' \left[\,\,\int \limits_{\partial {\cal D}} \mathbf{u}' \cdot \hat\mathbf{\Pi} 
\cdot d\mathbf{s} 
- \int\limits_{{\cal D}} \mathbf{u}' \cdot \mathbf{f} \, d^3\mathbf{r}\,\right] = 
\mu \left[\,\,\int \limits_{\partial {\cal D}} \mathbf{u} \cdot \hat\mathbf{\Pi}' 
\cdot d\mathbf{s} - \int\limits_{{\cal D}} \mathbf{u} \cdot \mathbf{f}' 
\,d^3\mathbf{r} \,\right]\,.
\end{equation}
Here the orientation of the surface element $d\mathbf{s} := dS \mathbf{e}_r$ 
is the one given by the \textit{inner} normal (i.e., pointing 
\textit{into} the fluid domain ${\cal D}$).

For the unprimed system we choose the fluid motion as the one 
corresponding to the self-propelled colloid moving with velocity 
$\mathbf{V} = V {\mathbf{e}}_z$ (in this case $\partial {\cal D}$ is the 
surface $r = R$ of the sphere and the fluid domain ${\cal D}$ is the volume 
${\cal V}$ of the solution):
\begin{eqnarray}
 \label{St_sol_1}
 &&\nabla \cdot \hat \mathbf \Pi = - \mathbf{f} := 
 - f {\mathbf{e}}_r\,, 
~~\nabla \cdot \mathbf u 
= 0\,,\nonumber\\
  &&\mathbf{u}(r = R) = V {\mathbf{e}}_z \,, ~~\mathbf{u}(r \to \infty) = 0\,,
\end{eqnarray}
where in the first equation we have used the fact that $\mathbf{f}$ has only a 
radial component $f := \mathbf{f} \cdot {\mathbf{e}}_r$ (due to $W_{A,B}$ 
being radially symmetric, see Eq. (\ref{force_dens})).

For the primed system we choose the one corresponding to the motion of 
the fluid (of the same viscosity, i.e., $\mu' = \mu$) due to a sphere of 
radius $R$ translating along the $z$-direction with uniform velocity 
$\mathbf{U}_0 = U_0 \, {\mathbf{e}}_z$:
\begin{eqnarray}
 \label{St_sol_2}
 &&\nabla \cdot \hat \mathbf \Pi' = 0 \,, ~~\nabla \cdot 
 \mathbf u' = 0\,,\nonumber\\
 &&\mathbf{u'}(r = R) = U_0 \, {\mathbf{e}}_z \,, ~~\mathbf{u'}(r \to \infty) = 0\,.
\end{eqnarray}
For the latter system the solution is known and the axially symmetric velocity field 
$\mathbf{u'}$ is given in spherical coordinates by (see, e.g., Ch. 4 in Ref. 
\cite{happel_brenner_book})
\begin{eqnarray}
 \label{flow_translate}
 u'_r &=& \frac{U_0}{2} \left[3 \left(\frac{R}{r}\right) - 
\left(\frac{R}{r}\right)^3 \right] \cos\theta \, := U_0 \,\xi(r) \cos\theta\,, 
\nonumber\\
 u'_\theta &=& -\frac{U_0}{4} \left[3 \left(\frac{R}{r}\right) + 
\left(\frac{R}{r}\right)^3 \right] \sin\theta\,.
\end{eqnarray}

By using Eqs. (\ref{St_sol_1}) and (\ref{St_sol_2}), and noting that in the 
latter case $\mathbf{f}' \equiv 0$ while $\mathbf{U}_0$ and $\mathbf{V}$ are 
constants with respect to the integration over the spherical surfaces, Eq. 
(\ref{Teubner}) leads to
\begin{equation}
\label{first_step}
 \mathbf{U}_0\, \cdot \int\limits_{r = R}  \, \hat\mathbf{\Pi} 
 \cdot \mathbf{e}_r \, dS  -  
\int\limits_{\cal V} \mathbf{u'} \cdot \mathbf{f} \, d^3 \mathbf{r} = 
\mathbf{V}\, \cdot 
\int\limits_{r = R} \hat\mathbf{\Pi}' \cdot d\mathbf{s}\,.
\end{equation}
%\SD{where $\mathbf{V} = V \hat \mathbf{e}_z$.}
The last integral is the drag force acting on the translating sphere and it is 
given by the well known Stokes formula (see, e.g., Ch. 4 in Ref. 
\cite{happel_brenner_book}), $\int\limits_{r = R} \hat\mathbf{\Pi}' \cdot 
d\mathbf{s} = - 6 \pi \mu R \mathbf{U}_0$. By noting that the surface integral 
on the left hand side of Eq. (\ref{first_step}) is the contribution 
$\mathbf{F}_{hyd}$ of the force acting on the colloid, and by using Eq. 
(\ref{force_balance}) to eliminate it in favor of the other two force 
contributions, we arrive at
\begin{equation}
 \label{velocity_force_connect}
 \mathbf{U}_0 \cdot \left[- \mathbf{F}_{ext} + 6 \pi \mu R \mathbf{V}\right] = 
 \int\limits_{\cal V} d^3 \mathbf{r} \left(\mathbf{u'} -  \mathbf{U}_0 \right) \cdot 
 \mathbf{f}\,.
 \end{equation}
The two cases of interest are immediately obtained form the general result in 
Eq. (\ref{velocity_force_connect}) as follows.\newline

%\pagebreak

\noindent\textbf{(i)} \textit{Velocity of the force free colloid}.\newline

\noindent With $\mathbf{F}_{ext} = 0$, and by noting that 
$\cos\theta = {\mathbf{e}}_z \cdot 
{\mathbf{e}}_r$, Eq. (\ref{velocity_force_connect}) renders
\begin{eqnarray}
 \label{vel_V}
 V & =& \frac{1}{6 \pi \mu R} \int\limits_{\cal V} \,d^3 \mathbf{r}\, 
\left[\xi(r) - 1 \right] f \cos\theta \nonumber\\
  &=& 
  \frac{1}{6 \pi \mu R} \int\limits_{\cal V} d^3 \mathbf{r} \,
\left[ \frac{3}{2} \left(\frac{R}{r}\right) - 
\frac{1}{2} \left(\frac{R}{r}\right)^3 - 1 \right] f \cos\theta\,.
\end{eqnarray}
(We recall that $f = \mathbf{f} \cdot {\mathbf{e}}_r$ with $\mathbf{f}$ 
given by Eq. (\ref{force_dens}).)\newline

As the next step, using the equality in the first line in Eq. (\ref{F_x}), we express the self-propulsion 
velocity $V$ 
in terms of
the coefficients with index $n = 1$ in 
the expansions of the densities of the reactants and products. For the simplest type of reaction
$A + {\rm CP} \to B + {\rm CP}$, one has
\begin{equation}
\label{Vas}
 V = \frac{2 n_0}{9 \mu R} \left(a_1 J_A + b_1 J_B\right) \,,
\end{equation}
where
\begin{equation}
\label{JA}
 J_A = \int^{\infty}_R dr \, r^2  g_1(r) W'_A(r) \left(1+ \frac{1}{2} 
\left(\frac{R}{r}\right)^3 - \frac{3}{2} \frac{R}{r}\right) e^{- \beta W_A(r)}
\end{equation}
and
\begin{equation}
\label{JB}
 J_B = \int^{\infty}_R dr \, r^2  j_1(r) W'_B(r) \left(1+ \frac{1}{2} 
\left(\frac{R}{r}\right)^3 - \frac{3}{2} \frac{R}{r}\right) e^{- \beta W_B(r)} \,.
\end{equation}
Similarly as the expressions $I_A$ and $I_B$ for
the force integrals, the expressions $J_A$ and $J_B$ for the velocity integrals are exact, too.

In turn, for a more general
catalytically-induced dissociation in Eq.(\ref{react_dissoc}) 
we obtain
\begin{equation}
\label{Vdis}
 V = \frac{2 n_0}{9 \mu R} \left(a_1 J_A + b_1 J_B + \gamma_1 J_C\right) \,,
\end{equation}
where $\gamma_1$ is defined in Eq. (\ref{cn}) and
\begin{equation}
\label{JC}
 J_C = \int^{\infty}_R dr \, r^2 h_1(r) W'_C(r) \left(1+ \frac{1}{2} 
\left(\frac{R}{r}\right)^3 - \frac{3}{2} \frac{R}{r}\right) e^{- \beta W_C(r)} \,.
\end{equation}

\noindent\textbf{(ii)} Stall force $F_{ext}$ needed to immobilize a colloid.
 \newline

\noindent
With $\mathbf{V} = 0$ and $\cos\theta = {\mathbf{e}}_z \cdot 
{\mathbf{e}}_r$, Eq. (\ref{velocity_force_connect}) yields
\begin{equation}
\label{force_to_stop}
 F_{ext} = - \int\limits_{\cal V} \,d^3 \mathbf{r}\, \left[\xi(r) - 1 \right] f \cos\theta\,.
\end{equation}
The stall force is proportional to the velocity $V_{free}$ (Eq. (\ref{vel_V})) 
of a force-free colloid, 
\begin{equation}
\label{stall}
 F_{ext} = - 6 \pi \mu R \, V_{free} \,,
\end{equation}
and has the same form as the Stokes formula. This result 
follows directly from Eq. (\ref{velocity_force_connect}). Therefore it is 
very general, in the sense that this result does not depend on the details of the 
chemical activity but only on the assumptions of a spherical shape, steady state 
motion, and (negligibly) small Re and Pe numbers.

Therefore, by using  Eqs. (\ref{Vas}) and (\ref{Vdis})
$F_{ext}$ can be expressed explicitly as
\begin{equation}
\label{stall1}
 F_{ext} = - \frac{4 \pi n_0}{3}  \left(a_1 J_A + b_1 J_B\right) \,, \,\,\, A + {\rm CP} \to B + {\rm CP}
\end{equation}
and as
\begin{equation}
\label{stall2}
 F_{ext} = - \frac{4 \pi n_0}{3}  \left(a_1 J_A + b_1 J_B + \gamma_1 J_C\right) \,, \,\,\, A + {\rm CP} \to B + C + {\rm CP} \,,
\end{equation}
respectively.

\section{\label{disc}Triangular-well potentials as a case study}

\begin{figure}[t]
\begin{center}
\centerline{\includegraphics[width = .6 \textwidth]{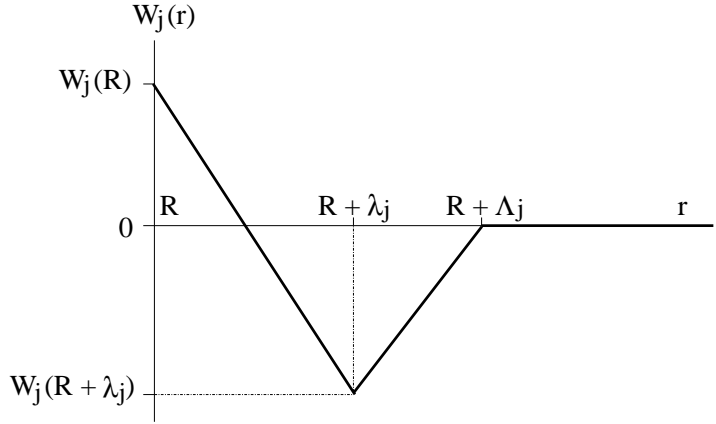}}
\caption{Illustration of the triangular-well potential $W_j(r)$ 
defined by Eqs. (\ref{<}) and (\ref{I}).}
\label{fig77}
\end{center}
\end{figure}

We illustrate the general results derived in the previous section by choosing
a particular form of the interaction potential between the colloid and the various 
molecular species involved in the reaction. This allows us to obtain sufficiently 
simple explicit expressions for the force contribution $\mathbf{F}_{chem}$ (Eq. 
(\ref{F_xABC})), the stall force $\mathbf{F}_{ext}$ (Eq. (\ref{stall})), 
or, equivalently, the velocity $\mathbf{V}$ (Eq. (\ref{vel_V})) of a force-free colloid, 
such that their dependences on the strength of the interactions, catalytic coverage, chemical rates, diffusion coefficients of 
the reactants and products,
and  radius of the colloid can be determined and discussed.
This choice is provided by the so-called triangular-well potentials 
(depicted in Fig. \ref{fig77}) which, despite their simplicity, capture well 
the thermodynamic properties of fluids with Lennard-Jones pair interactions, 
provided the defining parameters of the triangular-well potentials are tuned 
suitably (see, e.g., Refs. \cite{fowler1965,Barcenas2015,Sengupta2016} for 
a detailed discussion of this issue).  
For this choice of the potential one can derive \textit{explicit} formulae for the 
chemical and the stall force, which allows one to highlight several counter-intuitive 
effects concerning diffusiophoretic self-propulsion.

We stipulate that the potential $W_j(r)$ for the interactions of the colloid 
with the $j$-th species ($j=A,B,C$) is a piece-wise continuous function defined 
as follows (see Fig. \ref{fig77}):
\begin{equation}
\label{<}
 W_j(R \leq r \leq R + \lambda_j)  \, \equiv \,  W^{<}_j(r) = - \xi_1^{j} \, r + 
\left(W_j(R) + \xi_1^{j} R\right) \,,
\end{equation}
where
\begin{equation}
\label{xi1}
 \xi_1^{j} = \frac{W_j(R) - W_j(R + \lambda_j)}{\lambda_j} 
= \frac{\Delta \varepsilon_{j}}{\beta \lambda_{j}}
\end{equation}
with
\begin{equation}
 \Delta \varepsilon_{j} \, \equiv \,  \varepsilon_{w}^{j} - \varepsilon_{m}^{j} 
= \beta W_j(R) - \beta W_j(R+\lambda_j)
\end{equation}
in the inner region $r \in (R, R+\lambda_j)$;\newline
\begin{equation}
\label{I}
 W_j(R+ \lambda_j \leq r \leq R + \Lambda_j) \, \equiv \,  W^{I}_j(r) = \xi_2^{j} \, r - \xi_2^{j}(R + \Lambda_j)  \,,
\end{equation}
where
\begin{equation}
\label{xi2}
 \xi_2^{j} = - \frac{\varepsilon_{m}^{j}}{\beta \left(\Lambda_j - \lambda_j\right)} \,,
\end{equation}
in the intermediate region $r \in (R+\lambda_j,R+\Lambda_j)$. In the outer region 
$r \in (R + \Lambda_j,\infty)$ the potential vanishes: 
\begin{equation}
\label{>}
 W_j(r \geq R + \Lambda_j) \, \equiv \, W_j^{>}(r) \equiv 0 \,.
\end{equation}
The parameters (the value at the wall, the depth, and the location of 
the well) defining the interactions  of the colloid with the $j$-th species are, 
in general, different. In the following we shall focus the discussion of the 
results on the situation depicted in Fig. \ref{fig77}, i.e., $\Delta \varepsilon_j 
\geq 0$ and $\varepsilon_m^{j} \leq 0$. (Nonetheless, the derivation of the 
results is carried out for both signs of these parameters.)

Furthermore, we introduce the dimensionless parameters $q_j = 
\lambda_j/R$, $Q_j=\Lambda_j/R$, and $z_j = Q_j/q_j \geq 1$. The derivation of 
the radial functions $g_n(r)$ (Eq. (\ref{fp})) for arbitrary values of the 
parameters $q_A$ and $Q_A$, $\varepsilon^{A}_m$ and $\varepsilon_w^{A}$ is detailed 
in  \ref{append2}. (The corresponding results for the other 
species follow, as in Sec. \ref{conc_distr}, by simply switching label $A$ 
with $B$ or $C$.) Based on these functions, in \ref{append2} 
we also derive the general expressions for the integrals $I_j$ (Eq. 
(\ref{iab})) and $J_j$ ( Eqs. (\ref{JA}), (\ref{JB}), and 
(\ref{JC})) for $j = A, B, C$. Here we restrict the discussion to the 
physically meaningful limit $\lambda_j/R \ll 1$ (at fixed $z_j \geq 1$) and retain 
only the leading order term. For colloids of radii $R \gtrsim 1~\mathrm{\mu m}$ 
and typical surface interactions, the parameter $\lambda_j/R$ is expected to be of 
the order of $10^{-2}$ or smaller. Thus the corrections $\Or(\lambda_j/R)$
to the leading order term (which are included in \ref{append2}) are 
expected to be negligibly small.

\subsection{The contribution $\mathbf{F}_{chem}$ due to the direct 
interactions of the colloid with the molecular species}    

In leading order in $\lambda_j/R$, the force integrals (Eq. (\ref{iab})) are given 
by
 \begin{equation}
 \label{forceintegrals}
 I_j = - \frac{\left(1 - e^{- \varepsilon_w^j}\right)}{\beta} R^2 \,.
 \end{equation}
This shows that, somewhat surprisingly, the leading order value of $I_j$ is 
controlled only by the value of the potential at the wall 
and is independent of the depth $\varepsilon_m^j$ of the well.  
The leading order value of $I_j$ is negative for $\varepsilon_w^j > 0$ 
(repulsion from the wall) and positive for  $\varepsilon_w^j < 0$ (attraction to 
the wall). The corrections depend on $\varepsilon_m^j$ and may be positive 
or negative hinging on the relation between $\varepsilon_m^j$ and 
$\varepsilon_w^j$ (see Eq. (\ref{Ioverall}) in \ref{append2}).

In this limit the Debye radius (Eq. (\ref{raddeb})) is simply given by 
$R_D = R$ (the correction terms are given by Eq. (\ref{debyeapp}) in 
 \ref{append2}), the steric factor $f_{dc}(\theta_0)$ (Eq. (\ref{ster})) 
turns into $f_{dc}(\theta_0) = f_{dc}^{(sls)}(\theta_0)$ 
(Eq. (\ref{ster1})), and the derivative of the first radial function at 
the wall is
 \begin{equation}
 \label{lead1}
  g'_1(R) = - \frac{2 e^{\varepsilon_w^{A}}}{R} \,.
 \end{equation} 
Consequently, we find (Eqs. (\ref{inverse}), (\ref{sd}), and (\ref{K*_def}))
that in this limit the coefficient $a_1$ (Eq. (\ref{an})) is
\begin{equation}
 a_1 =  -\frac{3}{8} \frac{\kappa \, 
e^{-\varepsilon_w^{A}} \, R }{\left(\kappa \, 
e^{-\varepsilon_w^{A}} \, R \, \sin^2(\theta_0/2) + D_A \, f_{dc}^{(sls)}
(\theta_0)\right)} \sin^2(\theta_0) f_{dc}^{(sls)}(\theta_0)\,;
\end{equation}
therefore, the contribution of the $A$ molecules to the force 
$F_{chem}$ (Eq. (\ref{F_x})) is given by
\begin{equation}
\label{A}
F^{A}_{chem} = \frac{ \pi n_0}{2 \,\beta} \frac{\kappa \, e^{-\varepsilon_w^{A}} 
\left(1-  e^{-\varepsilon_w^{A}}  \right) \, R^3 }{\left(\kappa 
\, e^{-\varepsilon_w^{A}} 
\, R \sin^2(\theta_0/2) + D_A \, f_{dc}^{(sls)}(\theta_0)\right)} \, 
\sin^2(\theta_0) \,  f_{dc}^{(sls)}(\theta_0) \,.
\end{equation}
In this limit, the sign of $F^{A}_{chem}$ is determined fully by the sign of 
$\varepsilon_w^{A}$: for $\varepsilon_w^{A} > 0$, i.e., if the $A$ molecules 
are repelled from the surface of the colloid, $F^{A}_{chem}$ is positive. This is a 
simple consequence of the fact that, due to the reaction, the $A$ 
molecules are depleted in the region $z > 0$, as compared to the region 
at the other side of the catalytic patch, which effectively pushes the colloid 
into the positive $z$-direction. We further remark that $F^{A}_{chem}$ is a 
non-monotonic function of $\varepsilon_w^{A} $ and vanishes if either 
$\varepsilon_w^{A}  \to 0$ or $\varepsilon_w^{A}  \to \infty$. The latter can be 
understood by noting that an infinitely strong repulsion prevents the $A$ 
molecules from approaching the catalytic patch and hence completely suppresses 
the reaction, which renders a radially symmetric distribution.

We proceed with the calculation of the coefficients $b_1$ and $\gamma_1$ defined in 
Eqs. (\ref{gensolute}) and (\ref{cn}). Since the derivatives of the radial functions 
$j_1(r)$ and $h_1(r)$, defined by Eqs. (\ref{fp1}) and (\ref{fp4}), have the same 
form as $g'_1(R)$ but with $\varepsilon_w^{A}$ replaced by $\varepsilon_w^{B}$ or 
$\varepsilon_w^{C}$, in leading order in $\lambda_j/R$ the coefficients $b_1$ and 
$\gamma_1$ are given by
\begin{equation}
b_1 = \frac{3 D_A}{8 D_B} 
\frac{\kappa \, e^{-\varepsilon_w^{A}} \, R }{\left(\kappa \, e^{-\varepsilon_w^{A}} \, 
R \, \sin^2(\theta_0/2) + D_A \, f_{dc}^{(sls)}(\theta_0)\right)} \sin^2(\theta_0) 
f_{dc}^{(sls)}(\theta_0)\,,
\end{equation}
and
\begin{equation}
 \gamma_1 = \frac{3 D_A}{8 D_C} \frac{\kappa \, e^{-\varepsilon_w^{A}} \, R }{\left(\kappa \, 
e^{-\varepsilon_w^{A}} \, R \, \sin^2(\theta_0/2) + D_A \, f_{dc}^{(sls)}
(\theta_0)\right)} \sin^2(\theta_0) \, f_{dc}^{(sls)}(\theta_0) \,.
\end{equation}
Consequently, we arrive at the following explicit expressions for the force 
$F_{chem}$: \newline
\textbullet~the case $A + {\rm CP} \to B + {\rm CP}$ :
\begin{eqnarray}
\label{example}
\fl F_{chem} &=& \frac{\pi n_0 R^3 D_A}{2 \beta} 
\frac{\kappa \, e^{-\varepsilon_w^{A}}}{\left(\kappa \, e^{-\varepsilon_w^{A}} \, 
R \sin^2(\theta_0/2) + D_A \, f_{dc}^{(sls)}(\theta_0)\right)} \,  \nonumber\\
\fl  &\times& \left(\frac{\left(1-  e^{-\varepsilon_w^{A}}  \right)}{D_A} -  
\frac{\left(1-  e^{-\varepsilon_w^{B}}  \right)}{D_B}\right) \sin^2(\theta_0) \,  
f_{dc}^{(sls)}(\theta_0)\,;
\end{eqnarray}
\textbullet~the case  $A + {\rm CP} \to B + C + {\rm CP}$ (dissociation reaction):
\begin{eqnarray}
\label{example0}
\fl F_{chem} &=& \frac{\pi n_0 R^3 D_A}{2 \beta} 
\frac{\kappa \, e^{-\varepsilon_w^{A}}}{\left(\kappa \, e^{-\varepsilon_w^{A}} \, 
R \sin^2(\theta_0/2) + D_A \, f_{dc}^{(sls)}(\theta_0)\right)} \, \nonumber\\
\fl  &\times& \left(\frac{\left(1-  e^{-\varepsilon_w^{A}}  \right)}{D_A} -  
\frac{\left(1-  e^{-\varepsilon_w^{B}}  \right)}{D_B} - \frac{\left(1-  
e^{-\varepsilon_w^{C}}  \right)}{D_C}\right) \sin^2(\theta_0) \,  f_{dc}^{(sls)}
(\theta_0)\,.
\end{eqnarray}
If the values $\varepsilon_w^{j}$ are equal to each other, i.e.,
$\varepsilon_w^{A} = \varepsilon_w^{B} = \varepsilon_w^{C}$, 
Eqs. (\ref{example}) and (\ref{example0}) take the simple 
forms ($A + {\rm CP} \to B + {\rm CP}$)
 \begin{eqnarray}
\label{example01}
F_{chem} &=& \frac{\pi n_0 R^3 D_A}{2 \beta} \frac{\kappa \, e^{-\varepsilon_w^{A}} 
\left(1-  e^{-\varepsilon_w^{A}}  \right)}{\left(\kappa \, 
e^{-\varepsilon_w^{A}} \, 
R \sin^2(\theta_0/2) + D_A \, f_{dc}^{(sls)}(\theta_0)\right)} \, \nonumber\\
 &\times& \left(\frac{1}{D_A} -  \frac{1}{D_B}\right) \sin^2(\theta_0) \,
f_{dc}^{(sls)}(\theta_0)
\end{eqnarray}
and ($A + {\rm CP} \to B + C + {\rm CP}$)
\begin{eqnarray}
\label{example00}
F_{chem} &=& \frac{\pi n_0 R^3 D_A}{2 \beta} \frac{\kappa \, e^{-\varepsilon_w^{A}} 
\left(1-  e^{-\varepsilon_w^{A}}  \right)}{\left(\kappa \, e^{-\varepsilon_w^{A}} \, 
R \sin^2(\theta_0/2) + D_A \, f_{dc}^{(sls)}(\theta_0)\right)} \, \nonumber\\
 &\times& \left(\frac{1}{D_A} -  \frac{1}{D_B} - \frac{1}{D_C}\right) 
\sin^2(\theta_0) \,  f_{dc}^{(sls)}(\theta_0)\,,
\end{eqnarray} 
respectively. Thus, in this case the sign of $F_{chem}$ is determined by the 
relation between the diffusion coefficients of the reactive species.

Focusing here on the case of the reaction $A + {\rm CP} \to B + {\rm CP}$, we 
discuss the limiting behavior of the force $F_{chem}$ in Eq. (\ref{example}), 
rendering explicit results for several general conclusions presented above in 
Subsec. \ref{force_def}. To this end, we first consider the so-called 
diffusion-limited case and assume that the parameters of the model obey 
the following inequality: 
\begin{equation}
\label{diffusion-control}
 \kappa  e^{-\varepsilon_w^{A}} R \sin^2(\theta_0/2) 
\gg D_A  f_{dc}^{(sls)}(\theta_0) \,. 
\end{equation}
(We remark that, in contrast to the case discussed by Collins and Kimball, 
(Eq. (\ref{Collins_Kimball})), here the lhs of the 
inequality depends also on $\varepsilon_w^{A}$ and both sides depend on $\theta_0$. 
Therefore, this limit has to be taken with appropriate caution, if one intends to 
study the dependence of $F_{chem}$ on $\theta_0$ in the limit 
$\theta_0 \to 0$.) Since $\sin^2(\theta_0/2) \sim \theta_0^2$, while  $f_{dc}^{(sls)}
(\theta_0) \sim \theta_0$, for fixed $\kappa$, $R$, $D_A$, and 
$\varepsilon_w^{A}$, which obey the inequality in Eq. (\ref{diffusion-control}) for 
moderate $\theta_0$, the sign of the inequality changes for sufficiently 
small values of $\theta_0$. In other words, as we have already noted in the 
text after Eq. (\ref{ster1}), for $\theta_0 \to 0$ there is always the 
kinetically-controlled regime, but not the diffusion-controlled one. 
Similarly, the dependence of $F_{chem}$ on $\varepsilon_w^{A}$ (as well as 
the one of $F_{ext}$ or, equivalently, $V$, which will be discussed below) 
obtained for this regime cannot be extrapolated to arbitrarily large 
values of $\varepsilon_w^{A}$. The reason for this is that
the effective reduction of the reactivity due to an increase of the repulsion 
at the wall (and hence, on the catalytic patch) ultimately causes a reversal  
of the inequality in Eq. (\ref{diffusion-control}).  

In the diffusion-controlled limit Eq. (\ref{example}) attains the form
\begin{eqnarray}
\label{example-diff-cont}
F_{chem} &= &\frac{2 \pi n_0 R^2}{\beta}  \, 
\left(1 -   e^{-\varepsilon_w^{A}}  -  
\frac{D_A}{D_B} \left(1-  e^{-\varepsilon_w^{B}}  \right)\right) \nonumber\\
&\times& \cos^2(\theta_0/2) 
\,  f_{dc}^{(sls)}(\theta_0) \,.
\end{eqnarray}
In this regime, the force is proportional to $R^2$ (i.e., the \textit{area} of 
the colloid) and attains its maximal value for $\theta_0$ close to $\pi/2$ (but 
not exactly equal to it, see the discussion in Subsec. \ref{force_def}). The 
sign of the force depends on the diffusion coefficients of both species and on 
the amplitudes of their interaction potentials with the wall. For
\begin{equation}
 \frac{\left(1-  e^{-\varepsilon_w^{A}}  \right)}{D_A}  >  
\frac{\left(1-  e^{-\varepsilon_w^{B}}  \right)}{D_B}
\end{equation} 
the force $F_{chem}$ is positive, while it is negative if the inequality is 
reversed. 

In the limit of the kinetic control, if $D_A$ is sufficiently large so that 
the inequality in Eq. (\ref{diffusion-control}) is reversed, one has
\begin{equation}
\label{example-kin-cont}
F_{chem} = - \frac{\pi n_0 R^3}{2 \beta D_B} \kappa \, e^{-\varepsilon_w^{A}} \, 
\left(1-  e^{-\varepsilon_w^{B}}  \right)  \sin^2(\theta_0) \,.
\end{equation}
In this regime the force is proportional to the $R^3$ (i.e., \textit{volume} of 
the colloid), attains the maximal value for $\theta_0 \equiv \pi/2$, (i.e., 
in the case of Janus particles with the catalytic patch occupying 
exactly one half of the surface), and its sign is determined by the sign of 
$\varepsilon_w^{B}$. For positive values of $\varepsilon_w^{B}$, i.e., if 
the products $B$ are repelled from the surface of the colloid, the force 
$F_{chem}$ is negative, and is positive in the case of 
negative values of $\varepsilon_w^{B}$. In the limit $D_B \to \infty$ 
the force vanishes $\sim 1/D_B$.

\subsection{\label{sec:stall}Stall force $F_{ext}$ and self-propulsion velocity $V$ 
of a force-free colloid}

Next we turn to the analysis of the stall force $F_{ext}$ (Eqs. (\ref{stall1}) 
and (\ref{stall2})) and, equivalently (see Eq. (\ref{stall})), of the 
self-propulsion velocity $V$ of a force-free colloid. For the triangular well 
potentials, the exact expressions for the integrals $J_j$ (Eqs. (\ref{JA}), 
(\ref{JB}), and (\ref{JC})) entering the definition of $F_{ext}$ and $V$ are 
presented in \ref{append2}. Analogous to the previous section, 
here we focus on the leading order term in $\lambda_j/R$, which is given by
\begin{equation}
\label{expression1}
J_j = {\cal J}_j \frac{\lambda_j^2}{\beta}\,,
\end{equation}
with
\begin{eqnarray}
\label{expression2}
{\cal J}_j  = &-& \frac{3 e^{-\varepsilon_w^{j}}}{2 \Delta\varepsilon_j^2} 
\Bigg(e^{\Delta\varepsilon_j} \left(1+\left(\Delta\varepsilon_j-1\right)^2\right) 
- 2\Bigg)   + \nonumber\\
&+&  \frac{3}{2 (\varepsilon_m^{j})^2} \Bigg((z_j-1)^2 \left(e^{-\varepsilon_m^{j}} 
- 1\right) + \nonumber\\
&+&
\left(\varepsilon_m^{j} 
- z_j+1\right)^2 e^{-\varepsilon_m^{j}} -\left(1+ z_j \varepsilon_m^{j} 
- z_j\right)^2\Bigg) \,.
\end{eqnarray}
We note that, in contrast to $I_j$, $J_j$ depends on the depth 
$\varepsilon_m^{j}$ of the potential well even in leading order. Mathematically, 
this is a consequence of the additional factor $(1+ (R/r)^3/2 - 3 R/(2 r))$ in 
the integrands in Eqs. (\ref{JA}), (\ref{JB}), and (\ref{JC}). 
% , which 
% in turn stems from the coupling of the spatial distributions of the 
% reactive species to the velocity field $\mathbf{u}$. 
Consequently, the dependence 
of ${\cal J}_j$ on the parameters characterizing the interaction potentials (Eqs. 
(\ref{expression1}) and (\ref{expression2})) is more complex. 

\begin{figure}[ht]
\begin{center}
\includegraphics[width=.6\hsize]{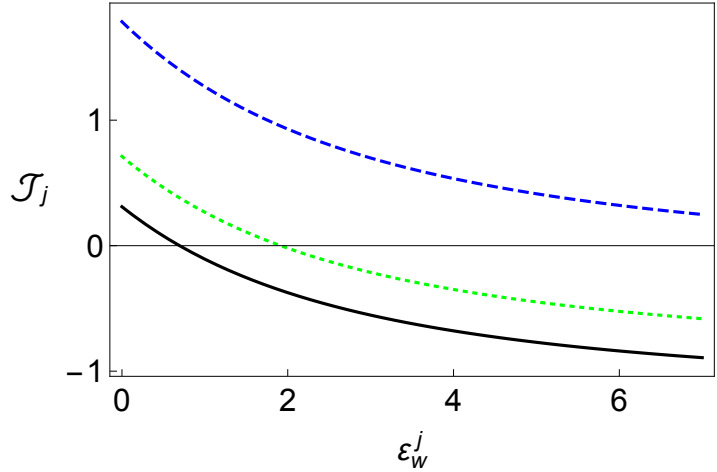} 
\caption{Dimensionless factor ${\cal J}_j$ (Eq. (\ref{expression2})) in 
the integral $J_j$ (Eq. (\ref{expression1})) as a function of 
$\varepsilon_w^{j}$ for $z_j = \Lambda_j/\lambda_j = 2$ (see Fig.  \ref{fig77}) and $\varepsilon_m^{\, j} = - 0.1$ (solid line), 
$\varepsilon_m^{\, j} = - 0.22$ (dotted line), and $\varepsilon_m^{\, j} 
= - 0.5$ (dashed line).}
\label{fig88}
\end{center}
\end{figure}

In Fig. \ref{fig88} we show ${\cal J}_j$ (Eq. (\ref{expression2})) as a 
function of $\varepsilon_w^{j}$ for several values of the depth of the well. 
We observe that even for the smallest value of $\varepsilon_m^{j}$ (here, 
$\varepsilon_m^{j}= - 0.1$), a significant repulsion at the wall 
($\varepsilon_w^{j} \approx 0.6$) is needed in order to have 
${\cal J}_j$ becoming negative. This is in contrast to the behavior exhibited 
by $I_j$, which is negative as soon as $\varepsilon_w^{j}$ is negative. For a 
slightly deeper well ($\varepsilon_m^{j}=-0.22$) a much stronger repulsion 
at the wall ($\varepsilon_w^{j} \approx 2$) is required for turning
${\cal J}_j$ negative. Furthermore, for $\varepsilon_m^{j}=-0.5$, 
${\cal J}_j$ remains positive for $\varepsilon_w^{j}$ as large as 
$\varepsilon_w^{j} \approx 7$. This leads to the conclusion that, while the sign of 
$I_j$ is dominated by the value of the interaction potential at the wall, the sign 
of $J_j$ depends strongly on the depth of the well. Thus, for sufficiently 
small values of $\varepsilon_w^{j}$, one can observe an intriguing 
behavior in that the contributions $\mathbf{F}_{chem}^{j}$ and 
$\mathbf{F}_{ext}^{j}$, respectively, of the molecular species $j$ to the 
chemical and stall force, respectively, 
have the same direction. Due to Eq. (\ref{stall}) this implies that 
the self-propulsion velocity $V_j$ and $F_{chem}^{j}$ have opposite signs. 
For small values of $\varepsilon_w^{j}$ and $|\varepsilon_m^{j}|$, the 
Taylor series in powers of $\varepsilon_w^{j}$ and $\varepsilon_m^{j}$ of the 
right-hand-side of Eq. (\ref{expression2}) can be truncated at first order. In 
this case, one obtains
\begin{equation}
\label{crit}
{\cal J}_j \approx - \frac{z_j (z_j + 1)}{2} \varepsilon_m^{j} - 
\frac{1}{2} \varepsilon_w^{j} \,,
\end{equation}
which, in the case of sufficiently small interaction parameters $\varepsilon_w^{j}$ 
and $|\varepsilon_m^{j}|$,  provides a simple criterion for the sign of ${\cal J}_j$, 
and hence of $F^{j}_{ext}$ and $V_j$.

Furthermore, Eqs. (\ref{F_x}), (\ref{F_xABC}), (\ref{Vas}), (\ref{Vdis}), 
(\ref{stall1}), and (\ref{stall2}) imply that the contribution to $F_{ext}^{j}$ (or, 
equivalently, to $V_j$) due to the interactions with the molecular species $j$
is related to $F_{chem}^{j}$ as
\begin{equation}
\label{rel2}
F^{j}_{ext} =  \left(\frac{\lambda_j}{R}\right)^2  
\frac{{\cal J}_j}{\left(1 - e^{-\varepsilon_w^{j}}\right)} F_{chem}^{j}
\end{equation}
and equivalently,
\begin{equation}
\label{rel1}
V_j = - \frac{1}{6 \pi \mu R} \left(\frac{\lambda_j}{R}\right)^2  
\frac{{\cal J}_j} {\left(1 - e^{-\varepsilon_w^{j}}\right)} F_{chem}^{j}\,.
\end{equation}

\begin{figure}
\begin{center}
\includegraphics[width = .6\hsize]{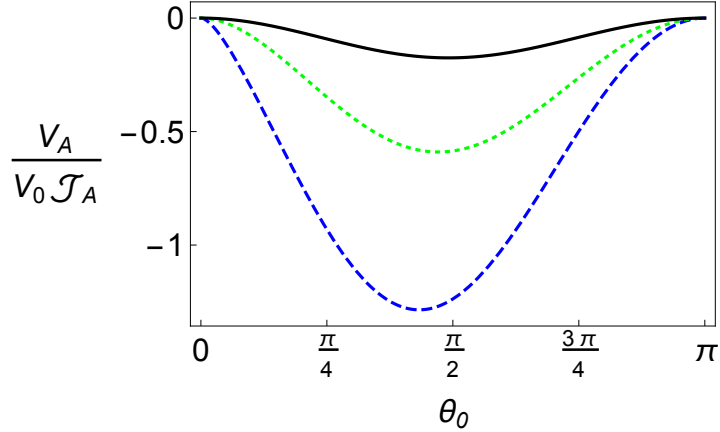}
\caption{Reduced velocity $V_A/(V_0 {\cal J}_A)$ (Eq. (\ref{redvel})) 
%taken with a \SD{minus} sign
 as a function of the catalytic coverage $\theta_0$ for several values of the parameter $D_A \left(\exp\left(\varepsilon_w\right)\right)/(\kappa R)$: $0.1$ (dashed line), $1$ (dotted line),  
 and $5 $ (solid line).
}
\end{center}
\label{FigA}
\end{figure}

Equations (\ref{rel2}) and (\ref{rel1}) lead to various conclusions. 
First, they show that $F_{ext}^{j}$ and $V_j$ are, by a factor 
$(\lambda_j/R)^2$, much smaller than the force contribution $F_{chem}^{j}$. Second, 
the dependence of $F_{ext}^{j}$ (and thus also that of $V_j$) on $\theta_0$, 
$\kappa$, and $D_j$ comes about only via $F_{chem}^{j}$. The factor 
${\cal J}_j/(1 - \exp(- \varepsilon_w^{j}))$ implies that only for 
$\varepsilon_w^{j}$ sufficiently large (ensuring that ${\cal J}_j > 0$)
the stall force contribution $F_{ext}^{j}$ and the velocity contribution $V_j$  
have signs which are opposite and equal, respectively, to $F_{chem}^{j}$, 
which one might expect \textit{a priori} on intuitive grounds. These properties 
are, as explained above, due to the fact that for not too large
$\varepsilon_w^{j}$ the integrals $J_j$, which determine $F_{ext}$ and $V$, 
are dominated by the depth of the well, while the integrals $I_j$, which determine 
$F_{chem}$, are always dominated by $\varepsilon_w^{j}$.

%\begin{figure}
%\centering
%\hfill
%\subfigure{\includegraphics[width = .6 \textwidth]{Fig4a.eps}}
%\hfill
%\subfigure{\includegraphics[width = .6 \textwidth]{Fig4b.eps}}
%\caption{Reduced velocity $V_A/V_0$ (Eq. (\ref{redvel})) 
%of a Janus colloid ($\theta_0 = \pi/2$) as a function 
%of $\varepsilon_w^{A}$ for $z_A=\Lambda_A/\lambda_A=2$ (see Fig. \ref{fig77}) and
%$\varepsilon_m^{A} = - 0.1$ (solid line), $\varepsilon_m^{A} = - 0.22$
%(dotted line), and $\varepsilon_m^{A} = - 0.5$ (dashed line). 
%The scaled velocity depends on the ratio $D_A/(\kappa R)$, which is $2$ 
%in (a) and $0.32$ in (b).}
%\label{fig80}
%\end{figure}

The overall dependence of $V_j$ (or, equivalently, of the contribution $F_{ext}^{j}$ 
to the stall force) on  $\varepsilon_w^{A}$ and $\varepsilon_m^{A}$ requires 
further analysis because, as explained earlier, upon a gradual increase 
of $\varepsilon_{w}^{A}$ the system, which is in the diffusion-limited regime 
for small values of this parameter, eventually crosses over into the  kinetic control 
regime. We illustrate this behavior for $j = A$. The general dependence on 
$\kappa$ and $D_A$ is given by
\begin{equation}
\label{redvel}
V_A  = - V_0 \, 
\left(\frac{\sin^2(\theta_0/2)}{f_{dc}^{(sls)}(\theta_0)} + 
\frac{D_A}{\kappa R} e^{\varepsilon_w^{A}}\right)^{-1}
 {\cal J}_A  \sin^2(\theta_0) \,,  \,\,\, 
 V_0 = \frac{n_0 \lambda_A^2 }{12 \beta \mu R}  \,.
\end{equation}
$V_0$ is independent of $\varepsilon_w^{A}$ and $\varepsilon_m^{A}$, while 
the ratio $D_A/(\kappa R)$
%, $D_A$ and $R$ 
can  be estimated from available experimental data. In particular, for the experimental situation studied in Ref. \cite{Howse2007} one has the ratio $D_A/(\kappa R) \approx 2$.
%\GO{In particular,  
%as 
%$D_A \sim 10^{-9}~\mathrm{m^2 s^{-1}}$ (molecular species in 
%water-like solvents at room temperature), $R \sim 10^{-6}~\mathrm{m}$, 
%and 
%$\kappa^v$ (in units where the concentration is in percentage) is 
%$\kappa^v \sim \times 10^{25}~\mathrm{m^{-2} s^{-1}}$, leading to 
%$D_A/(\kappa R) \simeq 2$ from \cite{Golestanian2012} and as $D_A/(\kappa R) \simeq 0.32$ from \cite{Howse2007}}.

In Fig. 4 we plot the rescaled velocity $ V_A/(V_0 {\cal J}_A)$ with $V_A$ and ${\cal J}_A$ defined in Eqs. (\ref{redvel}) and (\ref{expression2}), respectively, as a function of the catalytic coverage $\theta_0$. 
Interestingly one finds  that the maximum of the absolute value of the self-propulsion velocity is 
attained at coverages close to $\theta_0 = \pi/2$ (i.e., for Janus colloids), only if the parameter $D_A \left(\exp\left(\varepsilon_w\right)\right)/(\kappa R)$ is sufficiently large, which corresponds to the regime of  kinetic control.
For small values of  $D_A \left(\exp\left(\varepsilon_w\right)\right)/(\kappa R)$, one finds that the maximum is shifted to values of $\theta_0$ lower than $\pi/2$, which matches with the observations we made above 
in Subsec. \ref{force_def}.

\begin{figure}
\begin{center}
\includegraphics[width = .6\hsize]{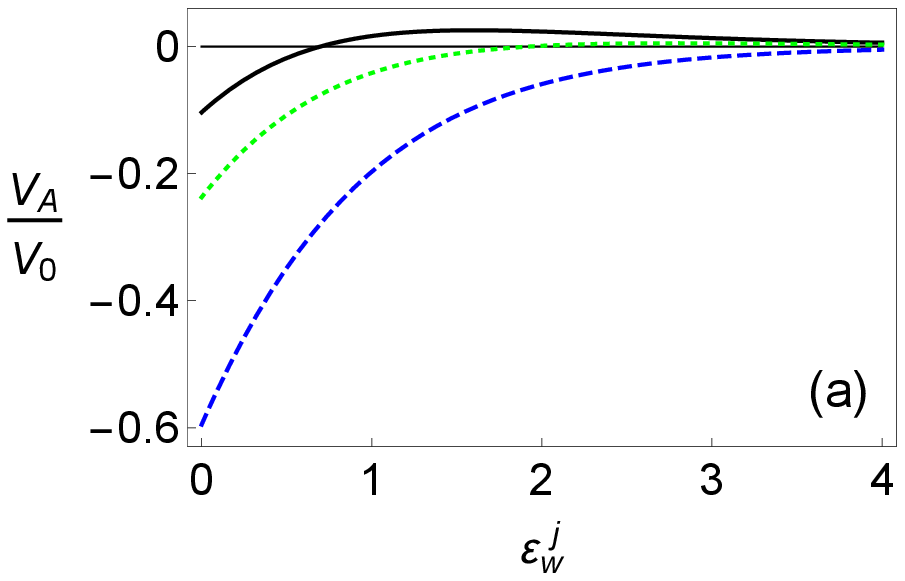}
\includegraphics[width = .6\hsize]{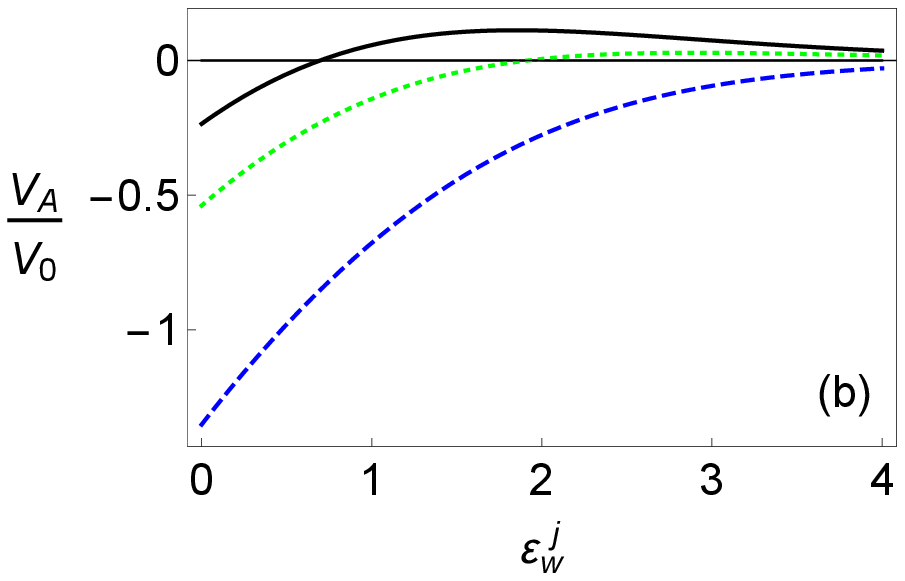}
\caption{Reduced velocity $V_A/V_0$ (Eq. (\ref{redvel})) 
of a Janus colloid ($\theta_0 = \pi/2$) as a function 
of $\varepsilon_w^{A}$ for $z_A=\Lambda_A/\lambda_A=2$ (see Fig. \ref{fig77}) and
$\varepsilon_m^{A} = - 0.1$ (solid line), $\varepsilon_m^{A} = - 0.22$
(dotted line), and $\varepsilon_m^{A} = - 0.5$ (dashed line). 
The scaled velocity depends on the ratio $D_A/(\kappa R)$, which is $2$ 
in (a) and $0.32$ in (b).
}
\end{center}
\label{fig80}
\end{figure}

Further on, we focus on the special case of Janus colloids (i.e., $\theta_0 = \pi/2$ and $f^{(sls)}_{dc}(\pi/2) \approx 0.706$) 
and analyze the behavior of $V_A$ in Eq. (\ref{redvel}) as a function of the parameters of the triangular-well interaction potential. In Fig. 5  we show $V_A/V_0$ as a function of $\varepsilon_w^{A}$ for three different values of 
$\varepsilon_m^{A}$ and for two values of the ratio $D_A/(\kappa R)$ (i.e., two 
values of $R$ with $D_A$ and $\kappa$ kept at fixed values). We observe 
that for sufficiently small values of $\varepsilon_m^{A}$, here 
$\varepsilon_m^{A} = - 0.1$, the contribution $V_A$ to the self-propulsion 
velocity $V$, generated by the interactions with the $A$ molecules, 
exhibits a non-monotonic behavior. For small $\varepsilon_w^{A}$, 
$V_A/V_0$ is negative (hence $V$ and $F_{chem}$ have opposite signs). Upon 
increasing $\varepsilon_w^{A}$ it crosses $0$, attains a peak value at 
$\varepsilon_w^{A} \approx 1.4$ (Fig. 5(a)) and 
$\varepsilon_w^{A} \approx 1.8$ (Fig. 5(b)), respectively, and
decreases to zero for large $\varepsilon_w^{A}$. 
According 
to Eq. (\ref{redvel}), this decrease is exponential in $\varepsilon_w^{A}$, 
which is due to the crossover to the regime of kinetic control. For larger 
value  of $\varepsilon_m^{A}$, (e.g., $\varepsilon_m^{A} = - 0.5$ in Fig. 5), the velocity remains negative 
over the whole interval $\varepsilon_w^{A} \in (0,4)$. In principle, for such values 
of $\varepsilon_m^{A}$, the reduced velocity will also attain positive values and 
will exhibit a non-monotonic behavior, but this behavior is shifted to very 
large values of $\varepsilon_w^{A}$, which can be unphysical. Moreover, the 
peak values of $V_A$ are small. We also observe that the maximal absolute 
values of $V_A/V_0$ are achieved for $\varepsilon_w^{A} = 0$, which signals 
that the most effective self-propulsion (due to interactions with the $A$ molecules)
takes place for potentials which have a deep minimum and no repulsion at the 
wall, beyond the hard-wall interaction. This conclusion remains valid 
if the multi-species interactions are the same for all species, i.e., 
${\cal J}_A={\cal J}_B={\cal J}_C = {\cal J}$, in which case the integral 
${\cal J}$ will factor out (see Eqs. (\ref{last1}) and (\ref{last2}) below). 
Of course, in the general case this may not hold because the contributions 
to $V$ stemming from different species (which appear with different signs) may
compensate each other. We note, as well, that even for ${\cal J}_A={\cal J}_B 
={\cal J}_C = {\cal J}$, the sign of $V$ is not determined by the sign of 
${\cal J}$ but depends also on the relation between the diffusion coefficients 
$D_A$, $D_B$, and $D_C$.

Next we discuss the dependence of the velocity $V$ and of $F_{ext}$ 
(Eqs. (\ref{Vas}), (\ref{Vdis}), and (\ref{stall})) on the size $R$ of the colloid. 
From our results in Eqs. (\ref{Vas}) and (\ref{Vdis}), we find that in leading order 
the self-propulsion velocity of the colloid for a reaction $A + {\rm CP} \to B + {\rm CP}$ 
is given by
\begin{eqnarray}
\label{last1}
 V  = &-& \frac{n_0 D_A}{12 \beta \mu} \frac{\kappa \, e^{-\varepsilon_w^{A}}}
{\left(\kappa \, e^{-\varepsilon_w^{A}} \, R \sin^2(\theta_0/2) + D_A \, 
f_{dc}^{(sls)}(\theta_0)\right)} \, \nonumber\\
 &\times& \left(\frac{\lambda_A^2 {\cal J}_A}{D_A} -  
\frac{\lambda_B^2 {\cal J}_B}{D_B}\right) \sin^2(\theta_0) \,  
f_{dc}^{(sls)}(\theta_0) \,,
\end{eqnarray}
while for a dissociation reaction $A + {\rm CP} \to B + C + {\rm CP}$ it is given by
\begin{eqnarray}
\label{last2}
 V  = & -& \frac{n_0 D_A}{12 \beta \mu} 
\frac{\kappa \, e^{-\varepsilon_w^{A}}}{\left(\kappa 
\, e^{-\varepsilon_w^{A}} \, R \sin^2(\theta_0/2) 
+ D_A \, f_{dc}^{(sls)}(\theta_0)\right)} \, \nonumber\\
 &\times& \left(\frac{\lambda_A^2 {\cal J}_A}{D_A} -  \frac{\lambda_B^2 {\cal J}_B}{D_B} 
-  \frac{\lambda_C^2 {\cal J}_C}{D_C}\right) \sin^2(\theta_0) \,  
f_{dc}^{(sls)}(\theta_0) \,.
\end{eqnarray}
In the diffusion controlled limit (Eq. (\ref{diffusion-control})),  
Eqs. (\ref{last1}) and (\ref{last2}) reduce to
\begin{equation}
\label{last3}
V  = - \frac{n_0 }{3 \beta \mu R}  \, 
\left(\lambda_A^2 {\cal J}_A -  \frac{D_A \lambda_B^2 {\cal J}_B}{D_B}\right) 
\cos^2(\theta_0/2) \,  f_{dc}^{(sls)}(\theta_0)
\end{equation}
and
\begin{equation}
\label{last4}
V  = - \frac{n_0 }{3 \beta \mu R}  \, \left(\lambda_A^2 {\cal J}_A -  \frac{D_A 
\lambda_B^2 {\cal J}_B}{D_B} -  \frac{D_A \lambda_C^2 {\cal J}_C}{D_C}\right) 
\cos^2(\theta_0/2) \,  f_{dc}^{(sls)}(\theta_0) \,,
\end{equation}
respectively. In this limit one has $V \sim 1/R$. Hence, in accordance 
with Eq. (\ref{stall}) and in leading order, the stall force $F_{ext}$ is 
independent of $R$. In the kinetic control limit, Eqs. (\ref{last1}) 
and (\ref{last2}) yield
\begin{equation}
\label{last5}
V  = - \frac{n_0}{12 \beta \mu} \kappa  \,  e^{-\varepsilon_w^{A}} \, 
\left(\frac{\lambda_A^2 {\cal J}_A}{D_A} -  
\frac{\lambda_B^2 {\cal J}_B}{D_B}\right) \sin^2(\theta_0)
\end{equation}
and 
\begin{equation}
\label{last6} 
V  = - \frac{n_0}{12 \beta \mu} \kappa  \, e^{-\varepsilon_w^{A}} 
\left(\frac{\lambda_A^2 {\cal J}_A}{D_A} -  \frac{\lambda_B^2 {\cal J}_B}{D_B}  
-  \frac{\lambda_C^2 {\cal J}_C}{D_C}\right) \sin^2(\theta_0) \,,
\end{equation}
respectively. Therefore, in the limit of kinetic control $V $ is 
\textit{independent} of $R$, and hence $F_{ext} \sim R$. 

The dependence on $R$ predicted by our analysis is consistent with previous 
theoretical \cite{Golestanian2012} arguments and experimental observations 
\cite{Howse2007,Golestanian2012} which tell
% In \cite{Golestanian2012}.
% , which 
% experimentally probed the size dependence of self-propelling 
% catalytically-decorated colloids, 
% with well-defined sizes ranging from $250 nm$ to $5 \mu m$ in radius,
that for $R$ larger than a certain threshold value the velocity $V$ decreases 
$\sim 1/R$ upon increasing
$R$  \cite{Golestanian2012}, while 
for $R$ smaller than this threshold value the velocity saturates at a value 
independent of $R$ \cite{Howse2007}. As follows, this size 
dependence can be understood from the inequality in Eq. (\ref{diffusion-control}), 
which distinguishes between the diffusion-controlled and the kinetically-controlled regimes. 
For fixed physical parameters $\kappa$, $\varepsilon_w^{A}$, 
$\theta_0$, and $D_A$, there is a threshold value of $R$ at which 
Eq. (\ref{diffusion-control}) turns into an equality. For $R$ below this threshold 
value, the system is in the kinetic control regime, in which $V$ is independent 
of $R$ (which is consistent with the observations in Ref. \cite{Howse2007}). 
Conversely, if $R$ exceeds the threshold value upon increasing $R$ further, 
the system gradually crosses over into the diffusion control regime and, 
accordingly, $V \sim 1/R$ (which agrees with the experimental results and  the
theoretical arguments reported in Ref. \cite{Golestanian2012}).

Finally, we compare the dependences on \textit{all} physical parameters of
our
predicted 
self-propulsion velocity with the results obtained in
Ref. \cite{Golestanian2012}
for the particular case of Janus colloids, and
 for a diffusion-reaction model 
 of the 
 hydrogen peroxide dissociation in water, 
 using the concept of the effective Derjaguin length and phoretic slip.  This provides the means 
 for \textit{a posteriori} checks of the accuracy of the self-consistent 
 approximation used for solving the diffusion equations (Sec. \ref{conc_distr}). 
 Before we proceed, we remark that 
 the reaction scheme considered in Ref. \cite{Golestanian2012} is slightly more complicated than the one used in our analysis: 
there, a two-step process has been assumed in which the hydrogen peroxide first forms a complex with the platinum (at a rate $K_1$) and then, at a rate $K_2$, it dissociates into a water and an oxygen molecule.  Our setting becomes identical to the one Ref.  \cite{Golestanian2012}, 
 if we consider the reaction in Eq. (\ref{reaction}) (rather than the one in Eq. (\ref{react_dissoc}), because one of the reaction products is the same as the solvent) and if we set $K_2 = \infty$ in the results of Ref. \cite{Golestanian2012}. With this, our general prediction for the self-propulsion velocity is given by Eq. (\ref{last1}).

First we consider the limit $R \gg D_A/K_1$. Calculating the Derjaguin length 
for the triangular-well potentials, we find that in the present notation
the self-propulsion velocity defined by Eq. (8) in Ref. \cite{Golestanian2012}
is given by
 \begin{equation}
\label{last8}
V  = - 0.1 \frac{n_0 }{\beta \mu R}  \, 
\left(\lambda_A^2 {\cal J}_A -  \frac{D_A \lambda_B^2 {\cal J}_B}{D_B}\right) \,,
\end{equation}
where ${\cal J}_A$ and ${\cal J}_B$ are the velocity integrals defined in Eq. (\ref{expression2}) above. 
In view of Eq. (\ref{diffusion-control}) with $\theta_0 = \pi/2$, this 
limit corresponds to the regime of diffusion control, and hence, 
to our result in Eq. (\ref{last3}) with $\theta_0 = \pi/2$. 
It turns out
that the expression in Eq. (\ref{last8}), obtained from Ref. \cite{Golestanian2012}, 
   and our prediction in Eq. (\ref{last3}) exhibit 
    the same dependences on the diffusion coefficients of $A$ and $B$ molecules, the radius $R$ of the colloid, the viscosity, and, most importantly, the same rather non-trivial dependences on the parameters of the interaction potentials. 
    The two expressions
    differ only slightly with respect to the numerical prefactor: $0.1$ in Eq. (\ref{last8}) 
    and $0.117$ in Eq. (\ref{last3}).  

Further on, within the opposite limit $R \ll D_A/K_1$, which  in our analysis corresponds 
to the limit 
of reaction control, in the present notation Eq. (8) in Ref. \cite{Golestanian2012} is 
given by  
 \begin{equation}
 \label{zr3}
 V = - \frac{n_0}{12 \beta \mu D_A} K_1 \left(\lambda_A^2 {\cal J}_A - \frac{D_A}{D_B} \lambda_B^2 {\cal J}_B\right) \,.
 \end{equation}   
Taking into account $\sin^2(\pi/2) = 1$ and 
identifying $K_1$ with $\kappa \, \exp(-\varepsilon_w^{A})$, we find  perfect agreement 
(even concerning the numerical prefactors) 
between our prediction in Eq. (\ref{last5}) for $\theta_0 = \pi/2$ and the prediction of 
Ref. \cite{Golestanian2012} given by Eq. (\ref{zr3}). 
% \MNP{This is a reassuring result, because in the limit $D_A \to \infty$ the mean-field 
% closure relation is exact rather than an approximation (see the discussion at the end of 
% Sec. \ref{az}) and thus 
% distribution 
% of reactant molecules $n_A$ is spatially homogeneous and therefore Eqs. (\ref{rbcc}) and 
% (\ref{1}) are exactly equivalent.}

\section{\label{concl} Summary and Conclusions}

We have studied theoretically the dynamics of a chemically active colloid, i.e., 
a particle which has parts of its surface decorated by a catalyst promoting a 
chemical reaction in the surrounding solution. We have focused on the steady 
state stall force needed to immobilize it or, equivalently, its self-phoretic 
velocity if it is free to move. As a model system we have analyzed
a spherical particle with a catalytic patch (CP) forming a spherical cap 
with opening angle $\theta_0$. The particle is immersed in an 
unbounded Newtonian liquid solution, initially containing solvent ($S$) 
and reactant ($A$) molecules. We have considered two types of 
catalytically-activated reactions in which the $A$ molecules enter upon approaching 
the CP: either the simple conversion into product molecules $B$, i.e., 
$A + {\rm CP} \to B + {\rm CP}$, or a dissociation reaction $A + {\rm CP} \to B + C + {\rm CP}$. An 
example of the latter is the platinum catalyzed dissociation of hydrogen 
peroxide into oxygen and water, which is often employed in 
experimental studies of chemically-induced self-propulsion. All molecular 
species ($S$, $A$, $B$, $C$) exhibit interactions with the surface of the 
colloid (in addition to entering into the chemical reactions), which in 
general differ from species to species. Furthermore the reactant molecules and the 
product molecules of various species diffuse in the solution with, in general
distinct, diffusion coefficients $D_A$, $D_B$, and $D_C$.

In order to derive the equations governing the steady-state dynamics, we 
have employed the framework of classical linear non-equilibrium thermodynamics 
\cite{Mazur_book}. The number densities of the 
reactant and product molecules are taken to be sufficiently small so 
that they can be considered as an ideal-gas and that the Reynolds number 
and the corresponding P{\'e}clet number are very small (which is valid for 
typical experimental setups). On this basis we have arrived (Sec. 
\ref{conc_distr}) at a Smoluchowski reaction-diffusion 
problem with external interaction potentials (relative to the solvent one) 
$W_k$, $k = A,B,C$ for each species, and a Stokes flow problem with body 
forces determined by the distributions of molecular species and the 
potentials $W_k$. By generalizing a self-consistent 
approximation, developed originally in Ref. \cite{shoup2} in order to reduce 
the boundary conditions to the simple form of a ``constant flux'' (Sec. 
\ref{conc_distr}.B), we have solved the reaction-diffusion equations via separation 
of variables in spherical coordinates. The steady-state density distributions 
$n_k(r,\theta)$ of the various molecular species around the colloid are thus 
obtained in the typical form of series involving products between certain radial 
functions and Legendre polynomials. Interpreting the results within the chemical 
kinetics framework allowed us to calculate an effective reaction rate $K_{eff}$ 
(Eq. (37)) and an effective steric factor $f_{dc}(\theta_0)$ (Eq. (42)) 
for particles exhibiting non-spherical chemical reactivity. This effective 
reaction rate $K_{eff}$ satisfies a generalized form (Eq. (39)) of the 
well known ``Collins and Kimball relation''.

The knowledge of the steady-state density distributions $n_k(r,\theta)$ 
has allowed for a straightforward calculation of the general expression 
for the force $\mathbf{F}_{chem}$ (Eq. (\ref{F_x})) experienced by the 
colloid due the interactions with the molecules in solution (i.e., species $A$, 
$B$, and $C$). This general expression is valid for arbitrary interaction 
potentials and diffusion coefficients. These density distributions correspond 
to an out-of-equilibrium state which is maintained by the chemical 
reaction. Since $\mathbf{F}_{chem}$ is only one of the contributions to the 
total force on the colloid (which includes also at least a hydrodynamic contribution), 
it is not directly measurable. But it can be accessed indirectly, e.g., 
by marking the molecular species and measuring their steady state 
distributions around the colloid with the chemical reaction 
switched on and off, respectively. This general expression allows one to infer various 
implications (Sec. \ref{force_def}) such as the following: (i) As a 
function of the size $\theta_0$ of the catalyst patch, $F_{chem}$ is 
non-monotonic and exhibits a maximum magnitude for an angle $\theta_0$ 
which in general differs from $\pi/2$.  (ii) The direction of 
$\mathbf{F}_{chem}$ depends on the relative magnitude of the various diffusion 
coefficients.

The use of the reciprocal theorem \cite{KiKa91,Teubner1982} facilitates 
to express the main quantities of interest, i.e., the stall force 
$F_{ext}$ and the self-propulsion velocity $V$ (which are both directly 
measurable), solely in terms of the contribution $n = 1$ of the series 
representations of the densities $n_k(r,\theta)$ (Eqs. (\ref{stall}) and 
(\ref{Vas})). The direct calculations of the hydrodynamic force 
exerted on the colloid (via determining the flow of the solution) turned out to 
be a rather challenging problem. Since this is not playing a 
crucial role in the present study, we leave this issue to future studies. 

The general theoretical framework established here is illustrated by 
choosing so-called triangular-well potentials as examples of the basic 
underlying interaction potentials $W_k(r)$. These potentials are 
characterized for each of the species by a fixed value $W_j(R)$ of the 
potential at the wall (i.e., the colloid surface), by the well depth 
$W_j(R+\lambda_j)$ at $r= R + \lambda_j$, and by being truncated at 
$r= R + \Lambda_j$, $j = A,B,C$. In the physically relevant 
limit $\lambda_j/R \ll 1$, with $\lambda_j/\Lambda_j$ kept fixed, this 
choice of the interaction potentials allows one to obtain analytic expressions 
for $F_{chem}$, $F_{ext}$, and $V$ in leading order in $\lambda_j/R$.

This analysis reveals that the behavior of $F_{chem}$ is dominated by 
the values $W_j(R)$ of the potentials at the wall, while the behavior of 
the self-propulsion velocity $V$ and of the stall force $F_{ext}$ depends 
primarily on the depths $W_j(R+\lambda_j)$ of the potential wells. 
These expressions predict  that within certain parameter ranges  
$F_{chem}$ and $V$ may have opposite signs; the same sign occurs
for sufficiently strong repulsive interactions with the wall. In general, 
it appears that not only the calculation of the values of $F_{chem}$ 
and $V$, but even determining their signs, is a rather complicated 
problem which involves all interaction parameters and all diffusion 
coefficients. Our analysis shows that, similarly to 
$F_{chem}$, $F_{ext}$ and hence $V$ attain, in general, as functions 
of $\theta_0$ their maximal values for $\theta_0$ close to but not exactly 
at $\pi/2$ (Janus particles); only in the limit of kinetic control the 
maximum of these properties is attained exactly at $\theta_0= \pi/2$. Finally, 
we have found that for sufficiently small colloids the propulsion velocity 
is independent of their size. For large colloids, $V$ decreases with 
their size as $1/R$. Our conclusions concerning the 
dependence of $V$ on $R$ are consistent with previous theoretical arguments 
and experimental observations \cite{Howse2007,Golestanian2012}. 
Lastly, we have shown that,
for short-ranged triangular-well potentials, 
 in two limiting situations (the reaction control and the diffusion control regimes)
our general results reproduce the results obtained in  Ref. 
\cite{Golestanian2012} 
by using the concepts of the effective Derjaguin length and phoretic slip.
%, providing another justification to this phenomenological approach.  

%\GO{We close by noting that the approach which we have developed here can be generalized 
%in several directions.  First, one can focus on the 
%reaction scheme considered in Ref. \cite{Golestanian2012} in which 
 %the hydrogen peroxide first forms, at some rate, 
% a complex with the platinum and then, at a different rate, dissociates into a water and 
%an 
%oxygen molecule.
%Second, recent observations \cite{Howse2014} that 
% the self-propulsion velocity is  sensitive 
%to even tiny traces of salt in the solution, makes it especially challenging to analyse 
%the reaction schemes in which 
%oppositely charged ions are formed in the course of the dissociation reaction.
%Last but not least, effects of the rotational diffusion on the self-propulsion velocity 
%remain 
%rather loosely understood and no theoretical predictions
%exist as yet. All these interesting problems can be rather straightforwardly studied 
%within 
%our approach.
%}\newline

There are several directions along which generalizations of the approach developed 
here are feasible and could provide significant physical insight. 
First, one can invoke more involved contributions of the catalyst to the reaction 
scheme, for example the formation of catalyst-bounded activated complexes with finite 
lifetimes, as considered in Ref. \cite{Golestanian2012}. Second, it is plausible 
that the dissociation reaction at the catalyst is accompanied by recombination or 
annihilation of some (or all) of the products in the bulk solution or in a different 
region on the surface of the particles; such systems can be expected to exhibit a 
significantly richer behavior resulting from the interplay between the different 
time- and length-scales of the various reactions. Third, since most of the experimental 
studies of chemically active colloids involve particles suspended in (weak) electrolytes 
and reactions which are expected to produce ionic radicals, it is natural to consider extending the 
approach developed here by adding to the present description the corresponding equations 
which account for charge conservation. 
Last but not least, the influence of rotational 
diffusion of an active particle on the stall force or on the self-propulsion 
velocity can be straightforwardly included into the framework presented here.

\ack

GO and SD acknowledge partial financial support by and the hospitality of the 
Kavli Institute for Theoretical Physics, UCSB, where this study was pursued during 
the Focused Working Group on Self-Propelled Micro-Objects hosted by KITP.

\section*{References} 

\providecommand{\newblock}{}

\appendix

\section{Spatial distribution of the molecular components of the mixture}
\label{derivation} 

We aim at describing the distribution of the molecular components in the mixture within
the standard theory of linear non-equilibrium thermodynamics
\cite{Mazur_book,Landau_book,Seifert2012,Koplik2013}. The solution as a whole is assumed
to behave as an incompressible Newtonian fluid. The reaction will be accounted for via
boundary conditions on the surface of the colloid. Therefore, the description of the 
dynamics of the mixture will involve (i) continuity equations (conservation of mass) for
the mixture and for the $A$ and $B$ species; (ii) linear relations between the conjugate 
thermodynamic fluxes and forces determining the entropy production rate; (iii) the
Navier-Stokes equations  obeyed by the barycentric velocity ${\mathbf{u}} (\mathbf{r})$ of
the mixture (i.e., conservation of momentum). 

The liquids involved in typical experimental studies have a very low 
compressibility; therefore they can be assumed to have a constant mass density 
$\rho(\mathbf{r},t) := \rho = const$. (For a discussion of the more general case 
in which this assumption is softened one can consult, e.g., Ref. \cite{Graaf}.)  
In terms of the number densities $n_{A,B,S}$ and the molecular masses, and accounting 
for the fixed number densities of $A$ and $S$ molecules far from the colloid, this can be 
expressed as 
\begin{eqnarray}
\label{mass_density}
\rho(\mathbf{r},t) = \rho &=& n_A(\mathbf{r},t)\, m_A + n_B(\mathbf{r},t)\, m_B +
n_S(\mathbf{r},t) \, m_S \nonumber\\
&=&  n_0^{(A)} \,m_A + n_0^{(S)} \, m_S\,
\end{eqnarray}
where the rhs corresponds to $\rho(\mathbf{r},t = 0)$. The local number density of the
solvent can be determined once $\rho$ and the number densities of the $A$ and $B$ 
species are known.

Considering total mass conservation in a small volume element $\delta V$ located at
$\mathbf{r}$ (which, however, is sufficiently large to contain a large number of $A$, $B$,
and $S$ molecules such that their number densities in $\delta V$ are well defined) leads
to the continuity equation for the mixture:
\begin{equation}
 \label{continuity_NS}
\frac{\partial \rho}{\partial t} + \nabla (\rho \mathbf{u}) = 0\,
\stackrel{\rho \,=\, const}{\Longrightarrow} \nabla \mathbf{u} = 0\,,
\end{equation}
where $\rho \mathbf{u}$ is the mass current density or the momentum density; $\mathbf{u}$
is the barycentric velocity, i.e., the momentum density divided by the mass density.

Assuming local equilibrium, the free energy $F(T,V,N_A,N_B,N_S)$ of the isothermal fluid 
mixture in contact with the colloidal particle can be written in terms of a local free
energy density (per volume) 
$f(T,n_A := \delta N_A/\delta V,n_B := \delta N_B/\delta V,n_S := \delta N_S/\delta V)$ as 
$F = \int_{\cal V} \,dV f$, where 
\begin{equation}
\label{free_energ} 
df = -s dT + \tilde\mu_A dn_A + \tilde\mu_B dn_B + \tilde\mu_S dn_S\,.
\end{equation}
In the expression above $s: = \delta S/\delta V$ is the local entropy density, 
$\delta N_{A\,(B,S)}$ is the number of $A$ ($B$, $S$) molecules, respectively, in the 
volume $\delta V$  and the interactions with the colloid have been accounted for by 
the modified chemical potentials
\cite{Mazur_book}
\begin{equation}
\label{gen_chem_pot}
\tilde \mu_{j}:= \mu_{j} + \Phi_{j}\,,~~j = A, B, S \,.
\end{equation}
Note that while the modified chemical potentials $\tilde \mu_{j}$ are taken to be constant
over the small volume element $\delta V$, they vary spatially. This is the case because
the interaction potentials $\Phi_{j}$ vary spatially and the chemical potentials
$\mu_{j}$ (in the absence of interactions) vary with position due to the fluid mixture
being in contact with the source (sink) for the $B$ ($A$) species at the colloid surface
and with the sink (source) for the $B$ ($A$) species far away from the colloid surface so
that the system is out of thermodynamic equilibrium. 

The incompressibility condition in Eq. (\ref{mass_density}) implies that only two of
the densities $n_A$, $n_B$, and $n_S$ are independent. In order to connect
straightforwardly to the continuity equations (i.e., mass conservation) in $\delta V$
for each of the molecular species (see the next paragraphs), it is advantageous to change
the variables from number densities to concentrations, i.e., mass fractions, which
are defined as 
%%%%%%%%%%
\begin{eqnarray}
 \label{mass_fraction}
c_j(\mathbf{r},t):&=&\frac{\rho_j(\mathbf{r},t)}{\rho} = \nonumber\\
&=&
\frac{m_j \,n_j(\mathbf{r},t)}
{m_A\,n_A(\mathbf{r},t)+m_B\,n_B(\mathbf{r},t)+m_S\,n_S(\mathbf{r},t)}\,,~
j = A, B, S\,.
\end{eqnarray}
%%%%%%%%%%%
Here $\rho_j(\mathbf{r},t) = m_j n_j(\mathbf{r},t)$ denotes the mass density of species $j
= A, B, S$. The concentrations obey the relation $c_A + c_B + c_S = 1$. With $dn_j =
(\rho/m_j)\,dc_j$ and $dc_S = -(dc_A + dc_B)$, Eq. (\ref{free_energ}) takes the
form
\begin{equation}
\label{free_energ_dens} 
df = -s dT + \bar\mu_A dc_A + \bar\mu_B dc_B\,.
\end{equation}
The chemical potentials $\bar \mu_j$ conjugated to the concentration variables are 
given by
\begin{equation}
\label{bar_gen_chem_pot}
\bar \mu_{j}:= \left(\frac{\tilde \mu_j}{m_j} - \frac{\tilde \mu_S}{m_S} \right) \, \rho 
= \left(\frac{\mu_j}{m_j} - \frac{\mu_S}{m_S}\right) \, \rho + \frac{W_j}{m_j}\,
\rho \,,~~~j = A, B \,.
\end{equation}
Therefore the mixture is described completely by $c_A(\mathbf{r})$ and $c_B(\mathbf{r})$; 
$c_S(\mathbf{r})$ follows from the incompressibility condition in Eq. (\ref{mass_density}).

The local mass conservation for the species $A$ and $B$ in the region \textit{outside} 
the impermeable sphere ($r > R$) implies that the concentrations $c_{A,B}(\mathbf{r},t)$ 
obey the continuity equations \cite{Mazur_book} (see Eq. (\ref{mass_fraction}))
\begin{equation}
\label{cont}
\frac{\partial (\rho c_j)}{\partial t} + \nabla \cdot \mathbf{J}_{j} = 0 \,,~~j =
A, B \,,
\end{equation}
where, in line with the usual non-equilibrium thermodynamics framework 
\cite{Mazur_book,Julicher2009}, the currents $\mathbf{J}_{A,B}$ are split into a
convective part due to the barycentric motion and a ''diffusion`` part $\mathbf{j}_{A,B}$
due to the motion relative to that of the local center of mass:
\begin{equation}
\label{split_current}
\mathbf{J}_j := \rho c_j \mathbf{u} + \mathbf{j}_j \,,~~j = A, B \,,
\end{equation}
where $\mathbf{u}$ is the center of mass (barycentric) velocity of the mixture in the 
volume element $\delta V$.

The advantage of the decomposition in Eq. (\ref{split_current}) is that within the framework 
of linear non-equilibrium thermodynamics the currents ${\bf j}_j$ are the  ''fluxes`` coupled 
to the thermodynamic forces which, in accordance with the expression for the entropy 
production (see Ref. \cite{Mazur_book}), are given by the spatial gradients of the chemical 
potentials $\bar \mu_j$ introduced in Eq. (\ref{bar_gen_chem_pot}). Therefore, each of the 
${\bf j}_j$ can be written as a linear combination of these gradients
\cite{Mazur_book}:
%%%%%%%%%%%%%%%%
\begin{eqnarray}
\label{lin_resp}
{\bf j}_A &=& -\frac{L_{AA}}{T} \nabla \bar \mu_{A} - 
\frac{L_{AB}}{T} \nabla \bar \mu_{B}\,, \nonumber \\
{\bf j}_B &=& - \frac{L_{BB}}{T} \nabla \bar \mu_{B} -
\frac{L_{BA}}{T} \nabla \bar \mu_{A}\,,
\end{eqnarray}
%%%%%%%%%%%%%%%%%%%
where $T$ is the temperature (assumed to be spatially constant) and the couplings $L_{AA} > 0$, 
$L_{BB} > 0$, and $L_{AB} = L_{BA}$ are the so-called Onsager coefficients. 

Furthermore, since we consider low concentrations, so that there are effectively no
interactions between the $A$ and $B$ species, we shall disregard the possibility of a 
direct influence of the dynamics of $B$ on that of $A$. Therefore we disregard the 
possibility that currents of $A$ are directly driven by gradients of the concentration 
and thus of the chemical potential $\bar \mu_{B}$ of the $B$ molecules. Thus we set 
$L_{AB} = 0$. (In the context of active colloids, the conditions under which 
this assumption is expected to hold are discussed in more detail in Ref. \cite{Graaf}.) Although these cross-terms are sometimes important even for dilute solutions
\cite{Mazur_book}, our use of this approximation is motivated by previous reports showing
that in many cases the diffusion coefficients (accounting for currents driven by
concentration gradients, which are the most relevant terms in our system) in ternary
mixtures are such that the cross-terms induced contributions are much smaller than the
ones due to self-diffusion  
\cite{Dunlop1959,Gupta1971,Mangiapia2012,Liu2012,Miller1986,Rard1996}. (But we note 
that recently for a ternary mixture cross-term diffusion coefficients have been reported
which are of the same order of magnitude as the self-diffusion ones \cite{Mialdun2013};
however, this was found for comparable concentrations and thus far from the dilute
solution limit we are considering here.)

Within this approximation one has
%%%%%%%%%%%
\begin{eqnarray}
\label{mass_cur_A}
\mathbf{j}_A = -  \frac{\rho L_{AA}}{T} 
\nabla \left(\frac{\mu_A}{m_A} - \frac{\mu_S}{m_S}\right) 
- \frac{\rho L_{AA}}{T} \nabla \left(\frac{W_A}{m_A}\right) \,
\end{eqnarray}
%%%%%%%%%%%%%%
and the analogous expression for $\mathbf{j}_B$ in which $A$ is replaced by $B$. This
expression shows that the influence of the external interaction potentials has been 
isolated into a single contribution (the last term on the right hand side of Eq.
(\ref{mass_cur_A})). The mass current is now decomposed into a contribution solely due to
the composition of the mixture (i.e., the first term on the right hand side of Eq.
(\ref{mass_cur_A})) and a convective part solely due to the difference $W_A$ ($W_B$)
between the interaction of the $A$($B$) molecules with the colloid and the
interaction, weighted by the corresponding mass ratio $m_{A,B}/m_S$, of the $S$ molecules
with the colloid.

Under the above assumption, the chemical potentials $\hat \mu_A$ and $\hat \mu_B$ with
\begin{equation}
\hat \mu_{A,B} := \frac{\mu_{A,B}}{m_{A,B}} - \frac{\mu_S}{m_S}
\end{equation}
are functions (see Eq. (\ref{free_energ_dens})) of the temperature $T$ (assumed to be
spatially constant for our system) and of the local concentration $c_A$ of the $A$
molecules ($\hat \mu_{A}$) or of the concentration $c_B$ of the $B$ molecules ($\hat \mu_{B}$), respectively. 
% , but not
% of both concentrations simultaneously. 
 Explicit expressions for these dependences would
require knowledge of the free energy density, i.e., a specific model of the mixture.
However, further progress can be made by arguing that, because the concentrations of
$A$ and $B$ molecules are assumed to remain very small at all times and the mixture is
incompressible, the spatial variations of the chemical potential of the solvent
$\mu_S$ can be neglected relative to those of the chemical potentials of the solutes. 
This is so because for an ideal dilute solution the chemical potential of the solute 
is proportional to the logarithm of the solute concentration, while that of the solvent is proportional to 
 the solute concentration (see Ch. IX in Ref. \cite{Landau_book}). This implies that 
$(\nabla \mu_{A,B})_T \sim \frac{\nabla c_{A,B}}{c_{A,B}}$, while $(\nabla \mu_S)_T \sim 
\nabla (c_A + c_B)$, and thus the magnitude of the former is greater than that of the 
latter by a factor $1/c_{A,B}$, which is very large. Therefore, for a dilute solution one 
can approximate $(\nabla \hat \mu_{A,B})_T \simeq  \frac{(\nabla \mu_{A,B})_T}{m_{A,B}} = 
\frac{k_B T}{m_{A,B}} \frac{\nabla c_{A,B}}{c_{A,B}}$ \cite{Landau_book}, where 
$k_B$ is the Boltzmann constant. Accordingly the decomposition in Eq. (\ref{mass_cur_A})
implies that the diffusion (relative motion) part of the particle current can be written
as the superposition of concentration gradients and convective terms (which are due to 
the external fields), for which one can directly formulate the usual expression
\cite{Mazur_book}
\begin{equation}
\label{dif_cur_A}
\mathbf{j}_{A,B} = - D_{A,B} \left[\nabla (\rho c_{A,B}) + \beta \rho c_{A,B} \nabla W_{A,B} \right]\,.
\end{equation}
Here $\rho$ is constant, $\beta = 1/(k_B T)$, and $D_{A,B} = (k_B \,
L_{AA,BB})/(m_{A,B}\, c_{A,B})$ are the heuristically defined diffusion coefficients
of the $A$ and $B$ species, respectively. They are expressed in terms of the \textit{unknown}
Onsager coefficients $L_{AA}$ and $L_{BB}$ and, in general, the partial derivatives of the
chemical potentials $\hat \mu_{A,B}$ with respect to the corresponding concentrations
at constant temperature. We note that the diffusion coefficients $D_{A,B}$ defined above 
and  entering in Eq. (\ref{dif_cur_A}) are the so-called ''collective diffusion`` 
coefficients, which, in general, are different from the single particle ones, introduced in 
Sec. \ref{model}, which are describing the mean-square displacement of the Brownian motion 
of $A$ ($B$) molecules. However, in the dilute limit, which is the case considered in this 
study, the two quantities coincide, thus the use of the same notation.

Using the expression in Eq. (\ref{dif_cur_A}), the total particle current in Eq.
(\ref{split_current}) has the form ($\rho_j = \rho c_j$, Eq. (\ref{mass_fraction}))
\begin{eqnarray}
\label{nondim_JA}
 {\bf J}_j &=& \rho_j {\bf u} - 
D_j \left(\nabla \rho_j + \rho_j \nabla (\beta W_j) \right) \nonumber\\
&=& \frac{D_j \rho }{R} \left(\frac{U_0 R}{D_j} \,c_j {\bf U} - \nabla_\xi 
c_j - c_j \nabla_\xi (\beta W_j) \right)\,,~j=A,\,B\,,
\end{eqnarray}
where $U_0$ denotes a typical velocity scale for the flow of the mixture, 
$\xi = r/R$, and ${\bf U} = {\bf u}/U_0$. In the following we make the additional 
assumption that for the systems of interest the Peclet number 
$\mathrm{Pe} = U_0 R/\min(D_A,\,D_B)$ is much smaller than 1 and thus the effects 
of the barycentric convection are negligible. As previously reported, this is 
indeed the case for catalytically active colloids, for which
typical Pe numbers are $Pe~\lesssim~10^{-2}$ 
\cite{Golestanian2005,Popescu2009,Seifert2012}. With
this final approximation, after inserting Eq. (\ref{nondim_JA}) into the
Eq.~(\ref{cont}), one obtains that the concentrations $c_{A,B}(\mathbf{r},t)$ 
obey the following diffusion equations:
\begin{eqnarray}
\label{conc_diffusion}
 \frac{\partial c_j}{\partial t} 
&=&  D_j \nabla \cdot \left(\nabla c_j + \beta c_j \nabla W_j(r) \right) \nonumber\\
&=&  D_j \nabla \cdot \left[e^{-\beta W_j(r)} \nabla\left(e^{\beta W_j(r)} c_j 
\right)\right]
\,,~j = A,\,B\,.
\end{eqnarray}
Due to $c_j = (m_j/\rho) n_j$ and owing to the assumption of negligible cross-term
diffusion (see Eqs. (\ref{lin_resp}) and (\ref{mass_cur_A})), the dynamics of the 
concentrations of the different species are effectively decoupled. Accordingly, with 
$c_j = (m_j/\rho) n_j$, Eq. (\ref{conc_diffusion}) can be transcribed in terms of the 
number density fields, 
which is a more convenient representation for the system under study (see, c.f., Sec.
\ref{force}), and yields  Eq. (\ref{final_diffusion}).

\section{\label{dissoc_reaction}Catalytically induced dissociation}

Here we present, using our results in Subsec. \ref{products}, the derivation 
of the spatially inhomogeneous, steady state distribution of the product molecules $C$, which emerge in 
the catalytically induced 
dissociation reaction described in Eq. (\ref{react_dissoc}).

As in Subsec. \ref{products},  we suppose that the product 
molecules $C$ of mass $m_C$ diffuse with the diffusion coefficient $D_C$ and interact with 
the colloid via a radially symmetric potential $W_C(r)$ (relative to that of the solvent 
molecules, similar to the definition in Eq. (\ref{def_rel_pot}) for the $A$ and $B$ 
molecules). Accordingly, the time evolution of the local density of the reaction product 
$C$ obeys the differential equation
\begin{eqnarray}
\label{gen11}
0 &=& \frac{D_C}{r^2} 
\frac{\partial}{\partial r}
\left(r^2 \frac{\partial n_C}{\partial r}\right) + \frac{\beta D_C}{r^2}
\frac{\partial}{\partial r} \left(r^2 n_C \frac{d W_C}{d r}\right) + \nonumber\\
&+& \frac{D_C}{r^2 (\sin \theta)} \frac{\partial }{\partial \theta} \left((\sin \theta)
\frac{\partial n_C}{\partial \theta}\right) \,,
\end{eqnarray}
which has the same form as Eqs. (\ref{gen}) and (\ref{gen1}). Similarly to Eq. (\ref{gen1}),
which describes the time evolution of the reaction product $B$, Eq. (\ref{gen11}) is to be 
solved subject to a sink boundary condition at macroscopic distances from the colloid (as 
in Eq. (\ref{infinity1})) and subject to the mixed boundary conditions imposed at
the surface of the colloid: the zero current condition at the inert part of the surface
and the reactive boundary condition at the catalytically active patch (as in Eq. (\ref{6})
with $B$ replaced by $C$).

According to the assumptions of the present model the dynamics of $C$ is decoupled from
that of $B$. (Interactions between $B$ and $C$ and between $A$ and $B$ as well as $A$ and
$C$ would change this.) Therefore all results for the $B$ species obtained in the previous
subsection can be simply transcribed to the present case with the replacements
$B \to C$ and $j_n \to h_n$ (see below).

In line with Eq. (\ref{sol1}), the steady state distribution of the product $C$ has
the form
\begin{equation}
\label{sol3}
 n_C(r,\theta) = n_0 \, e^{- \beta W_C(r)} \,
\sum_{n = 0}^{\infty} \gamma_n \, h_n(r) \, P_n\left(\cos \theta\right) \,,
\end{equation}
where the functions $h_n(r)$ are those solutions of (compare Eq. (\ref{fp1}))
\begin{equation}
\label{fp4}
 h''_n(r) + \left(\frac{2}{r} - \beta W'_C(r) \right) h'_n(r) -
\frac{n (n + 1)}{r^2} h_n(r) = 0\,,
\end{equation}
which vanish for $r \to \infty$ and are normalized such that $h_n(r = R) = 1$. The 
dimensionless coefficients $\gamma_n$ are given by (see Eq. (\ref{gensolute}))
\begin{eqnarray}
\label{cn}
\gamma_n &=& - \frac{Q \,  e^{\beta W_C(R)}}{2 \, D_C \, n_0} \, 
\frac{\phi_n(\theta_0)}{h'_n(R)} \nonumber\\
 &=& - \frac{K_{eff}}{4 \, \pi \, D_C \, R} \, 
\frac{e^{\beta W_C(R)}}{R \, h'_n(R)} \, 
\frac{\phi_n(\theta_0)}{\phi_0(\theta_0)} \,,~n \geq 0\,.
\end{eqnarray}
As in Eq. (\ref{soluteprofile2}), in the diffusion-controlled limit one has
\begin{equation}
\label{soluteprofile4}
 \gamma_n(K^* \to \infty)  = - \frac{D_A }{D_C} \, \frac{R_D}{R} \, 
\frac{e^{\beta W_C(R)}}{ R \, h'_n(R)} \frac{\phi_n(\theta_0)}{\phi_0(\theta_0)} 
\,  f_{dc}(\theta_0) \,, \,~ n \geq 0\,,
\end{equation}
while in the limit of kinetic control the coefficients $\gamma_n$ are given by 
(Eq. (\ref{gensolute19}))
\begin{eqnarray}
 \gamma_n(D_A \to \infty) = - \frac{K^*}{4 \, \pi \, D_C \, R} \, 
\frac{e^{\beta W_C(R)}}{R \, h'_n(R)} \frac{\phi_n(\theta_0)}{\phi_0(\theta_0)} \,, 
\,\, n \geq 0 \,,
\end{eqnarray}
and, in this limit, are independent of $D_A$.

\section{The generalized reciprocal theorem of Teubner}
\label{reciproc_theorem} 

For an easier understanding, here we include a brief derivation of the 
generalized reciprocal theorem due to Teubner \cite{Teubner1982}, which we use in 
order to determine the velocity of the self-propelled colloid (Eq. 
(\ref{vel_V})). For a detailed discussion, as well as various applications of 
this result, the interested reader is referred  to the original paper 
\cite{Teubner1982} or the textbook by Kim and Karrila \cite{KiKa91}. 

By applying the equivalent of Gauss' theorem for tensor fields \cite{happel_brenner_book} 
and selecting the orientation of the surface elements entering into the surface 
integral to be the one given by the inner normals, for 
arbitrary tensor fields $\hat\mathbf{\Pi}$ and arbitrary vector fields $\mathbf{u}'$ one obtains
\begin{equation}
\fl
\label{rel_aux_1}
\int\limits_{{\cal D}} \nabla \left(\mathbf{u}' \cdot \hat\mathbf{\Pi}\right) 
\,d^3 \mathbf{r} 
= 
- \int \limits_{\partial {\cal D}} \mathbf{u}' \cdot \hat\mathbf{\Pi} \cdot d 
\mathbf{s} - \int \limits_{S_\infty} \mathbf{u}' \cdot \hat\mathbf{\Pi} \cdot 
d\mathbf{s}_\infty 
= - \int \limits_{\partial {\cal D}} \mathbf{u}' \cdot \hat\mathbf{\Pi} \cdot 
d\mathbf{s}
\,.
\end{equation}
${\cal D}$ is that domain in $\mathbb{R}^3$ which on the inner side is bounded by 
a closed surface $\partial {\cal D}$ (such that ${\cal D}$ is the exterior of 
$\partial {\cal D}$) and on the
outer side by the surface $S_\infty$ of an enclosing large sphere with radius 
$R_\infty$, which extends to infinity. The integral over $S_\infty$ vanishes 
for $R_\infty \to \infty$ if $\mathbf{u}'$ and $\hat\mathbf{\Pi}$ decay 
sufficiently rapidly upon increasing the distance from ${\cal D}$; this is the 
case for the hydrodynamic flows we are interested in, for 
which $|\mathbf{u}'(r \gg 1)| \sim 1/r$ or faster. A similar relation follows 
from swapping the primed 
and unprimed fields. In the following, $\mathbf{u}$ and $\hat\mathbf{\Pi}$, 
as well as the corresponding
primed quantities, are taken to be the velocity and pressure fields, respectively, 
entering into the Stokes
equations (Eqs. (\ref{St_eq}) and (\ref{pressure_tensor_Pi})).

In order to prove the proposition in Eq. (\ref{Teubner}) we first consider its 
left hand side which can then be transformed as follows (note that here 
Einstein's convention of summation over repeated indices is used):
\begin{eqnarray}
\label{proof_Teubner}
\fl {\cal A} & := & \mu' \left[\,\int \limits_{\partial {\cal D}} \mathbf{u}' 
\cdot \hat\mathbf{\Pi} \cdot d\mathbf{s} - \int\limits_{{\cal D}} \mathbf{u}' 
\cdot \mathbf{f} \,d^3 \mathbf{r} \right] 
 \stackrel{Eq.\,(\ref{rel_aux_1})}{=} 
\mu' \int\limits_{{\cal D}} \left[- \nabla \cdot \left(\mathbf{u}' 
\cdot \hat\mathbf{\Pi}\right) - \mathbf{u}' \cdot \mathbf{f} \,\right]\,d^3 \mathbf{r} 
\nonumber\\
\fl & = & \mu' \,\int \limits_{{\cal D}}\,\left[ -(\partial_j u'_i) \Pi_{ij} 
- \mathbf{u}' \cdot \left(\nabla \cdot \hat\mathbf{\Pi} + \mathbf{f} \right) \right] 
\,d^3 \mathbf{r}
\ \stackrel{Eq.\,(\ref{St_eq})}{=} 
- \mu' \int \limits_{{\cal D}}\, (\partial_j u'_i) \Pi_{ij} \,d^3 \mathbf{r} \nonumber\\
\fl & \stackrel{i \leftrightarrow j}{=} & 
-\frac{1}{2} \,\mu' \int \limits_{{\cal D}}\, \left[(\partial_j u'_i) \Pi_{ij} + 
(\partial_i u'_j) \Pi_{ji} \right]\,d^3 \mathbf{r} 
\stackrel{\Pi_{ij} = \Pi_{ji}}{=} 
- \frac{1}{2} \int \limits_{{\cal D}} \left[\mu' (\partial_j u'_i + 
\partial_i u'_j) \right] \,\Pi_{ij}\,d^3 \mathbf{r} \nonumber\\
\fl &\stackrel{Eq.\,(\ref{pressure_tensor_Pi})}{=}& 
- \frac{1}{2} \int \limits_{{\cal D}} \Pi'_{ij} \Pi_{ij} \,d^3 \mathbf{r} 
- \frac{1}{2} \int \limits_{{\cal D}} P' \delta_{ij} 
\left[\mu (\partial_i u_j + \partial_j u_i) - P \delta_{ij} \right] \,d^3 \mathbf{r} 
\nonumber\\
\fl &=& 
- \frac{1}{2} \int \limits_{{\cal D}} \hat\mathbf{\Pi'} \,: \, \hat\mathbf{\Pi} 
\,d^3 \mathbf{r} 
- \int \limits_{{\cal D}} \mu P' \,\nabla \, \cdot \, \mathbf{u} \,d^3 \mathbf{r} 
+ \frac{3}{2} \int \limits_{{\cal D}} P' P \,d^3 \mathbf{r} \nonumber\\
\fl & \stackrel{Eq.\,(\ref{St_eq})}{=} &
\frac{1}{2} \int \limits_{{\cal D}} 
\left[- \hat\mathbf{\Pi'} \, : \,\hat\mathbf{\Pi} + 3 P' P \right]\,d^3 \mathbf{r}\,.
\end{eqnarray}

As stated above, Eq. (\ref{rel_aux_1}) also holds after swapping the primed and 
unprimed fields (because they are defined in the same domain and are assumed to 
decay sufficiently rapidly at infinity). Therefore a similar sequence of 
transformations as in Eq. (\ref{proof_Teubner}) can be applied to the rhs of 
Eq. (\ref{Teubner}). Since the last line in Eq. (\ref{proof_Teubner}) is invariant 
with respect to interchanging the primed and unprimed quantities, one concludes 
that the lhs and the rhs are equal, so that Eq. (\ref{Teubner}) holds.

\section{\label{append2} Triangular-well interaction potentials}

In this appendix we summarize the derivations of the results presented in Sec. \ref{disc} 
for
the particular choice of triangular-well interaction 
potentials, which are defined by Eqs. (\ref{<}) and (\ref{I}) and depicted in 
Fig. \ref{fig77}. 
By focusing on the interaction between the colloid and the molecules 
of species $A$ we determine the corresponding Debye radius 
$R_D$ (Eq. (\ref{raddeb})),  the radial functions $g_0(r)$ and $g_1(r)$ 
(Eq. (\ref{fp})), as well as their derivatives at the wall (i.e., at the colloid surface $r = R$). From these 
quantities we obtain the force integral $I_A$ (Eq. (\ref{iab})) and the velocity integral 
$J_A$ (Eq. (\ref{JA})). The corresponding quantities associated with the 
interactions between the colloid and the species $B$ and $C$  are  obtained by 
 simply replacing the  labels: $A \to B$ or $A \to C$. Finally, in the 
leading order in $\lambda_A/R \ll 1$, we determine the expression for the first derivative at the 
wall of the radial functions $g_n(r)$ for arbitrary $n > 1$.  This allows us to 
infer the asymptotic  behavior of the steric factor 
$f_{dc}(\theta_0)$ (Eq. (\ref{ster})) for $\lambda_A/R \ll 1$.

\vspace{0.2in}
{\bf D1: Debye radius} \newline

For triangular-well interaction potentials Eq. (\ref{raddeb}) yields
\begin{eqnarray}
\label{raddeb3}
\fl R_D^{-1} &=& 
\exp\left(\varepsilon_w^{A} + \Delta\varepsilon_A \frac{R}{\lambda_A}\right) 
\int^{R+\lambda_A}_R \frac{d r}{r^2} \exp\left(- \Delta\varepsilon_A \frac{r}
{\lambda_A}\right) + \nonumber\\
\fl &+&\exp\left(\frac{(R+\Lambda_A)}{(\Lambda_A-\lambda_A)} \varepsilon_m^{A}\right) 
\int^{R+\Lambda_A}_{R+\lambda_A} \frac{dr}{r^2} \exp\left(- \frac{\varepsilon_m^{A}}
{(\Lambda_A- \lambda_A)} r\right) + \int_{R +\Lambda_A}^{\infty} \frac{dr}{r^2} \,.
\end{eqnarray}
The integrals in Eq. (\ref{raddeb3}) can be calculated analytically
and lead to
a rather cumbersome combination of exponential integrals. Focusing on the
physically relevant limit $\lambda_A \ll R$ (i.e., $q_A = \lambda_A/R \ll1$) and 
$\Lambda_A \ll R$ (i.e., $Q_A = \Lambda_A/R \ll 1$) while the ratio $z_A = 
\Lambda_A/\lambda_A > 1$ is fixed (see Fig. \ref{fig77}), from Eq. (\ref{raddeb3}) one finds that in 
leading and first sub-leading order in $q_A$ the Debye radius is given by   
\begin{equation}
\label{debyeapp}
 \frac{R_D}{R} = 1 - \frac{\left[\Delta\varepsilon_A - \varepsilon_w^{A} 
e^{\varepsilon_m^{A}} + \varepsilon_m^{A} e^{\varepsilon_w^{A}} - 
\Delta\varepsilon_A \, z_A\, \left(1 + \varepsilon_m^{A} - 
e^{\varepsilon_m^{A}}\right)\right]}{\varepsilon_m^{A} \Delta\varepsilon_A} q_A\,.
\end{equation}
In Eq. (\ref{debyeapp}) the first
sub-leading term can be 
positive or negative, depending on whether the attractive or the repulsive part 
of the interaction potential dominates. In particular, for small values of
$\varepsilon_m^{A}$ and $\varepsilon_w^{A}$, 
the expansion of  the exponentials in 
Eq. (\ref{debyeapp}) in
terms of
 power series up to second order in these parameters 
renders
 \begin{equation}
\label{debyeapp2}
\frac{R_D}{R} = 1 - \frac{\varepsilon_w^{A}  + 
z_A \varepsilon_m^{A}}{2} q_A \,.
\end{equation}
This implies that, for small values of $\varepsilon_w^{A}$ and $\varepsilon_m^{A}$, one has $R_D < R$ 
if $\varepsilon_w^{A} > z_A |\varepsilon_m^{A}|$, i.e., the repulsive part of 
the interaction potential dominates, while one has $R_D > R$ for 
$\varepsilon_w^{A} < z_A |\varepsilon_m^{A}|$, i.e., if the attractive part of the 
interaction potential prevails.

\pagebreak
%\vspace{0.1in} 
{\bf D2: The solution of Eq. (\ref{fp}) for $n=0$ and its derivative at the surface } \newline

Since the potential $W_A(r)$ is a piece-wise continuous function of $r$, 
which has 
different functional forms in the inner ($r \in (R,R+\lambda_A)$), intermediate ($r \in 
(R+\lambda_A,R+ \Lambda_A)$), and outer ($r \in (R+\Lambda_A,\infty)$) regions, 
respectively, we solve Eq. (\ref{fp}) for each region and connect the pieces 
by requiring the continuity of the solution and of its first derivative at the two 
connecting points between the three intervals. In the following we denote the 
solutions of Eq. (\ref{fp}) (for general $n$) in the corresponding intervals 
as $g_n^{<}(r)$, $g_n^{I}(r)$, and $g_n^{>}(r)$ for the inner, intermediate, 
and  outer interval, respectively.

For $n=0$ and with $W_A(r)$ given by Eqs. (\ref{<}), (\ref{I}), 
and (\ref{>}) Eq. (\ref{fp}) leads to
\begin{eqnarray}
\label{g0}
\fl g_0^{<}(r) = C_0^{(1)} F_{1,0}(r) + C_0^{(2)} \,, \qquad
g_0^{I}(r) = C_0^{(3)} f_{1,0}(r) + C_0^{(4)} \,, \qquad
 g_0^{>}(r) = C_0^{(5)} \frac{R}{r} \,,
\end{eqnarray}
where we have introduced
\begin{eqnarray}
\label{F0}
f_{1,0}(r) &=&  - \frac{\varepsilon_m^{A}}{\left(\Lambda_A - \lambda_A\right)} 
{\rm Ei}\left(- \frac{\varepsilon_m^{A}}{\left(\Lambda_A - \lambda_A\right)} r\right) - 
\frac{1}{r} \exp\left(- \frac{\varepsilon_m^{A}}{\left(\Lambda_A - 
\lambda_A\right)} r\right) \,,  \nonumber\\
F_{1,0}(r) &=&  \frac{\Delta\varepsilon_A}{\lambda_A} {\rm Ei}\left(- 
\frac{\Delta\varepsilon_A}{\lambda_A} r\right) + \frac{1}{r}
\exp\left(-  \frac{\Delta\varepsilon_A}{\lambda_A} r\right) \,,  
\end{eqnarray} 
with ${\rm Ei}(\cdot)$ denoting the exponential integral \cite{abram}.

The constants $C_0^{(k)}$, $k = 1,2,\ldots,5$, are determined from the boundary 
and the continuity conditions: $g_0(R) = 1$ with $g_0(r)$ and $g'_0(r)$ 
continuous at $r = R + \lambda_A$ and $r=R + \Lambda_A$. These conditions lead to
\begin{eqnarray}
\fl C_0^{(1)} &=& 
\frac{\left. f'_{1,0}(r)\right|_{r=R+\lambda_A}}{L_0(\lambda_A,\Lambda_A)} 
\,, \quad C_0^{(2)} = 1 - \frac{F_{1,0}(R) \, \left. f'_{1,0}(r)\right|_{r=R+\lambda_A}}
{L_0(\lambda_A,\Lambda_A)} \,, \quad
C_0^{(3)} = \frac{\left. F'_{1,0}(r)\right|_{r=R+\lambda_A}}{L_0(\lambda_A,
\Lambda_A)} \,,\nonumber\\
\fl C_0^{(4)} &=& -  \frac{f_{1,0}(R + \Lambda_A) 
\left. F'_{1,0}(r)\right|_{r=R+\lambda_A}}{L_0(\lambda_A,\Lambda_A)}  - 
\frac{R (1+ Q_A) \left. f'_{1,0}(r)\right|_{r=R+\Lambda_A} 
\left. F'_{1,0}(r)\right|_{r=R+\lambda_A}}{L_0(\lambda_A,\Lambda_A)} \,, 
\nonumber\\
\fl C_0^{(5)} &=& - \frac{R (1+ Q_A)^2 \left. f'_{1,0}(r)\right|_{r=R+\Lambda_A} 
\left. F'_{1,0}(r)\right|_{r=R+\lambda_A}}{L_0(\lambda_A,\Lambda_A)} \,, 
\end{eqnarray}
where $Q_A = \Lambda_A/R$ and
\begin{eqnarray}
\fl &&L_0(\lambda_A,\Lambda_A) = \left(F_{1,0}(R) - 
F_{1,0}(R+ \lambda_A)\right) \left. f'_{1,0}(r)\right|_{r=R+\lambda_A} \nonumber\\
\fl &+& \left(f_{1,0}(R+\lambda_A) - f_{1,0}(R+\Lambda_A) - R (1+Q_A) \left. 
f'_{1,0}(r)\right|_{r=R+\Lambda_A}\right)\left. F'_{1,0}(r)\right|_{r=R+\lambda_A} \,.
\end{eqnarray}
The derivative of $g_0(r)$ at the wall (i.e., $r = R$) is thus given by
\begin{equation}
\label{g0prime}
\fl  g'_0(R) = 
\left. \left(\frac{d}{dr} g_0^{<}(r)\right)\right|_{r=R} 
= \frac{\left. f'_{1,0}\right|_{r=R+\lambda_A} 
\left. F'_{1,0}(r)\right|_{r=R}}{L_0(\lambda_A,\Lambda_A)} 
= -   \frac{ \Delta\varepsilon_A \, \varepsilon_m^{A} \, e^{\Delta\varepsilon_A}}
{R  \, {\cal L}_0}
\end{equation}
where $q_A = \lambda_A/R$, $z_A = Q_A/q_A$, $Q_A = \Lambda_A/R$, and
\begin{eqnarray}
 {\cal L}_0 &=& \frac{\left(\left(e^{\Delta\varepsilon_A}
(1+q_A)^2-1\right)\varepsilon_m^{A}+(z_A-1)\Delta\varepsilon_A\right) q_A}{(1+q_A)^2} 
\nonumber\\
\fl &+& e^{-\varepsilon_m^{A}} 
\frac{\left(q_A + \varepsilon_m^{A} + \left(\varepsilon_m^{A}-1\right) z_A q_A\right) 
\Delta\varepsilon_A}{(1+ z_A q_A)^2}\,.
\end{eqnarray}
For $q_A \ll 1$,  in leading and first sub-leading order Eq. (\ref{g0prime}) renders 
\begin{eqnarray}
\label{g0der}
 g'_0(R) &=& - \frac{e^{\varepsilon_w^{A}}}{R} +
\Bigg[\varepsilon_m^{A} e^{\varepsilon_m^{A}} \left(e^{\Delta\varepsilon_A} 
- 1\right) + \Delta\varepsilon_A \left(1 - e^{\varepsilon_m^{A}}\right) - \nonumber\\
 &-&z_A 
\Delta\varepsilon_A \left(1 - e^{\varepsilon_m^{A}}+ \varepsilon_m^{A}\right)\Bigg]
\frac{e^{\varepsilon_w^{A}}}{\Delta\varepsilon_A \, \varepsilon_m^{A} R}  q_A  \,.
\end{eqnarray}
The second term in Eq. (\ref{g0der}) vanishes for 
$\Delta\varepsilon_A = 0$ as well as for $\varepsilon_m^{A} = 0$.

\vspace{0.2in} 
{\bf D3: The solution of Eq. (\ref{fp}) for $n \geq 1$} \newline

For $n \geq 1$, the solutions $g_n^{<}(r)$, $g_n^{I}(r)$, 
and $g_n^{>}(r)$ of Eq. (\ref{fp}) corresponding to the inner, intermediate, and outer region, 
respectively, are given by
\begin{eqnarray}
\label{g<}
 g_n^{<}(r) &=& C_n^{(1)} F_{1,n}(r) + C_n^{(2)} F_{2,n}(r) \,, \quad
g_n^{I}(r) = C_n^{(3)} f_{1,n}(r) + C_n^{(4)} f_{2,n}(r) \,, \nonumber\\
 g_n^{>}(r) &=& C_n^{(5)} \left(\frac{R}{r}\right)^{n+1} \,,
\end{eqnarray}
where
\begin{eqnarray}
\label{F}
 F_{1,n}(r) &=& r^n \exp\left(- \frac{\Delta\varepsilon_A}{\lambda_A} r\right) 
U\left(n+2, 2 n+2 ; \frac{\Delta\varepsilon_A}{\lambda_A} r\right) \,, \nonumber\\
 F_{2,n}(r) &=& r^n \exp\left(- \frac{\Delta\varepsilon_A}{\lambda_A} r\right) 
M\left(n+2, 2 n+2 ; \frac{\Delta\varepsilon_A}{\lambda_A} r\right) \,, \nonumber\\
 f_{1,n}(r) &=& r^n U\left(n, 2 n + 2 ; -\frac{\varepsilon_m^{A}}{\left(\Lambda_A - 
\lambda_A\right)} r\right) \,, \nonumber\\
 f_{2,n}(r) &=& r^n M\left(n, 2 n + 2 ; -\frac{\varepsilon_m^{A}}{\left(\Lambda_A - 
\lambda_A\right)} r\right) \,,
\end{eqnarray}
with $M(\cdot)$ and $U(\cdot)$ as Kummer's and Tricomi's hypergeometric 
function, respectively \cite{abram}.

The coefficients $C_n^{(k)}$ are spatially constant and determined by the boundary and the continuity 
conditions (similar to  the case $n = 0$). 
% Taking into account that, by definition, $g^{<}_n(R) \equiv 1$, and also 
% requiring the 
% continuity of the radial functions and of their first derivatives at points 
% $r = R +  \lambda_A$ and $r = R + \Lambda_A$, 
We find that the constants $C_n^{(k)}$ ($n \geq 1$) are given by the following explicit 
(albeit rather lengthy) expressions:
\begin{eqnarray}
\fl C_n^{(1)} &=& \frac{f_{1,n}(R + \lambda_A) \, l_{\Lambda}^{(2)} \, 
\left(L_{\lambda}^{(2)} - l_{\lambda}^{(1)}\right)}{F_{1,n}(R+ \lambda_A) 
f_{1,n}(R + \Lambda_A) L_n(\lambda_A,\Lambda_A)} - \frac{f_{2,n}(R + \lambda_A) 
\, l_{\Lambda}^{(1)}  \, \left(L_{\lambda}^{(2)} - l_{\lambda}^{(2)}\right)}{F_{1,n}(R+ 
\lambda_A) f_{2,n}(R + \Lambda_A) L_n(\lambda_A,\Lambda_A)} \,, \nonumber\\
\fl C_n^{(2)} &=& \frac{f_{2,n}(R + \lambda_A) \, l_{\Lambda}^{(1)}  \, 
\left(L_{\lambda}^{(1)} - l_{\lambda}^{(2)}\right)}{F_{2,n}(R+ \lambda_A) f_{2,n}(R + 
\Lambda_A) L_n(\lambda_A,\Lambda_A)} 
- \frac{f_{1,n}(R + \lambda_A) \, l_{\Lambda}^{(2)} \, \left(L_{\lambda}^{(1)} - 
l_{\lambda}^{(1)}\right) }{F_{2,n}(R+ \lambda_A) f_{1,n}(R + \Lambda_A) L_n(\lambda_A,
\Lambda_A)} \,, \nonumber\\
\fl C_n^{(3)} &=& \frac{l_{\Lambda}^{(2)} \, \left(L_{\lambda}^{(2)} - 
L_{\lambda}^{(1)}\right)}{f_{1,n}(R+\Lambda_A) L_n\left(\lambda_A,\Lambda_A\right)} \,,  \quad
 C_n^{(4)} = - \frac{l_{\Lambda}^{(1)} \, \left(L_{\lambda}^{(2)} - 
L_{\lambda}^{(1)}\right)}{f_{2,n}(R+\Lambda_A) L_n\left(\lambda_A,\Lambda_A\right)} \,, 
\nonumber\\
\fl C_n^{(5)} &=& \frac{\left(1 + Q_A\right)^{n+1} \left(L_{\lambda}^{(2)} - 
L_{\lambda}^{(1)} \right) \left(l_{\Lambda}^{(2)} - l_{\Lambda}^{(1)} \right)}
{L_n\left(\lambda_A,\Lambda_A\right)} \,,
\end{eqnarray}
where
\begin{eqnarray}
 L_n\left(\lambda_A,\Lambda_A\right) &=& \frac{F_1(R)}{F_1(R+\lambda_A)} \Bigg(\frac{ 
f_1(R+\lambda_A)\,  l_{\Lambda}^{(2)} \, \left(L_{\lambda}^{(2)} - 
l_{\lambda}^{(1)}\right) }{ f_1(R+\Lambda_A)} - \nonumber\\
 &-& \frac{f_2(R+\lambda_A) \, l_{\Lambda}^{(1)} \, \left(L_{\lambda}^{(2)} 
- l_{\lambda}^{(2)}\right) }{f_2(R+\Lambda_A)} \Bigg) \nonumber\\
 &-& \frac{F_2(R)}{F_2(R+\lambda_A)}  \Bigg(\frac{f_1(R+\lambda_A) \, l_{\Lambda}^{(2)} 
\,  \left(L_{\lambda}^{(1)} - l_{\lambda}^{(1)}\right) }{ f_1(R+\Lambda_A)} - \nonumber\\
 &-& \frac{ f_2(R+\lambda_A) \, l_{\Lambda}^{(1)} \, \left(L_{\lambda}^{(1)} - 
 l_{\lambda}^{(2)}\right) }{f_2(R+\Lambda_A)} \Bigg)\,,
\end{eqnarray}
while $l_{\lambda}^{(i)}$, $l_{\Lambda}^{(i)}$, and $L_{\lambda}^{(i)}$ ($i=1,2$) 
are given as the logarithmic derivatives of the expressions in Eq. (\ref{F}) :
\begin{eqnarray}
\fl l^{(i)}_{\Lambda} &=& \frac{n+1}{R (1+Q_A)} + \left. \frac{d}{d r} 
\ln\left( f_{i,n}(r)\right)\right|_{r=R + \Lambda_A} \,, \,\, 
 l^{(i)}_{\lambda} = \frac{n+1}{R (1+Q_A)} + \left. \frac{d}{d r} 
\ln\left(f_{i,n}(r)\right)\right|_{r=R + \lambda_A} \,,  \nonumber\\
\fl  L^{(i)}_{\lambda} &=& \frac{n+1}{R (1+Q_A)} + \left. \frac{d}{d r} 
 \ln\left(F_{i,n}(r)\right)\right|_{r=R + \lambda_A} \,. 
\end{eqnarray}

\vspace{0.2in} 
{\bf D4: The radial function $g_1(r)$ and its derivative 
at the surface} \newline

We discuss in more detail the radial function $g_1(r)$ because it enters into 
the force and the velocity integrals (see Sec. \ref{force}) while its derivative  
$g'_1(R)$ at the surface determines the expansion coefficient $a_1$ (Eq. (\ref{an})). 
For $n=1$ Eq. (\ref{F}) yields
\begin{eqnarray}
\label{FF}
\fl F_{1,1}(r) &=& \frac{(\lambda_A)^3}{(\Delta \varepsilon_A)^3 r^2} \exp\left(- \Delta\varepsilon_A \frac{r}{\lambda_A}\right) \,, \,\,\, f_{1,1}(r) = r  \, U\left(1, 4 ; -\frac{\varepsilon_m^{A}}{(z_A - 1)} \frac{r}{\lambda_A}\right) \,, \nonumber\\
\fl F_{2,1}(r) &=& \frac{(\lambda_A)^3}{(\Delta \varepsilon_A)^3 r^2} \left(6 \left(1 -  \exp\left(- \Delta\varepsilon_A \frac{r}{\lambda_A}\right)\right) 
+ \frac{3 \Delta\varepsilon_A \, r}{\lambda_A} \left(\frac{\Delta\varepsilon_A \, r}{\lambda_A} - 2\right)\right) \,, \nonumber\\
\fl  f_{2,1}(r) &=& \frac{(z_A - 2)^3 \, (\lambda_A)^3}{\left(\varepsilon_m^{A}\right)^3 r^2} \Bigg(6 \left(1 - \exp\left(- \frac{\varepsilon_m^{A}}{(z_A - 2)} \frac{r}{\lambda_A}\right) \right) - \nonumber\\
\fl &-& \frac{3 \varepsilon_m^{A} \, r}{(z_A - 2) \, \lambda_A} \left(2 - \frac{\varepsilon_m^{A} \, r}{(z_A - 2) \, \lambda_A}\right)
\Bigg) \,.
\end{eqnarray}

We 
consider the actually interesting limits
$q_A = \lambda_A/R \ll 1$ and $Q_A = \Lambda_A/R \ll 1$ (while $z_A= Q_A/q_A > 1$ is fixed). 
The coefficients $C_{n=1}^{(k)}$, which are spatially constant, contain a non-analytic part, 
diverging -- in the limit $q_A \to 0$ --  either exponentially or as a power-law, 
and a part which is an analytic function of $q_A$. We keep the non-analytic 
contribution, but expand the analytic part into a Taylor series in powers of $q_A$ 
and retain from it only the leading term (independent of $q_A$) and the first sub-leading 
term (linear in $q_A$). This leads to the following approximate expressions for 
$C_{n=1}^{(k)}$:    
\begin{eqnarray}
\label{cs}
\fl C_1^{(1)} &=& \frac{2 (\Delta \varepsilon_A)^2}{R \, q_A^2} 
\exp\left(\varepsilon_w^{A} + \frac{\Delta \varepsilon_A}{q_A}\right)  
\Bigg(1 - \Bigg[e^{-\varepsilon_m^{A}} \Big(z_A \Delta \varepsilon_A -
\varepsilon_w^{A} \Big) - 3 (z_A-1) \Delta\varepsilon_A + 2 \varepsilon_m^{A}  \nonumber\\
\fl &+& 2 e^{\varepsilon_m^{A}} \Big( e^{\Delta \varepsilon_A} 
\varepsilon_m^{A} + z_A \Delta \varepsilon_A - \varepsilon_w^{A}\Big) 
 - (z_A-2)  \varepsilon_m^{A}
\Delta\varepsilon_A\Bigg] \frac{q_A}{\varepsilon_m^{A} \Delta\varepsilon_A }\Bigg) \,, 
\nonumber\\
\fl C_1^{(2)} &=& \frac{\Delta \varepsilon_A}{3 \, R \, q_A} \left(1 - \frac{2 
\left(e^{\varepsilon_{w}^{A}} - 1\right)}{\Delta \varepsilon_A} q_A\right) \,, \,\,
 C_1^{(3)} = - \frac{\varepsilon_m^{A}}{R \, q_A \, (z_A-1)} \Bigg(1 - \nonumber\\
 \fl &-& \frac{2 
\bigg[e^{\varepsilon_m^{A}} \left(e^{\Delta \varepsilon_A}  \varepsilon_m^{A} + z_A 
\Delta \varepsilon_A - \varepsilon_w^{A}\right) - (z_A - 1) \Delta 
\varepsilon_A\bigg]}{\varepsilon_m^{A} \, \Delta \varepsilon_A} q_A
\Bigg) \,, \\
\fl C_1^{(4)} &=& - \frac{\left(\varepsilon_m^{A}\right)^2}{3 \, R \, q_A^2 \, (1-z_A)^2} 
\exp\left(\frac{(1+ q_A z_A) \varepsilon_m^{A}}{(z_A - 1) q_A}\right) 
\nonumber\\
\fl &\times& \left(1 + \frac{\left(z_A \varepsilon_m^{A} + z_A - 1\right) \Delta 
\varepsilon_A - 2 e^{\varepsilon_m^{A}} \left(e^{\Delta \varepsilon_A}  
\varepsilon_m^{A} + z_A \Delta \varepsilon_A - \varepsilon_w^{A}\right)}
{\varepsilon_m^{A} \, \Delta \varepsilon_A} q_A\right) \,, \nonumber\\
\fl C_1^{(5)} &=& 1 + \frac{2 \left(\left(z_A \varepsilon_m^{A} + z_A - 1\right) \Delta 
\varepsilon_A - e^{\varepsilon_m^{A}} \left(e^{\Delta \varepsilon_A}\varepsilon_m^{A} + 
z_A \Delta \varepsilon_A - \varepsilon_w^{A}\right)\right)}{\varepsilon_m^{A} \, \Delta 
\varepsilon_A} \, q_A\,.\nonumber
\end{eqnarray}

Since $g'_1(r=R)$ is given by
\begin{equation}
 g'_1(R) = \left. \left(\frac{d}{dr} g_1^{<}(r)\right)\right|_{r = R} = 
\left. C_1^{(1)} F'_{1,1}(r)\right|_{r = R } + \left. C_1^{(2)} 
F'_{2,1}(r)\right|_{r = R } \,,
\end{equation}
by using Eqs. (\ref{FF}) and (\ref{cs}) we obtain the following approximation in 
the limit $q_A \ll 1$ 
\begin{eqnarray}
\label{g1der}
 g'_1(R) &=& - \frac{2 e^{\varepsilon_w^{A}}}{R}  +  \Bigg[\left(e^{-\varepsilon_m^{A}} + 2 
e^{\varepsilon_m^{A}}\right)\bigg(\varepsilon_m^{A}  \, e^{\Delta\varepsilon_A} + z_A 
\Delta\varepsilon_A - \varepsilon_w^{A}\bigg) - \nonumber\\ 
\fl &-& \bigg((z_A-2) \varepsilon_m^{A} + 3 
(z_A-1)\bigg) \Delta\varepsilon_A \Bigg] \frac{2 e^{\varepsilon_w^{A}}}{\varepsilon_m^{A} \Delta\varepsilon_A R}  q_A \,.
\end{eqnarray}
The leading term in this expansion has been used in Eq. (\ref{lead1}) in Sec. \ref{disc}. 
For $\varepsilon_m^{A} = \varepsilon_w^{A} = 0$ the sub-leading term vanishes.

\vspace{0.2in} 
{\bf D5: Force and velocity integrals} \newline

First we consider the integral $I_A$ in Eq. (\ref{iab}). To this end we write $I_A = I^<_A + 
I^{I}_A$, where the first term and the second term result from integrations over the inner 
(repulsive) and the intermediate (attractive) regions of the pair potential, respectively. (The 
contribution due to the integration over the outer region vanishes because there
$W_A^>(r) \equiv 0$.)

By using the above results, we find that in leading and sub-leading order in $q_A$ 
the integral $I_A^<$ is given by
\begin{eqnarray}
\label{i<}
\fl I^{<}_A &=& \int^{R+ \lambda_A}_R dr \, r^2  \frac{W^{<}_A(r)}{dr} g_1^{<}(r) \exp\left(- \beta W^{<}_A(r)\right) \nonumber\\
\fl &=& -  \frac{e^{-\varepsilon_w^{A}} \left(e^{\Delta\varepsilon_A} - 1\right) R^2}{\beta} 
  \left(1 - \frac{2}{\Delta\varepsilon_A} \left[1+ e^{\varepsilon_w^{A}}- \frac{\Delta\varepsilon_A e^{\Delta\varepsilon_A} \left(e^{\varepsilon_m^{A}}+1\right)}{e^{\Delta\varepsilon_A}-1}\right] q_A\right)\,.
\end{eqnarray}
The leading order term in Eq. (\ref{i<}) is negative for $\Delta\varepsilon_A > 0$ 
and vanishes if $\Delta\varepsilon_A = 0$ (i.e., if there is no interaction), 
and it is proportional to $R^2$. Similarly, in leading and sub-leading order in 
$q_A$ the integral $I_A^I$ is given by
\begin{eqnarray}
\fl I^{I}_A &=& \int^{R+ \Lambda_A}_{R+\lambda_A} dr \, r^2 \frac{W^{I}_A(r)}{dr} g_1^{I}(r) \exp\left(- \beta W^{I}_A(r)\right) \nonumber\\
\fl &=& \frac{\left(e^{- \varepsilon_m^{A}} - 1\right) R^2}{\beta}  \Bigg(1 - \frac{2}{\varepsilon_m^{A} \Delta\varepsilon_A \left(e^{-\varepsilon_m^{A}}-1\right)} \Bigg[\varepsilon_m^{A} e^{\varepsilon_w^{A}} \left(e^{-\varepsilon_m^{A}}-1\right) - 
\left(\varepsilon_w^{A} + z_A \Delta\varepsilon_A\right) e^{\varepsilon_m^{A}} \nonumber\\
\fl &+& \Delta\varepsilon_A \left(z_A-1-\varepsilon_m^{A}\right) e^{-\varepsilon_m^{A}}  + \left(\left(2 z_A - \right) \Delta\varepsilon_A - 1\right) \varepsilon_m^{A}\Bigg] q_A\Bigg)\,.
\end{eqnarray}
The leading term in this expression is positive for all $\varepsilon_m^{A} \leq 0$, 
and, as $I^{<}_A$, it is proportional to $R^2$. 

It follows that the integral $I_A$ is, in leading and sub-leading order in $q_A$, 
given by
\begin{eqnarray}
\label{Ioverall}
 I_A &=& - \frac{\left(1 - e^{- \varepsilon_w^{A}}\right) \, R^2}{\beta} - \Bigg[\left(\varepsilon_w^{A} - z_A \Delta\varepsilon_A\right) \sinh\left(\varepsilon_m^{A}\right) - \nonumber\\
 &-& \varepsilon_m^{A} \left(\sinh\left(\varepsilon_w^{A}\right)- z_A \Delta\varepsilon_A\right)\Bigg] \frac{4 \lambda_A R}{\beta \varepsilon_m^{A} \Delta\varepsilon_A} \,.
\end{eqnarray}
The leading order term in Eq. (\ref{Ioverall}) is used in Eq. (\ref{forceintegrals}) in 
Sec. \ref{disc}. It depends only on $\varepsilon_w^{A}$, it is negative for 
$\varepsilon_w^{A} >0$, and it is proportional to $R^2$. If $\varepsilon_w^{A} = 0$,  
the first term in Eq. (\ref{Ioverall}) vanishes; in this case, in leading order in 
$q_A$ the force integral $I_A$ takes the form 
\begin{equation}
 I_A =  z_A \left(\frac{\sinh\left(\varepsilon_m^{A}\right)}{\varepsilon_m^{A}} 
- 1\right) \frac{4 \lambda_A R}{\beta} \,,
\end{equation}
which implies that it is positive and depends only linearly on $R$.

Next we consider the integral $J_A$ (Eq. (\ref{JA})), which enters into the definition of 
the velocity $V$ (Eqs. (\ref{Vas}) and (\ref{Vdis})) and of the stall force 
$F_{ext}$ (Eqs. (\ref{stall1}) and (\ref{stall2})). As for the integral $I_A$, 
we write $J_A = J_A^< + J_A^{I}$ (the contribution from the outer region 
vanishes in this case, too), where
\begin{eqnarray}
\label{jaless}
\fl J_A^< &=& \int^{R+ \lambda_A}_R dr \, r^2  \frac{W^{<}_A(r)}{dr} g_1^{<}(r) \left(1 + \frac{1}{2} \left(\frac{R}{r}\right)^3 - \frac{3}{2} \frac{R}{r}\right)   \exp\left(- \beta W^{<}_A(r)\right)  \nonumber\\
\fl &=& - \frac{3 e^{-\varepsilon_w^{A}}}{2 \left( \Delta\varepsilon_A\right)^2} \left(e^{\Delta\varepsilon_A} \left(1+\left(\Delta\varepsilon_A-1\right)^2\right) - 2\right) \frac{(\lambda_A)^2}{\beta} + \nonumber\\
\fl &+&
 \Bigg[ 6 e^{-\varepsilon_w^{A}} + 2 \left(3 e^{\Delta\varepsilon_A} \left(1 +\left(\Delta\varepsilon_A - 1\right)^2\right)
- \Delta\varepsilon_A^3 - 6\right)  + \nonumber\\
\fl &+& e^{-\varepsilon_m^{A}} \left(\Delta\varepsilon_A \left(6 - \Delta\varepsilon_A\left(3 - \Delta\varepsilon_A\right)\right) 
- 6\right)
\Bigg]  \frac{(\lambda_A)^2}{2 \beta (\Delta\varepsilon_A)^3} q_A \,,
\end{eqnarray}
and 
\begin{eqnarray}
\label{jainner}
\fl J_A^{I} &=&   \int^{R+ \Lambda_A}_{R+\lambda_A} dr \,  r^2  \frac{W^{I}_A(r)}{dr} g_1^{I}(r) \left(1 + \frac{1}{2} \left(\frac{R}{r}\right)^3 - \frac{3}{2} \frac{R}{r}\right) \exp\left(- \beta W^{I}_A(r)\right)  \nonumber\\
\fl &=&
\frac{3}{2 \left(\varepsilon_m^{A}\right)^2} \Bigg((z_A-1)^2 \left(e^{-\varepsilon_m^{A}} - 1\right) + \left(\varepsilon_m^{A} - z_A+1\right)^2 e^{-\varepsilon_m^{A}} -\left(1+ z_A \varepsilon_m^{A} - z_A\right)^2\Bigg) \frac{(\lambda_A)^2}{\beta} \nonumber\\
\fl &-& \frac{1}{\left(\varepsilon_m^{A}\right)^3 \Delta\varepsilon_A} \Bigg[
6 \varepsilon_m^{A} \left(\left(\varepsilon_m^{A}\right)^2 - 2 (z_A-1) \varepsilon_m^{A} + 2 (z_A - 1)^2  \right) e^{\Delta\varepsilon_A} - \nonumber\\
\fl &-& \Delta\varepsilon_A \left(-  \left(\varepsilon_m^{A}\right)^3 + 3 (z_A-1) \left(\varepsilon_m^{A}\right)^2 - 6 (z_A-1)^2 \varepsilon_m^{A} + 6 (z_A-1)^3\right) e^{- \varepsilon_m^{A}} - \nonumber\\
\fl &-& 6 \varepsilon_m^{A} \left(    2 + 2  z_A (\varepsilon_m^{A} - 2) + (z_A)^2 \left(2 + \varepsilon_m^{A} (\varepsilon_m^{A} - 2)\right)\right) e^{\varepsilon_w^{A}} + \nonumber\\
\fl &+& 6 \left(  \varepsilon_m^{A} - (z_A - 1) \Delta\varepsilon_A \right) \left(2 + 2 z_A (\varepsilon_m^{A} - 2) + (z_A)^2 \left(2 + \varepsilon_m^{A} (\varepsilon_m^{A}- 2)\right)\right) e^{\varepsilon_m^{A}} - \nonumber\\
\fl &-& \bigg(\left(6 + 2 \Delta\varepsilon_A - (z_A)^3 \varepsilon_A\right) \left(\varepsilon_m^{A}\right)^3 - 3 (z_A-1) \left(4 + \Delta\varepsilon_A (2 + (z_A)^2)\right)  \left(\varepsilon_m^{A}\right)^2 + \nonumber\\
\fl &+& 6 (z_A - 1)^2 \left(2 + \Delta\varepsilon_A (2 + z_A)\right) \varepsilon_m^{A}- 18 \Delta\varepsilon_A (z_A-1)^3
\bigg)
\Bigg] \frac{(\lambda_A)^2}{2 \beta} q_A \,.
\end{eqnarray}
One finds that for any $\Delta\varepsilon_A > 0$ the leading order term in 
$J_A^<$ is negative (similarly to the behavior exhibited by $I_A^<$ (Eq. (\ref{i<}))), 
and is independent of $R$. Equations (\ref{jaless}) and (\ref{jainner}) lead to 
\begin{eqnarray}
\label{jar}
 J_A = {\cal J}_A \frac{(\lambda_A)^2}{\beta} + S_A \frac{(\lambda_A)^2}{2 \beta} q_A \,.
\end{eqnarray}
The contribution from the leading term ${\cal J}_A (\lambda_A)^2/\beta$ (see Eq. (\ref{expression2}))
is used in Eq. (\ref{expression1}) in Sec. \ref{disc}. The sub-leading contribution $S_A$ in Eq. (\ref{jar}) collects all terms in square brackets in Eqs. (\ref{jaless}) and (\ref{jainner}).

\vspace{0.2in} 
{\bf D6: Asymptotic behavior of $f_{dc}(\theta_0)$ in the limit 
$q_A \to 0$} \newline

Finally we consider the asymptotic behavior of the steric factor $f_{dc}
(\theta_0)$ in the limit $q_A \to 0$. 
According to Eq. (\ref{ster})
this amounts to the analysis of the behavior 
of the ratio $g'_0(R)/g'_n(R)$. 

This behavior contains the following subtle issue.  Due to Eqs. (\ref{g<}) and (\ref{F}), the radial 
functions $g_n(r)$ are defined explicitly via Kummer's and 
Tricomi's hypergeometric functions $M(\cdot)$ and $U(\cdot)$. In the limit $q_A \to 0$ the 
arguments of both functions become large, which facilitates the derivation of the asymptotic 
behavior. On the other hand, the sub-dominant terms in the  expansions 
of these hypergeometric functions 
for large arguments
also depend on whether the parameters of the 
hypergeometric function are fixed and finite, or are allowed to vary and become 
large. Accordingly, the calculation of the correction terms to the leading order 
behavior poses a mathematically very involved problem. We therefore focus only on 
the leading order behavior of $g'_n(r=R)$ (with $g_n(r)$ defined by Eqs. (\ref{g<}) and (\ref{F})):
\begin{equation}
\label{on_colloid}
g'_n(R) = C_n^{(1)} \left. F'_{1,n}(r)\right)|_{r=R} + C_n^{(2)} 
\left. F'_{2,n}(r)\right)|_{r=R}\,.
\end{equation} 
In the limit $q_A \ll 1$, the leading order behavior of the terms in Eq. 
(\ref{on_colloid}) is
\begin{eqnarray}
 &&C_n^{(1)} \sim \frac{(n+1) \Delta\varepsilon_A }{q_A} 
\left(\frac{\Delta\varepsilon_A}{q_A R}\right)^n \exp\left(\varepsilon_w^{A} + 
\frac{\Delta\varepsilon_A}{q_A}\right) \,, \nonumber\\
&& F'_{1,n}(R) \sim - \frac{q_A}{R \Delta\varepsilon_A} \left(\frac{R q_A}
{\Delta\varepsilon_A}\right)^n \exp\left(- \frac{\Delta\varepsilon_A}{q_A}\right) \,,
\end{eqnarray}
and 
\begin{eqnarray}
 &&C_n^{(2)}\sim \frac{\Gamma(n+1)  \Delta\varepsilon_A}{\Gamma(2 n+2)} 
\left(\frac{\Delta\varepsilon_A}{q_A R}\right)^n\,, \nonumber\\
 &&F'_{2,n}(R) \sim (-1)^{n+1} \frac{\Gamma(2 n +2) q_A}{\Gamma(n) \Delta\varepsilon_A} 
\left(\frac{q_A R}{\Delta\varepsilon_A}\right)^n 
\exp\left(- \frac{\Delta\varepsilon_A}{q_A}\right)\,.
\end{eqnarray}
These expressions imply that the first term in Eq. (\ref{on_colloid}) provides the dominant 
contribution, which is independent of $q_A$, while the second term is proportional 
to  $q_A$. Therefore one finds for the leading behavior
\begin{equation}
g'_n(R) \sim - \frac{(n+1)}{R} e^{\varepsilon_w^{A}} \,,
\end{equation}
which is consistent with Eqs. (\ref{g0der}) and (\ref{g1der}) and implies 
$g_n'(R)/g'_0(R) \sim (n+1)$, so that $f_{dc}(\theta_0) \sim f_{dc}^{sls}(\theta_0)$ 
(see Eq. (\ref{ster1})). \newline

 \section{Notations and definitions}

%\vspace{0.2in} 
%{\bf C1: Latin letters} \newline

\noindent $A$, $B$, $C$ (near Fig. 1): reactant and product molecular species;\newline
$D_j$ (near Fig. 1), $j = A,B,C$: diffusion constant of the molecules of species $j$;\newline
$\mathbf{F}_{chem}$, $\mathbf{F}_{hyd}$ (after Eq. (71); Eq. (72)): contributions to the force on the colloid due 
to the direct interactions with the molecules of various species and due to the flow of the 
solution, respectively;\newline
$\mathbf{F}_{ext}$ (Eqs. (71), (95)-(98)): external force acting on the colloid; stall force, if it corresponds to $\mathbf{V} = 0$;\newline
$F_{ext}^j$, $V_j$, $j = A,B,C$ (Eqs. (122), (123)): that part of the stall force and 
of the velocity, respectively, which is due to the species $j$;\newline
$I_j$, $j = A,B,C$ (Eq. (73)): force integrals in the definition of $F_{chem}$;\newline
$J_j$, $j = A,B,C$ (Eqs. (91) - (94)): velocity integrals in the definition of 
$V$;\newline
${\cal J}_j$, $j = A,B,C$ (Eq. (119)): dimensionless factor in the velocity integral $J_j$ for 
species $j$;\newline
$\mathbf{J}_j$ and $\mathbf{j}_j$, $j = A,B,C$ (Eqs. (10)-(12)): particle current in the laboratory frame 
and relative to the (local) center of mass motion, respectively; \newline
$K = W_0 V_a$ (near Eq. (23)): total number of reaction events per unit time (rate of reaction $W_a$) in a volume $V_a$;\newline
$K_{eff}$ (Eq. (37)): effective reaction rate;\newline
$K^{(0)}_{eff}$ (Eq. (43)): effective reaction rate in the absence of 
interactions between molecules and the colloid;\newline
$K_S$ (after Eq. (38)): Schmoluchowski constant; \newline
$K^*$ (Eq. (41)): effective reaction rate for an elementary reaction act;\newline
$K_{SD}$ (Eq. (40)): Schmoluchowski-Debye constant; \newline
$L_{ij}$, $i,j = A,B,C$ (Eq. (12)): Onsager coefficients; \newline
$P$ (Eqs. (67): hydrostatic pressure (the isotropic part of the stress tensor);\newline
Pe, Re (near Eqs. (16) and (66)): P{\'e}clet and Reynolds numbers, respectively; \newline
$Q$ (Eqs. (\ref{7}) and (\ref{1})): constant negative particle current over the catalytic patch;\newline 
$R$ (Fig. 1): radius of the colloid;\newline
$R_D$ (Eq. (31)): Debye radius;\newline
$S$: solvent molecular species;\newline
$T$: absolute temperature;\newline
$\mathbf{V}$, $V = |\mathbf{V}|$: velocity of the colloid and its magnitude, 
respectively;\newline
${\cal V}$: volume of the system \newline
$W_j = \Phi_j(r) - \frac{m_j}{m_S} \Phi_S(r)$, $j = A,B,C$ (Eq. (2)): effective (relative to 
the solvent) interaction potential of molecules of type $j$ with the colloid;\newline

\noindent $a$ (near Eq. (1)): minimal distance between a molecule of species $A$ and the colloid surface;\newline
$a_n$, $b_n$, $\gamma_n$ (Eqs. (28), (33), (54), (56), (62), (63)): coefficients in the series expansion of the radial functions of 
integer index $n \geq 0$ describing the dependence on $r$ of the number densities 
$n_A$, $n_B$, and $n_C$, respectively;\newline
$c_j$, $j = A,B,C$: concentration (mass fraction) of molecular species $j$;\newline
$f_g$ (Eq. (23)): geometric steric factor;\newline
$f_{dc}$ (Eqs. (39), (42)): effective (diffusion-controlled) steric factor;\newline
$f_{dc}^{(sls)}$ (Eq. (44)): effective steric factor in the absence of interactions between molecules and the colloid;\newline
$\tilde{f}$, $f$ (Eq. (68)): force density on the fluid with (tilde) or without the contribution due 
to the gradient of the interactions between the solvent molecules and the colloid, respectively;\newline
$g_n$, $j_n$, $h_n$ (Eqs. (29), (55), (62)): radial functions of integer index $n \geq 0$ describing the 
dependence on $r$ of the number densities $n_A$, $n_B$, and $n_C$, respectively;\newline
$k_B$: Boltzmann constant ;\newline
$m_j$, $j = A,B,C,S$ (near Fig. 1): molecular mass of species $j$;\newline
$n_j$  (near Eq. (1)), $j = A,B,C$: number density of molecular species $j$;\newline
$n_0$ (near Eq. (1)): bulk value of the number density of species $A$;\newline
$n_0^{(S)}$ (near Eq. (1)): mean number density of solvent molecules;\newline
$p$ (Eq. (1)): probability of the conversion $A \rightarrow B$;\newline
$q_j$, $Q_j$, $z_j$ (after Eq. (104)): dimensionless parameters of the triangular-well 
potential for species $j$;\newline
$\mathbf{r}$, $r = |\mathbf{r}|$ (Fig. 1): position vector and radial coordinate measured from the center 
of the colloid, respectively;\newline
$\tilde r = r - R$ (Fig. 1): distance from the surface of the colloid;\newline
$\mathbf{u}(\mathbf{r})$ (near Eq. (4)): hydrodynamic flow of the solution;\newline
%\newline

\noindent $\hat{\boldsymbol{\Pi}}$ (Eqs. (66), (67)): Newtonian fluid stress tensor; \newline
$\tilde \Phi_j(\tilde r)$ and $\Phi_j(r) = \tilde \Phi_j(r-R)$, $j = A,B,C,S$ (near Eq. (2)): 
interaction potential between a molecule of species $j$ and the colloid;\newline 
$\Omega^{(AB)}$ (Eq. (75)), $\Omega^{(ABC)}$ (Eq. (81)): contribution to parts 
in the force integrals which are independent of $\theta_0$;\newline

\noindent $\beta = 1/(k_B T)$ (Eq. (15)); \newline
$\delta {\cal V}$ (near Eq. (4)): small volume element  in the solution;\newline
$\theta_0$ (Fig. 1): opening polar angle (in spherical coordinates) of the catalytic patch;
\newline
$\theta$ (Fig. 1): polar angle in spherical coordinates;\newline
$\kappa$  (Eqs. (22), (23)): effective velocity defining the particle currents due to reactions at the 
catalytic patch;\newline
$\mu$ (near Eq. (1)): viscosity of solution;\newline
$\mu_j$ and ${\tilde \mu}_j = \mu_j + \Phi_j$, $j = A,B,C,S$ (Eqs. (5), (6)): chemical potential of 
molecular species $j$;\newline
${\hat \mu}_j = ({\mu}_j/m_j)-({\mu}_S/m_S)$, $j = A,B,C$ (Eq. (14)): chemical potential 
relative to that of the solvent;\newline
${\bar \mu}_j = [({\tilde \mu}_j/m_j)-({\tilde \mu}_S/m_S)] \rho$, $j = A,B,C$ (Eq. (9)): 
chemical potential of the molecular species $j$ relative to that of solvent molecules, 
including the contribution from the interactions with the colloid;\newline
$\xi_{1,2}^j$, $\lambda_j$, $\Lambda_j$,$\Delta \epsilon_j$, $\epsilon_w^j$, and
$\epsilon_m^j$, $j = A,B,C$ 
(Eqs. (99)-(103)): parameters of the triangular-well potential for species $j$;\newline
$\rho$ (Eq. (3)): mass density of the solution;\newline
$\sigma_j$, $\epsilon_j$, $j = A,B,C,S$ (after Eq. (2)): parameters of the Lennard-Jones pair 
potential between molecules of species $j$ and molecules composing the colloid;\newline

%\bibliographystyle{unsrt}

%\bibliography{citations_force_active_colloid}

\end{document}